\documentclass[aps,showpacs,preprintnumbers,amsmath,amssymb,twocolumn,superscriptaddress,floatfix,nofootinbib,10pt]{revtex4-1}

\usepackage{amsfonts}
\usepackage{amsmath}
\usepackage{amssymb}
\usepackage{bm,bbm}
\usepackage{booktabs}
\usepackage{chngcntr}
\usepackage{centernot}
\usepackage{colortbl}
\usepackage{comment}
\usepackage{dcolumn}
\usepackage{epstopdf}
\usepackage{float}
\usepackage[bottom]{footmisc}
\usepackage{graphicx}
\usepackage{hyperref}
\usepackage{cleveref}
\usepackage[linesnumbered,lined,boxed,commentsnumbered,ruled,vlined]{algorithm2e}
\usepackage{algpseudocode}
\usepackage{mathrsfs}
\usepackage{mathtools}
\usepackage{multirow}
\usepackage{physics}
\usepackage{ragged2e}
\usepackage{soul}
\usepackage{subfigure}
\usepackage{tabularx}
\usepackage{tikz}
\usepackage{verbatim}
\usepackage{xcolor}
\usepackage{xkcdcolors}
\usepackage{float}
\usepackage{graphicx}
\usepackage{subcaption}
\usepackage{booktabs}

\DeclareCaptionJustification{justified}{\justifying}
\hypersetup{colorlinks=true, citecolor=orange, urlcolor=blue, linkcolor=magenta}

\definecolor{forest}{rgb}{0.0,0.27,0.13}
\definecolor{yellowcream}{rgb}{1.0,1.0,0.7}
\definecolor{verdeacqua}{rgb}{0.55,0.83,0.78}
\definecolor{lilla}{rgb}{0.75,0.73,0.85}

\newcommand{\beq}{\begin{equation}}
\newcommand{\eneq}{\end{equation}}

\begin{document}

\title{The hidden structure of innovation networks}

\author{Lorenzo Emer}
\affiliation{Institute of Economics and L'EMbeDS, Scuola Superiore Sant'Anna, P.zza Martiri della Libertà 33, 56127 Pisa (Italy)}
\affiliation{Department of Computer Science, University of Pisa, L.go Bruno Pontecorvo 3, 56126 Pisa (Italy)}
\author{Anna Gallo}
\affiliation{IMT School for Advanced Studies, P.zza San Francesco 19, 55100 Lucca (Italy)}
\affiliation{Center for Study and Research `Enrico Fermi' (CREF), Via Panisperna 89A, 00184 Rome (Italy)}
\affiliation{Institute for Applied Computing `Mauro Picone' (IAC-CNR), Via Madonna del Piano 10, 50019 Sesto Fiorentino (Florence, Italy)}
\affiliation{INdAM-GNAMPA Istituto Nazionale di Alta Matematica `Francesco Severi', P.le Aldo Moro 5, 00185 Rome (Italy)}
\author{Mattia Marzi}
\affiliation{IMT School for Advanced Studies, P.zza San Francesco 19, 55100 Lucca (Italy)}
\affiliation{Lorentz Institute for Theoretical Physics, University of Leiden, Niels Bohrweg 2, 2333 CA Leiden (The Netherlands)}
\affiliation{INdAM-GNAMPA Istituto Nazionale di Alta Matematica `Francesco Severi', P.le Aldo Moro 5, 00185 Rome (Italy)}
\author{Andrea Mina}
\affiliation{Institute of Economics and L'EMbeDS, Scuola Superiore Sant'Anna, P.zza Martiri della Libertà 33, 56127 Pisa (Italy)}
\affiliation{Centre for Business Research, University of Cambridge, Trumpington Street 11-12, CB2 1QA Cambridge (UK)}
\author{Tiziano Squartini}
\email{tiziano.squartini@imtlucca.it}
\affiliation{IMT School for Advanced Studies, P.zza San Francesco 19, 55100 Lucca (Italy)}
\affiliation{INdAM-GNAMPA Istituto Nazionale di Alta Matematica `Francesco Severi', P.le Aldo Moro 5, 00185 Rome (Italy)}
\author{Andrea Vandin}
\affiliation{Institute of Economics and L'EMbeDS, Scuola Superiore Sant'Anna, P.zza Martiri della Libertà 33, 56127 Pisa (Italy)}
\affiliation{DTU Technical University of Denmark, Anker Engelunds Vej 101, 2800 Kongens Lyngby (Denmark)}

\date{\today}

\begin{abstract}
Innovation emerges from complex collaboration patterns - among inventors, firms, or institutions. However, not much is known about the overall mesoscopic structure around which inventive activity self-organizes. Here, we tackle this problem by employing patent data to analyze both individual (\textit{co-inventorship}) and organization (\textit{co-ownership}) networks in three strategic domains (\textit{artificial intelligence}, \textit{biotechnology} and \textit{semiconductors}). We characterize the mesoscale structure (in terms of clusters) of each domain by comparing two alternative methods: a standard baseline - modularity maximization - and one based on the minimization of the Bayesian Information Criterion, within the Stochastic Block Model and its degree-corrected variant. We find that, across sectors, inventor networks are denser and more clustered than organization ones - consistently with the presence of small recurrent teams embedded into broader institutional hierarchies - whereas organization networks have a neater role-based structures, with few bridging firms coordinating the most peripheral ones; still, both are characterized by the presence of local core-periphery structures. We also find that the discovered meso-structures are connected to innovation output. In particular, Lorenz curves of forward citations show a pervasive inequality in technological influence: across sectors and methods, both inventor (especially) and organization networks consistently show high levels of concentration of citations in a few of the discovered clusters. Our results demonstrate that the baseline modularity-based method may not be capable of fully capturing the way collaborations drive the spreading of inventive impact across technological domains. This is due to the presence of local hierarchies that call for the more refined tools of Bayesian inference.
\end{abstract}


\maketitle

\section{Introduction}\label{sec:introduction}

Innovation is an outcome of the complex patterns characterizing the collaborations among inventors, firms and institutions~\cite{powell1996interorganizational,freeman1995national, uzzi1997social} - which affect the way knowledge is created, recombined and diffused. While the detection of communities within scientific collaboration networks has a well-established tradition in the literature~\cite{girvan2002community,newman2006modularity,blondel2008fast}, much less attention has been devoted to studying the mesoscopic structure of \textit{innovation networks}, the architecture of which not only reveals who collaborates with whom but also how knowledge circulates through the innovation system's modular organization~\cite{Acemoglu2016InnovationNetwork}.

\begin{figure*}[t!]
\centering
\includegraphics[width=\linewidth]{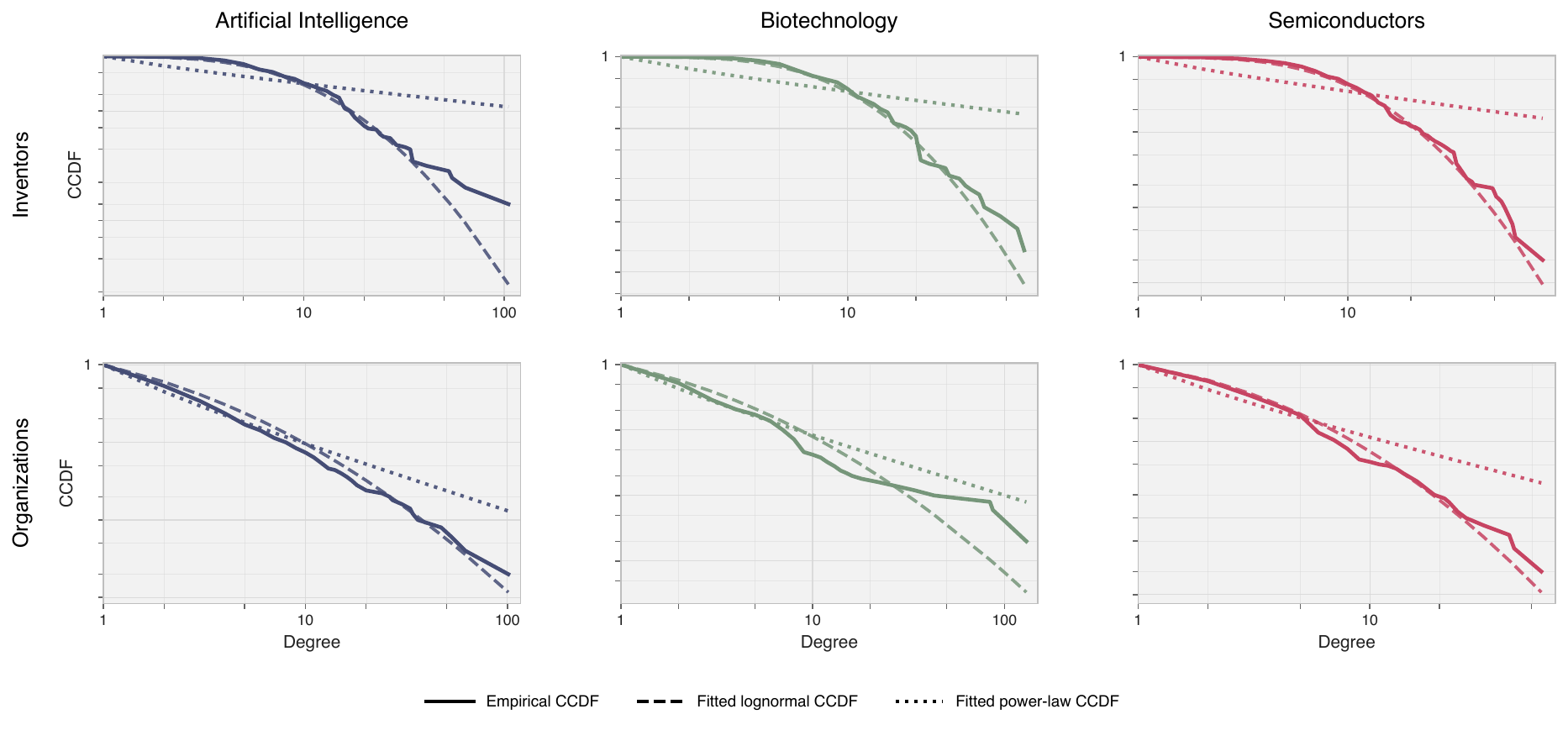}
\caption{\textbf{Node connectivity of innovation networks.} Complementary cumulative degree distributions (CCDDs) of inventor and organization networks in the three, strategic sectors of artificial intelligence, biotechnology and semiconductors (only the top-$500$ actors have been considered). Solid lines show empirical CCDDs, while dashed and dotted lines show the fitted lognormal and power-law CCDDs, respectively: while we cannot reject the hypothesis that inventor networks are compatible with lognormal distributions, for what concerns organization networks, instead, this holds true only for AI and SC.}
\label{fig:degree}
\end{figure*}

Here, we examine the mesoscale of innovation networks in the three strategic sectors of \textit{artificial intelligence} (AI), \textit{biotechnology} (BT) and \textit{semiconductors} (SC) - domains that are recognized as transformative and enabling the technologies that underpin Europe's, as well as the world's, innovation capacity and technological sovereignty~\cite{stoa2021kets,stanford2025setr}. Although each of the aforementioned sectors contributes to the emergence of new technological paradigms, different coordination logics are obeyed: AI advances through software-driven experimentations and data infrastructures~\cite{Chellappa2025,Safitra2024Advancements}, BT evolves through academic-industrial alliances and life-science discoveries~\cite{Powell2005,Oliver2004Biotechnology} and SC rely on capital-intensive, hierarchical production systems~\cite{Kapoor2014Unmasking,Huggins2023,Browning1995}.

In order to compare these heterogeneous `innovation ecosystems', we consider the ORBIS IP patent data across the years $2020$-$2024$ and construct the individual-level (co-inventorship) and organization-level (co-ownership) networks induced by the interconnections among the top-$500$ most influential actors in each sector, ranked by forward citations. This is a standard measure of technological impact~\cite{hall2000market}. The chosen threshold captures the densest and most influential segment of collaborations, while maintaining interpretability and cross-sector comparability\footnote{We also show that the obtained results are robust to alternative cut-offs (top-$300$ and top-$700$).}.

\begin{figure*}[t!]
\centering
\begin{minipage}[t]{\textwidth}
\centering
\includegraphics[width=\textwidth]{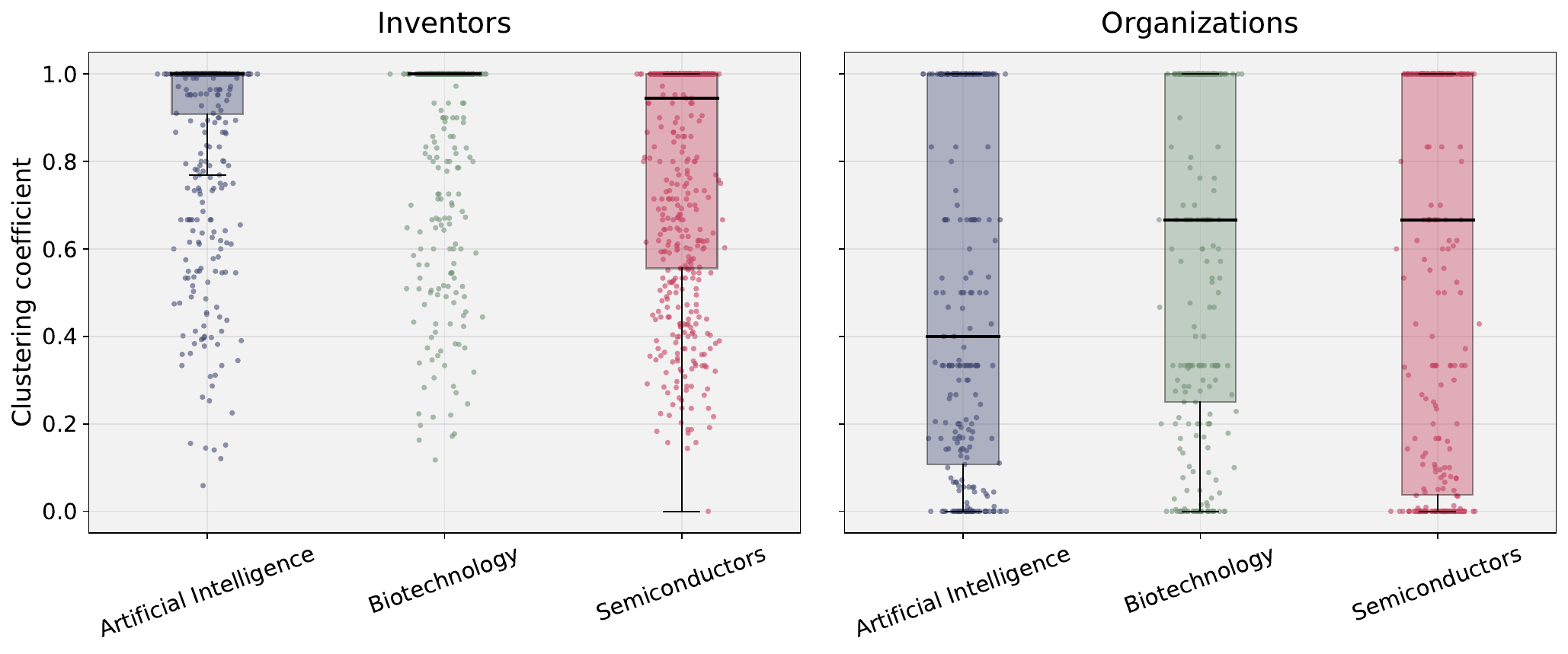}
\end{minipage}
\hfill
\begin{minipage}[t]{\textwidth}
\centering
\includegraphics[width=\textwidth]{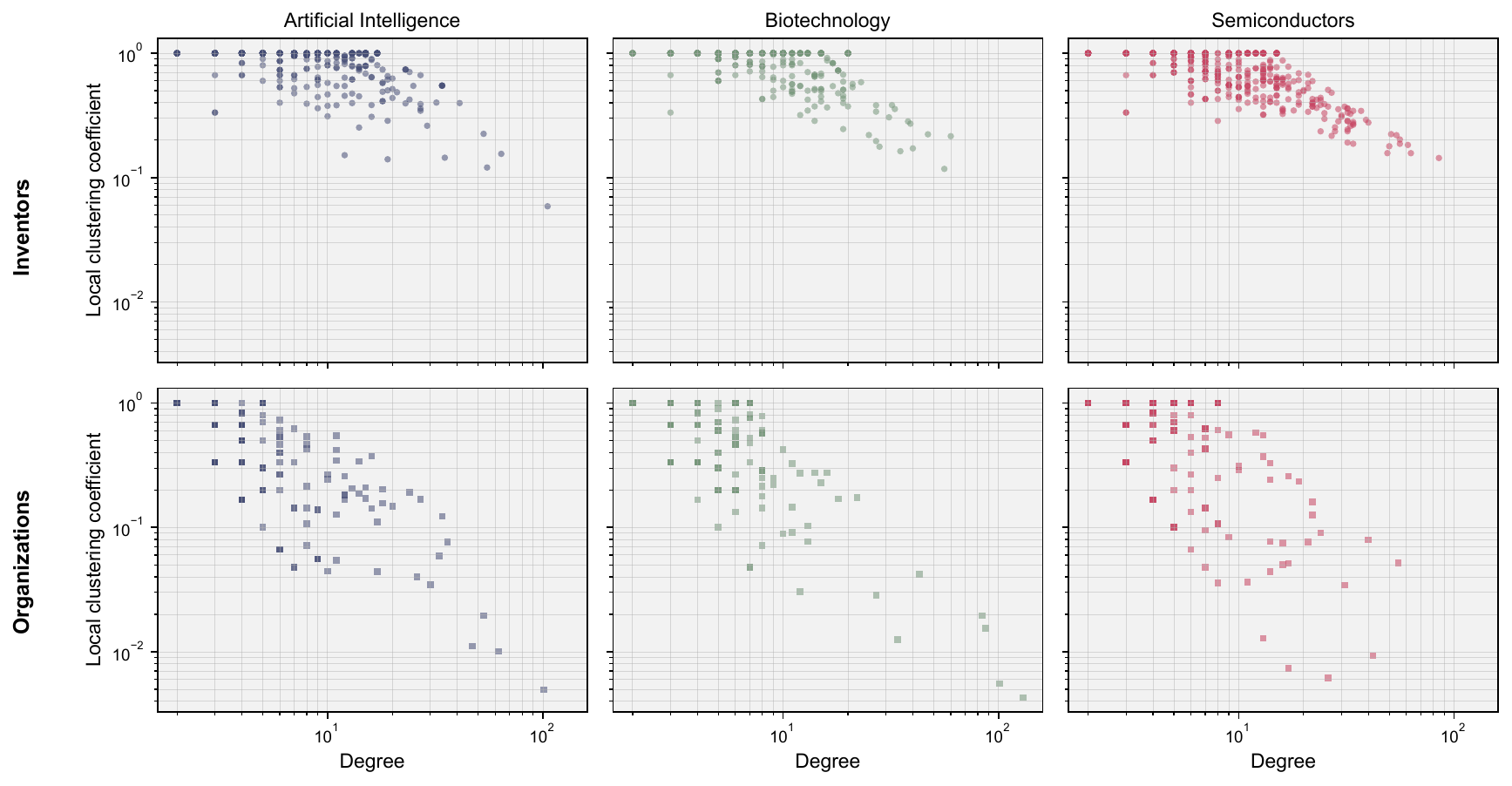}
\end{minipage}
\caption{\textbf{Local cohesion and node connectivity.} Clustering patterns of inventor and organization networks in the three, strategic sectors of artificial intelligence, biotechnology and semiconductors (only the top-$500$ actors have been considered). Top panels report the distribution of the clustering coefficient across sectors and actors: while inventor networks display higher clustering values, consistently with the presence of dense collaboration neighborhoods, organization networks show broader, and generally lower, clustering values reflecting less binding patterns. Bottom panels relate clustering to degree: across sectors, clustering tends to decline with degree, especially in organization networks, a result suggesting that highly connected actors often span multiple neighborhoods rather than belonging to a single, tightly closed, group.}
\label{fig:clustering}
\end{figure*}

Network theory provides a rigorous framework to represent and analyze such systems, where nodes and edges denote agents and their collaborations, respectively~\cite{newman2018networks}. A defining feature of social networks - of which innovation networks are instances - is their tendency to exhibit assortative mixing and a high clustering~\cite{newman2003social}, clues that \textit{i)} individuals with a similar degree tend to connect with each other; \textit{ii)} individuals who collaborate with the same partners tend to form tightly knit groups. These structural signatures reflect the social and institutional embeddedness of knowledge creation~\cite{wasserman1994social,watts1999small}.

Between local interactions and global network properties lies a mesoscale level of organization, characterized by groups of nodes arranged in hierarchies of varying strength that collectively govern information flows, specialization and inequality~\cite{newman2003social,fortunato2016community,khan2017network}. Detecting these structures requires to go beyond the too simplistic notions of `dense' and `central' subgraph: state-of-the-art approaches treat such patterns as emergent properties and model them via probabilistic frameworks like the one of Exponential Random Graphs (ERGs)~\cite{park2004statistical,bianconi2007entropy,fronczak2012exponential,Squartinia}. The adoption of statistically-grounded inference methods has allowed mesoscale structures detection to be rephrased in the jargon of model selection, and carried out by optimizing information criteria like the Minimum Description Length (MDL)~\cite{karrer2011stochastic,fronczak2012exponential,Squartinia,peixoto2019bayesian}, thus allowing genuine regularities - be they clusters, core-periphery structures, nested arrangements - to be distinguished from statistical noise. We apply and systematically compare two alternative approaches to mesoscale structure detection - modularity maximization, a widely used baseline for community detection, and Bayesian Information Criterion (BIC) minimization, within the Stochastic Block Model (SBM) framework and its degree-corrected variant (dcSBM) - across six collaboration networks spanning three technological domains (artificial intelligence, biotechnology and semiconductors) and two organizational levels (inventors and organizations).

We obtain three, major results: \textit{i)} across all domains, inventor (co-inventorship) networks are consistently denser, more assortative and more clustered than organization (co-ownership) networks, indicating the presence of relatively tight and recurrent collaboration circles. In contrast, organization networks exhibit a sparser, locally tree-like topology, consistent with hierarchical and role-differentiated structures; \textit{ii)} when inspecting the mesoscale organization through the lens of the inference-based framework (by minimizing BIC and related variants), additional hierarchical and role-based patterns emerge that are not fully captured by modularity alone: more specifically, both classes of networks are characterized by the presence of local core-periphery structures. We are, thus, led to conclude that (modularity-based) community detection is not capable of capturing these patterns, which are instead recoverable by employing a genuinely inferential framework: focusing exclusively on community structure can, therefore, obscure organizational features that are central to understanding collaboration patterns in innovation networks; \textit{iii)} differences in mesoscale organization are reflected in the distribution of inventive impact.

Inequality in patent influence, measured using Lorenz curves of forward citations, is particularly pronounced in inventor networks across all sectors, indicating that a small fraction of inventors accounts for a disproportionate share of technological influence; inequality is also evident at the organization level - most notably in the AI sector, where forward citations are concentrated in a limited number of dominant corporate-academic alliances - while BT and SC display comparatively more balanced patterns.

\section{Results}\label{sec:results}

The three results discussed in the introduction are treated in detail in each of the following subsections, respectively: in Section~\ref{sec:innomicro} we describe innovation networks using aggregate, network-level metrics and highlight systematic differences between inventor- and organization-level collaborations; in Section~\ref{sec:innomeso} we investigate the mesoscale organization of these networks by applying two alternative approaches, namely modularity maximization and BIC minimization, within the SBM framework and its degree-corrected variant; in Section~\ref{sec:innoimpact} we relate the detected structures to innovation outcomes, assessing how technological impact (measured through citations) is distributed across clusters.

\subsection{Innovation networks at the macroscale}\label{sec:innomicro}

Across the three domains, inventor networks are consistently denser than organization networks: while the density of the former reads $\rho^i_\text{AI}\simeq0.018$, $\rho^i_\text{BT}\simeq0.018$ and $\rho^i_\text{SC}\simeq0.022$, the density of the latter reads $\rho^o_\text{AI}\simeq0.007$, $\rho^o_\text{BT}\simeq0.007$ and $\rho^o_\text{SC}\simeq0.007$; besides, the degree distributions of all networks are heavy-tailed (see Fig.~\ref{fig:degree}), a feature indicating the coexistence of highly connected nodes with many, poorly connected ones - and consistent with the bursty, heterogeneous dynamics typical of human systems~\cite{Barabasi2005}. Finally, the average clustering coefficient of inventor networks is higher than that of organization networks (see Fig.~\ref{fig:clustering}).

\begin{figure}[t!]
\centering
\includegraphics[width=\linewidth]{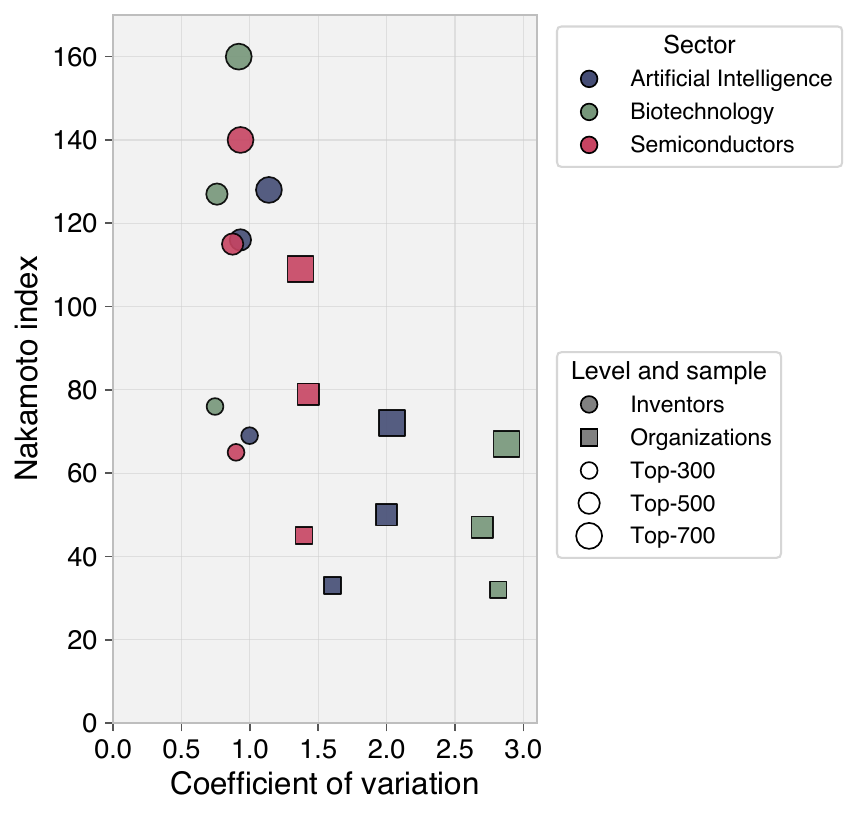}
\caption{\textbf{Network-level heterogeneity across sizes.} Nakamoto index versus the coefficient of variation for inventor (denoted with circles) and organization (denoted with squares) networks in the three, strategic sectors of artificial intelligence, biotechnology and semiconductors (only the top-$500$ actors have been considered). Inventor (organization) networks systematically occupy a region characterized by a lower (higher) degree heterogeneity and a higher (lower) Nakamoto index.}
\label{fig:nakamoto}
\end{figure}

These observations are confirmed by the metrics reported in Table~\ref{tab:network_metrics}: inventor networks display a moderate dispersion of degree values ($\text{CV}\lesssim 1$) and a high Nakamoto index~\cite{srinivasan2017quantifying, Lin2022WeightedLightning}, indicating that collaborations are quite evenly distributed across individuals; organization networks, instead, exhibit a higher heterogeneity ($\text{CV}>1$) and a markedly lower Nakamoto index (approximately half the one of inventor networks), signaling that an overall small subset of firms accounts for the majority of collaborative ties (see Fig.~\ref{fig:nakamoto}).

Taken together, the aforementioned results depict \textit{i)} inventor networks as characterized by small teams originating distributed collaborations; \textit{ii)} organization networks as less cohesive and more hierarchical, with few firms maintaining many partnerships that span institutional boundaries, while most remain specialized or, at least, regionally confined.

\subsection{Innovation networks at the mesoscale}\label{sec:innomeso}

In order to assess how the structure of collaborations shapes the technological influence, we implement two algorithms for detecting mesoscale structures, i.e. the maximization of the modularity and the minimization of BIC, instantiated with the SBM and the dcSBM.

\begin{figure*}[t!]
\centering
\includegraphics[width=\linewidth]{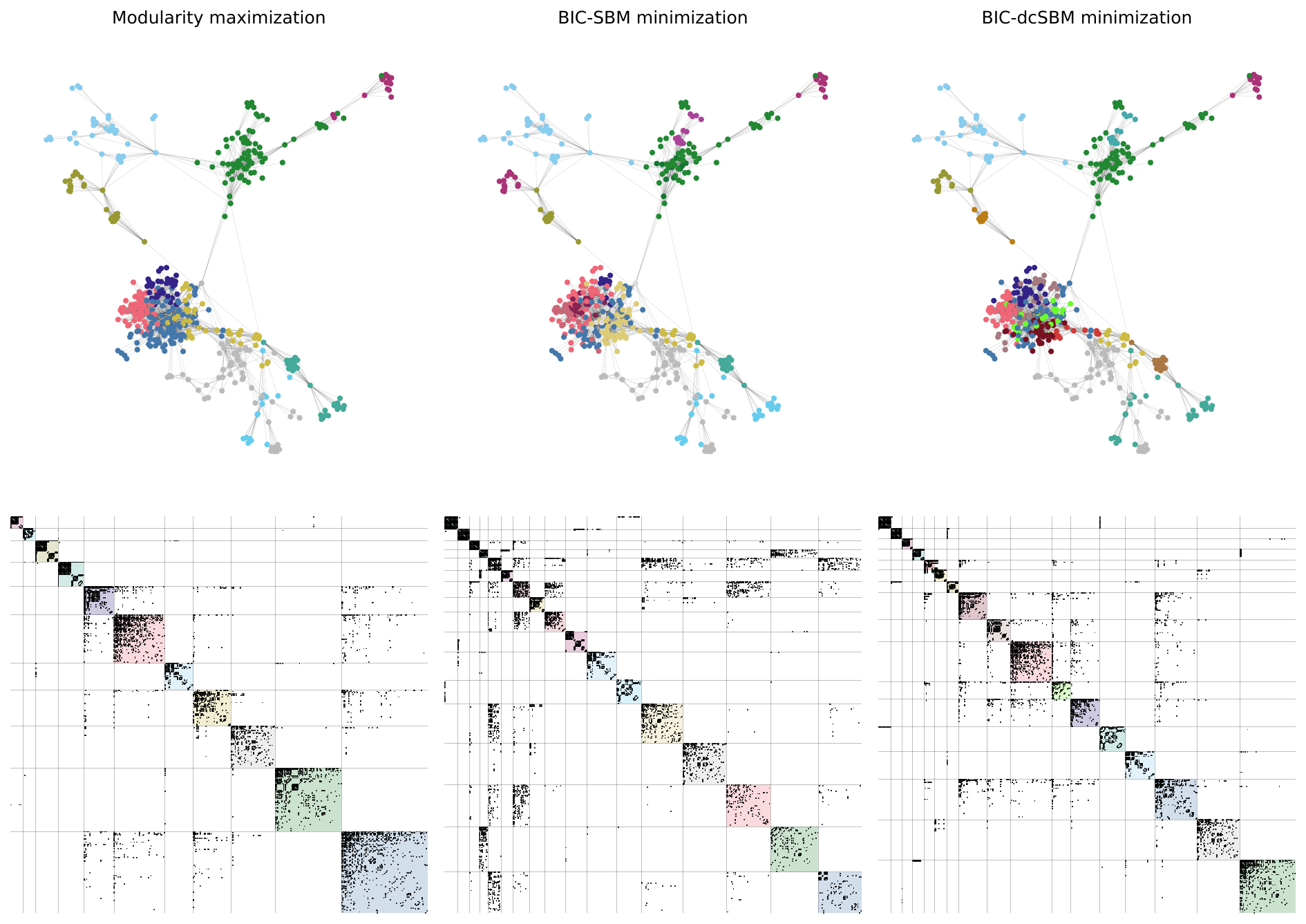}
\caption{\textbf{Mesoscale structure of the co-inventorship network in the semiconductor sector.} The higher density and average clustering coefficient characterizing inventor networks reflect into a nested architecture, with dense modules acting as cores of other sparser ones, that points out field-specific specializations leading to project-based collaborations: this, in turn, suggests that knowledge recombines through phases of concentrated teamwork within communities.}
\label{fig:inventors_top500}
\end{figure*}

As Fig.~\ref{fig:inventors_top500} shows, modularity maximization leads to identify clusters of densely-interconnected actors; the minimization of BIC instantiated with the SBM, on the other hand, leads to a sharper distinction between hubs and peripheral actors, a picture that is further refined by instantiating BIC with the dcSBM. As the adjacency matrices visually confirm, employing inferential techniques leads to spot a nested architecture, with dense modules acting as cores of other, sparser ones.

\begin{figure*}[t!]
\centering
\includegraphics[width=\linewidth]{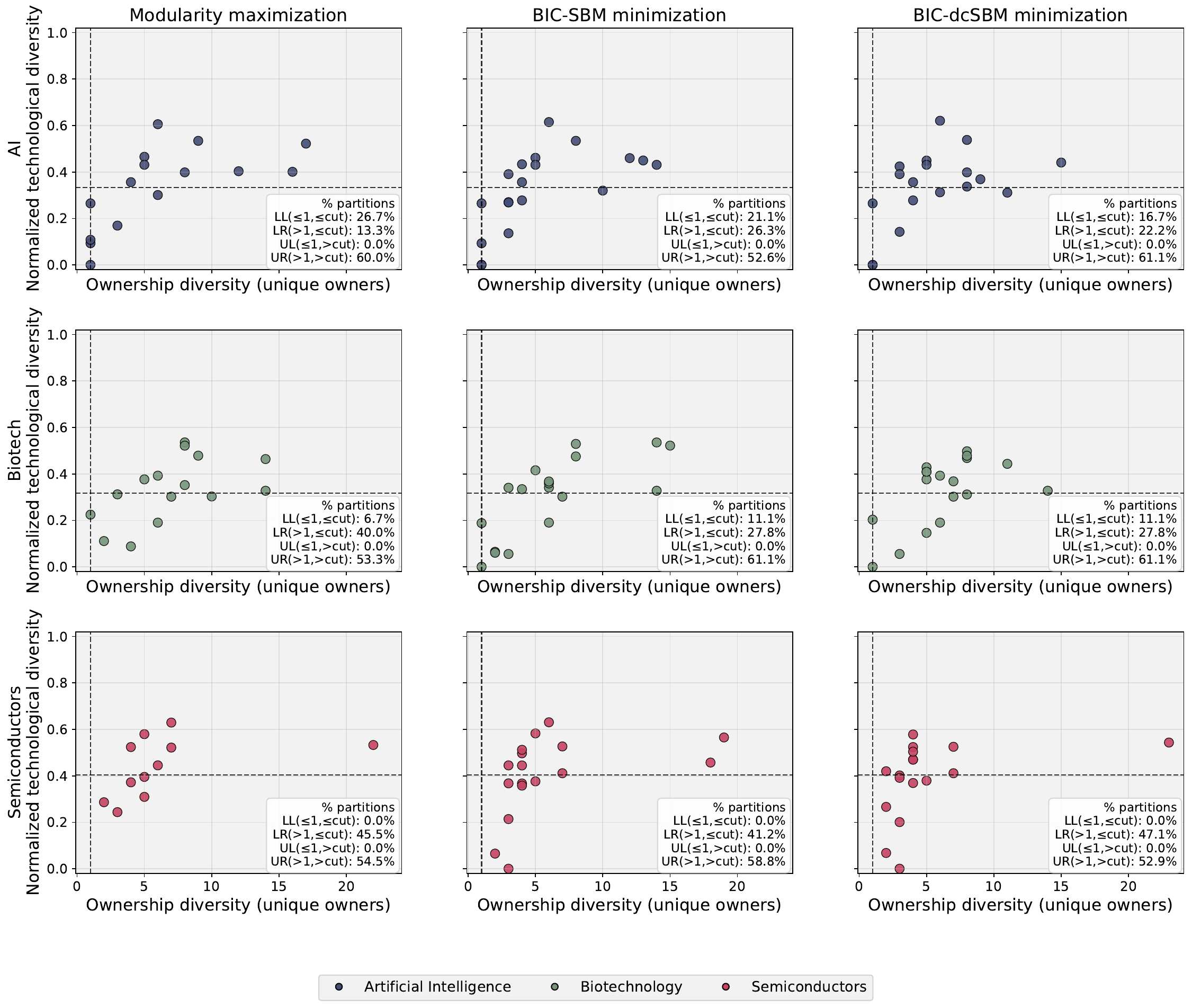}
\caption{\textbf{Diversity of the inventors' modules across sectors and models.} Each point represents a cluster of the top-$500$ inventor networks in the three, strategic sectors of artificial intelligence (top row), biotechnology (middle row) and semiconductors (bottom row), according to modularity maximization (left column), BIC minimization instantiated with the SBM (middle column) and BIC minimization instantiated with the dcSBM (right column). Ownership diversity (i.e. the number of distinct patent owners), distinguishing single-company from multi-company clusters, is reported on the $x$-axis; technological diversity (i.e. the normalized Shannon entropy of IPC codes), distinguishing specialized from generalist clusters, is reported on the $y$-axis. Dashed lines mark the relevant thresholds: Shannon entropy is split at its mean value; organizational diversity is split at $D_{\text{own}}=1$; the bottom-right boxes report the percentage of points/clusters falling into each quadrant. Across all sectors, modules of inventors are predominantly inter-company and generalist, a feature indicating that inventive collaborations typically involve multiple organizations. To be noticed that modularity maximization leads to individuate broader, more diverse clusters than BIC minimization.}
\label{fig:diversity}
\end{figure*}

Such a picture is further refined by the results shown in Fig.~\ref{fig:diversity}: by scattering the normalized Shannon entropy of International Patent Classification (IPC) codes versus the number of distinct patent owners, one is able to assess the internal composition of the inventors' modules in terms of both organizational and technological diversity, i.e. whether they correspond to intra-firm/cross-firm collaborations (horizontal dimension) and whether their inventive activity is technologically specialized/generalist (vertical dimension). The share of single-company modules is small (below $25\%$ in AI and BT and virtually absent in SC), a feature reflecting the cross-institutional character of innovation; in terms of technological scope, most modules fall above the mean, thus corresponding to generalist groups spanning several technological subfields. To be noticed that the clusters returned by modularity maximization tend to be concentrated in the upper-right quadrant, thus revealing inter-company and technologically diverse collaborations; the clusters returned by BIC minimization, instead, display a higher variability, with a subset of smaller, specialized ones alongside larger, multi-firm groups.

\begin{table}[t!]
\centering
\scriptsize
\setlength{\tabcolsep}{3.2pt}
\renewcommand{\arraystretch}{1.15}
\begin{tabular}{c|c|c|c|c|c|c|c|c}
\hline
\hline
\textbf{Sector} & \textbf{Level} & $N$ & $L$ & $\bar{k}$ & $\tilde{k}$ & $\sigma_k$ & $\sigma_k/\bar{k}$ & $N_{51}$ \\
\hline
\hline
\multirow{3}{*}{AI}
& \multirow{3}{*}{Inv.}
& 300 & 1364 & 9.09 & 7 & 8.79 & 0.96 & 73 \\
& & 500 & 2252 & 9.01  & 7  & 8.39  & 0.93 & 119 \\
& & 700 & 5367 & 15.33 & 9  & 17.47 & 1.14 & 133 \\
\hline
\multirow{3}{*}{BT}
& \multirow{3}{*}{Inv.}
& 300 & 1318 & 8.79 & 6 & 6.55 & 0.75 & 78 \\
& & 500 & 2300 & 9.20 & 7 & 6.98 & 0.76 & 131 \\
& & 700 & 3497 & 9.99 & 7 & 9.18 & 0.92 & 166 \\
\hline
\multirow{3}{*}{SC}
& \multirow{3}{*}{Inv.}
& 300 & 1640 & 10.93 & 8  & 9.83 & 0.90 & 67 \\
& & 500 & 2803 & 11.21 & 9  & 9.80 & 0.87 & 119 \\
& & 700 & 4801 & 13.72 & 10 & 12.78 & 0.93 & 145 \\
\hline
\hline
\multirow{3}{*}{AI}
& \multirow{3}{*}{Org.}
& 300 & 665  & 4.43 & 2 & 7.11 & 1.60 & 35 \\
& & 500 & 923  & 3.69 & 2 & 7.38 & 1.99 & 52 \\
& & 700 & 1434 & 4.10 & 2 & 8.35 & 2.04 & 76 \\
\hline
\multirow{3}{*}{BT}
& \multirow{3}{*}{Org.}
& 300 & 673  & 4.49 & 2 & 12.63 & 2.82 & 34 \\
& & 500 & 888  & 3.55 & 2 & 9.59  & 2.70 & 49 \\
& & 700 & 1496 & 4.27 & 2 & 12.29 & 2.88 & 71 \\
\hline
\multirow{3}{*}{SC}
& \multirow{3}{*}{Org.}
& 300 & 488  & 3.25 & 2 & 4.54 & 1.40 & 47 \\
& & 500 & 870  & 3.48 & 2 & 4.96 & 1.43 & 82 \\
& & 700 & 1229 & 3.51 & 2 & 4.81 & 1.37 & 114 \\
\hline
\hline
\end{tabular}
\caption{\textbf{Network-level inequality metrics.} Indicators of structural inequality for inventor and organization networks in the three strategic sectors of artificial intelligence, biotechnology and semiconductors (only the top-$500$ actors have been considered). Columns include the total number of nodes $N$, the total number of edges $L$, the average degree $\bar{k}$, the median degree $\tilde{k}$}, the degree standard deviation $\sigma_k$, the coefficient of variation $\sigma_k/\bar{k}$ and the Nakamoto index $N_{51}$, defined as the minimum number of nodes accounting for at least the $51\%$ of the total number of links. While inventor networks are moderately unequal, organization networks are much more so: many individuals collaborate, whereas few firms concentrate the majority of ties. At the sector level, inequality is more pronounced in AI and BT.
\label{tab:network_metrics}
\end{table}

As Fig.~\ref{fig:owners_top500} shows, organization networks display a mesoscale architecture constituted of tree-like structures, mirroring a level of core-periphery -ness that is even more pronounced than the one characterizing inventor networks and suggesting a role-based partition of the system under consideration.

The metrics reported in Table~\ref{tab:within_cluster_metrics} further detail such a description. Under modularity maximization, the clusters display the strongest internal cohesion - the IC/EC ratio is, in fact, the largest. Minimizing BIC instantiated with the SBM yields less cohesive clusters and the lowest degree standard deviation, thus revealing the presence of peripheries centered around few, influential core-nodes; employing the degree-corrected variant of the SBM reconciles the two pictures above, as less clusters than modularity are recovered, however showing a more pronounced hierarchical structure (see also Fig.~\ref{fig:bic_evolution3}, further clarifying the relationship between the structures detected by BIC-SBM and those detected by BIC-dcSBM).

Table~\ref{tab:communities} provides a comparative overview of the structures identified by our detection algorithms. In particular, \textit{i)} organizations within the AI sector tend to form hub-driven, national systems dominated by Chinese and Korean conglomerates (Huawei, KAIST, Samsung); \textit{ii)} organizations within the BT sector feature regionally-cohesive industrial complexes, such as the Sinopec-CNPC-CNOOC and the Sichuan-Zhejiang biomedical clusters; \textit{iii)} organizations within the SC sector display hierarchical supply chains linking major equipment producers with integrated device manufacturers (IBM, Soitec-CEA-STMicro, TEL).

\subsection{Inequality of the innovation impact}\label{sec:innoimpact}

In order to quantify the distribution of the innovation impact across collaboration structures, let us plot the Lorenz curves of forward citations, a well-established measure of inequality that reveals the extent to which technological influence is concentrated. As Fig.~\ref{fig:lorenz} shows, all inventor networks are characterized by the same trend, indicating that a small subset of communities is responsible for most of the technological influence - regardless of the sector. At the level of organizations, however, sectoral differences emerge: AI consistently exhibits the steepest Lorenz curves - i.e. the strongest concentration of forward citations - followed by BT and SC; this, in turn, suggests that (more than in other sectors) innovation is dominated by few alliances, acting as `attractors' of citations. These patterns are fully consistent with the Gini coefficients reported in appendices and quantify the same underlying inequality with a single scalar.

Importantly, the `magnitude' of inequality is affected by the algorithm employed to detected mesoscale structures. Modularity maximization, for example, yields flatter curves for BT and sharper curves for SC: these patterns are due to the fact that modularity tends to split networks into more homogeneous clusters, thus dispersing high-impact patents across several groups of nodes; BIC minimization, on the contrary, leads to gather highly-connected, influential actors into cohesive cores, thereby capturing hierarchical, role-based patterns of knowledge accumulation that, in turn, causes the appearance of steeper curves. 

\section{Discussion}\label{sec:discussion}

As revealed by our results, collaborations shape innovation networks according to two, complementary dynamics that operate at different levels: this duality reflects the nested architecture typical of innovation ecosystems, where micro-level creative recombination unfolds within meso-level institutional structures that govern coordination and diffusion \cite{Jackson2011}. Such a multi-scale organization is a hallmark of complex socio-technical systems, in which inventive activity emerges from the interplay between cohesive teams and hierarchical organizational backbones: this interpretation also connects our network evidence to the broader literature on co-invention systems, which has shown that patent collaboration networks evolve through clustered, uneven and path-dependent growth processes~\cite{pinto2021exploring}.

\begin{figure*}[t!]
\centering
\includegraphics[width=\linewidth]{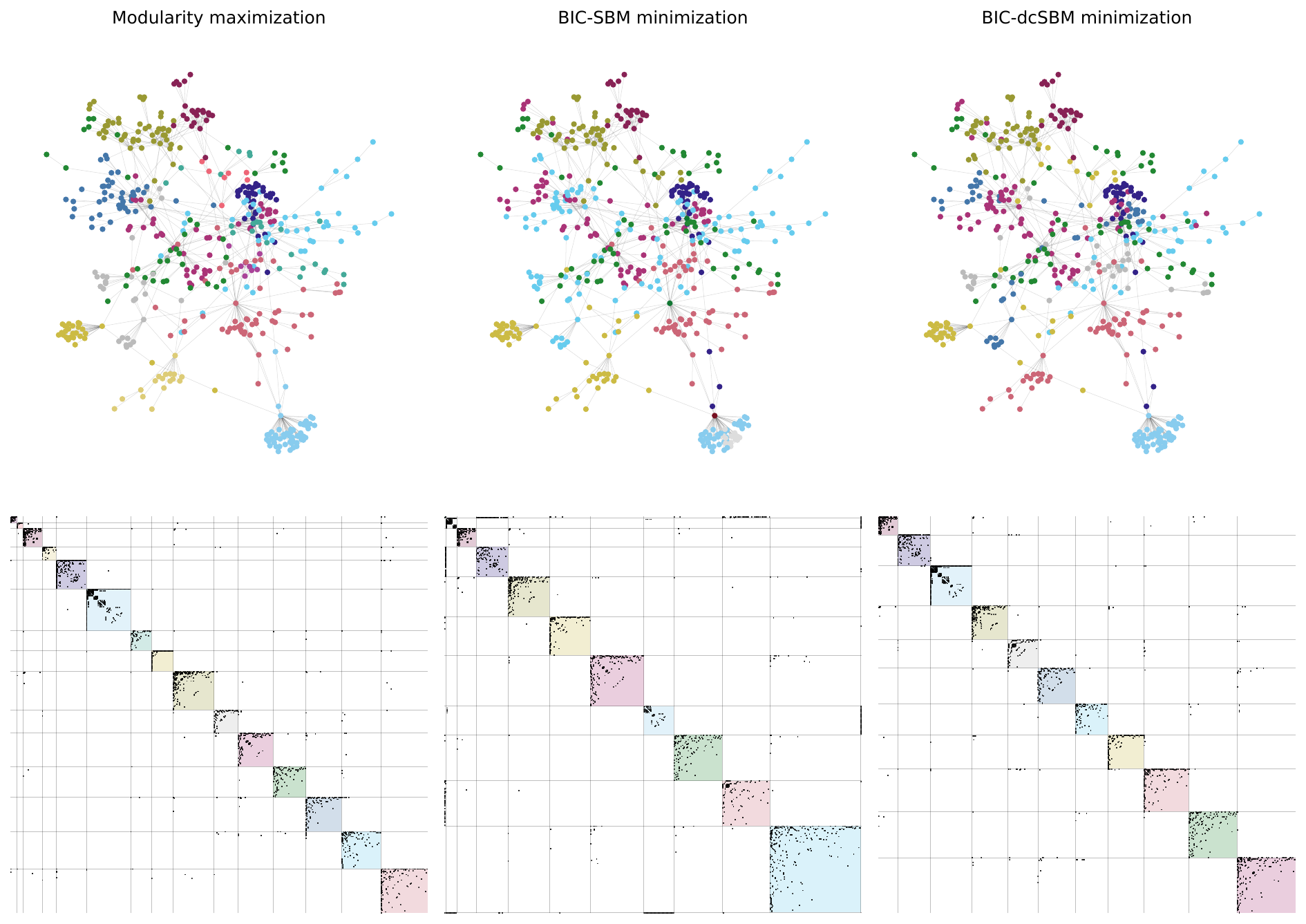}
\caption{\textbf{Mesoscale structure of the co-ownership network in the semiconductor sector.} Differently from inventor networks, organization networks are much less modular, appearing as combinations of tree-like structures, mirroring a level of core-periphery -ness that is even more pronounced than the one characterizing inventor networks: while few firms maintain many cross-institutional partnerships, most actors remain specialized or, at least, regionally confined; this, in turn, suggests that knowledge flows across communities in a hierarchical fashion.}
\label{fig:owners_top500}
\end{figure*}

At the individual level, knowledge production occurs through cohesive, recurrent teams, as indicated by the higher clustering and more distributed connectivity of inventor networks (reported in Fig.~\ref{fig:clustering} and Table~\ref{tab:network_metrics}, respectively). These structures suggest the importance of small, trusted groups for experimentation and rapid problem-solving - an attitude that is consistent with the project-based nature of inventive activity. These findings align with prior research showing that existing social ties, team cohesion and repeated collaborations enhance coordination and commercialization outcomes in inventive work~\cite{Bercovitz2011Mechanisms,Toth2021Repeated,GuimeraUzziSpiroAmaral2005}; besides, inventor teams that balance internal trust with disciplinary diversity tend to generate more valuable patents~\cite{Inoue2015,Petruzzelli2015Determinants}. Moreover, the high clustering observed in inventor networks is consistent with sociological evidence showing that small-world structures enhance collaborative creativity by balancing local cohesion and global reach~\cite{UzziSpiro2005}. This pattern assumes relevance in the light of some broader shifts in knowledge production, where teams, rather than individuals, have proven to become the dominant drivers of innovation~\cite{WuchtyJonesUzzi2007}. It is also consistent with evidence that breakthrough invention is rarely the product of isolated inventors alone but is often associated with collaborative structures that combine diverse knowledge inputs~\cite{singh2010lone}; this interpretation is further supported by Fig.~\ref{fig:diversity}, which shows that inventor modules are multi-owner and relatively technologically diverse - a suggestive sign of project-based recombination across organizational and technological boundaries.

At the organization level, instead, networks are sparser and more hierarchical (as reported in Fig.~\ref{fig:owners_top500} and Table~\ref{tab:network_metrics}, respectively): few, central agents (firms and institutions) connect otherwise weakly-linked actors, thus constituting the structural backbone upon which technological coordination rests. The aforementioned distinction is in line with previous studies indicating that co-assignment and R\&D alliances often reflect strategic coordination across institutional boundaries rather than direct, interpersonal ties \cite{Fritsch2020Identifying, ahuja2000collaboration, owensmith2004knowledge}. These hierarchical structures resonate with the well-documented dynamics through which star-firms and elite academic institutions exert disproportionate influence on emerging technological domains, shaping both the direction and the pace of innovation~\cite{ZuckerDarby1996}: this is particularly relevant for AI, where recent patent-based evidence also points to the evolution of global collaborative innovation networks structured around uneven international and organizational positions~\cite{ye2025evolutionary}; in this sense, organizational cores in AI and BT function as attractors of knowledge flows, mirroring the role of star-scientists and star-organizations in the formation of high-impact technological clusters.

This dual structure (cohesive inventor teams embedded within sparser, hierarchical organizations) creates a nested architecture for innovation: at the team level, dense, interpersonal ties support exploitation, thus enabling rapid iteration and refinement of ideas, i.e. efficient problem-solving; at the organization level, central hubs connect otherwise separate groups, thus facilitating exploration through cross-cluster search, recombination of knowledge and access to diverse, technological inputs \citep{march1991exploration,UzziSpiro2005,schilling2007interfirm, rosenkopf2003overcoming}.

\begin{table*}[t!]
\begin{tabular}{c|c|l|c|c|c|c|c|c|c}
\hline
\hline
\multicolumn{1}{c|}{\textbf{Sector}} &
\multicolumn{1}{c|}{\textbf{Level}} &
\multicolumn{1}{c|}{\textbf{Algorithm}} &
\multicolumn{1}{c|}{\textbf{\# clusters}} &
\multicolumn{1}{c|}{$\bar{k}_{\text{within}}$} &
\multicolumn{1}{c|}{$\sigma_{k}^{\text{within}}$} &
\multicolumn{1}{c|}{\textbf{IC/EC}} &
\multicolumn{1}{c|}{\textbf{BIC-SBM}} &
\multicolumn{1}{c|}{\textbf{BIC-dcSBM}} &
\multicolumn{1}{c}{\textbf{$Q$}} \\
\hline
\hline
\multirow{6}{*}{AI} & \multirow{3}{*}{Inventors} 
& Modularity maximization  & 15 & 8.41  & 6.11 & 11.42 & 12878.44 & 15473.80  & 0.8404 \\
& & BIC-SBM minimization   & 20 & 18.09 & 5.48 & 3.88  & 11202.85 & -         & 0.7145 \\
& & BIC-dcSBM minimization & 17 & 9.77  & 6.34 & 5.10  & -        & 14600.52  & 0.7890 \\
\cline{2-10}
& \multirow{3}{*}{Organizations}
& Modularity maximization  & 9 & 3.32  & 6.36 & 11.56 & 8234.95  & 10803.32  & 0.7264 \\
& & BIC-SBM minimization   & 12 & 22.75 & 4.15 & 3.46  & 6404.86  & -         & 0.2830 \\
& & BIC-dcSBM minimization & 9 & 3.62  & 7.18 & 7.35  & -        & 10787.43  & 0.6942 \\
\hline
\multirow{6}{*}{BT} & \multirow{3}{*}{Inventors} 
& Modularity maximization  & 15 & 9.42  & 5.82 & 11.94 & 12406.82 & 15400.75 & 0.8498 \\
& & BIC-SBM minimization   & 19 & 1.49 & 2.78 & 0.72  & 5739.71 & -        & 0.3200 \\
& & BIC-dcSBM minimization & 18 & 9.85  & 5.93 & 7.22  & -        & 14957.47 & 0.8232 \\
\cline{2-10}
& \multirow{3}{*}{Organizations}
& Modularity maximization  & 9 & 3.96  & 8.06 & 2.64  & 8532.56  & 10116.84 & 0.6856 \\
& & BIC-SBM minimization   & 14 & 26.31 & 5.24 & 0.52  & 6141.66  & -        & 0.2616 \\
& & BIC-dcSBM minimization & 9 & 3.68  & 8.66 & 2.13  & -        & 10105.29 & 0.6532 \\
\hline
\multirow{6}{*}{SC} & \multirow{3}{*}{Inventors} 
& Modularity maximization  & 11 & 10.45 & 8.18 & 13.58 & 17665.27 & 30009.80 & 0.7073 \\
& & BIC-SBM minimization   & 17 & 14.06 & 6.89 & 2.49  & 15524.97 & -        & 0.5160 \\
& & BIC-dcSBM minimization & 19 & 11.32 & 8.61 & 3.95  & -        & 18869.03 & 0.6662 \\
\cline{2-10}
& \multirow{3}{*}{Organizations}
& Modularity maximization  & 15 & 3.47  & 4.43 & 7.34  & 7725.65  & 11250.47 & 0.8309 \\
& & BIC-SBM minimization   & 12 & 11.19 & 3.12 & 3.40  & 7128.25  & -        & 0.6933 \\
& & BIC-dcSBM minimization & 11 & 3.65  & 4.79 & 6.49  & -        & 11132.57 & 0.8150 \\
\hline
\hline
\end{tabular}
\caption{\textbf{Clusters diversity across sectors, levels and models.} As the ratio between the partition-specific numbers of intra-cluster and inter-cluster edges (IC/EC) confirms, the maximization of modularity $Q$ leads to recover clusters displaying the highest internal cohesion.} Implementing the BIC-SBM, instead, leads to recover the least-cohesive clusters, in turn featuring the lowest within-cluster degree standard deviation $\sigma_{k}^{\text{within}}=R^{-1}\sum_{r=1}^R\sigma_k^r$ - a signature of the presence of peripheries centered around few cores. Finally, the BIC-dcSBM reconciles these two pictures, leading to recover fewer clusters than modularity, however displaying a more pronounced hierarchical structure.
\label{tab:within_cluster_metrics}
\end{table*}

\begin{figure*}[!t]
\centering
\includegraphics[width=\linewidth]{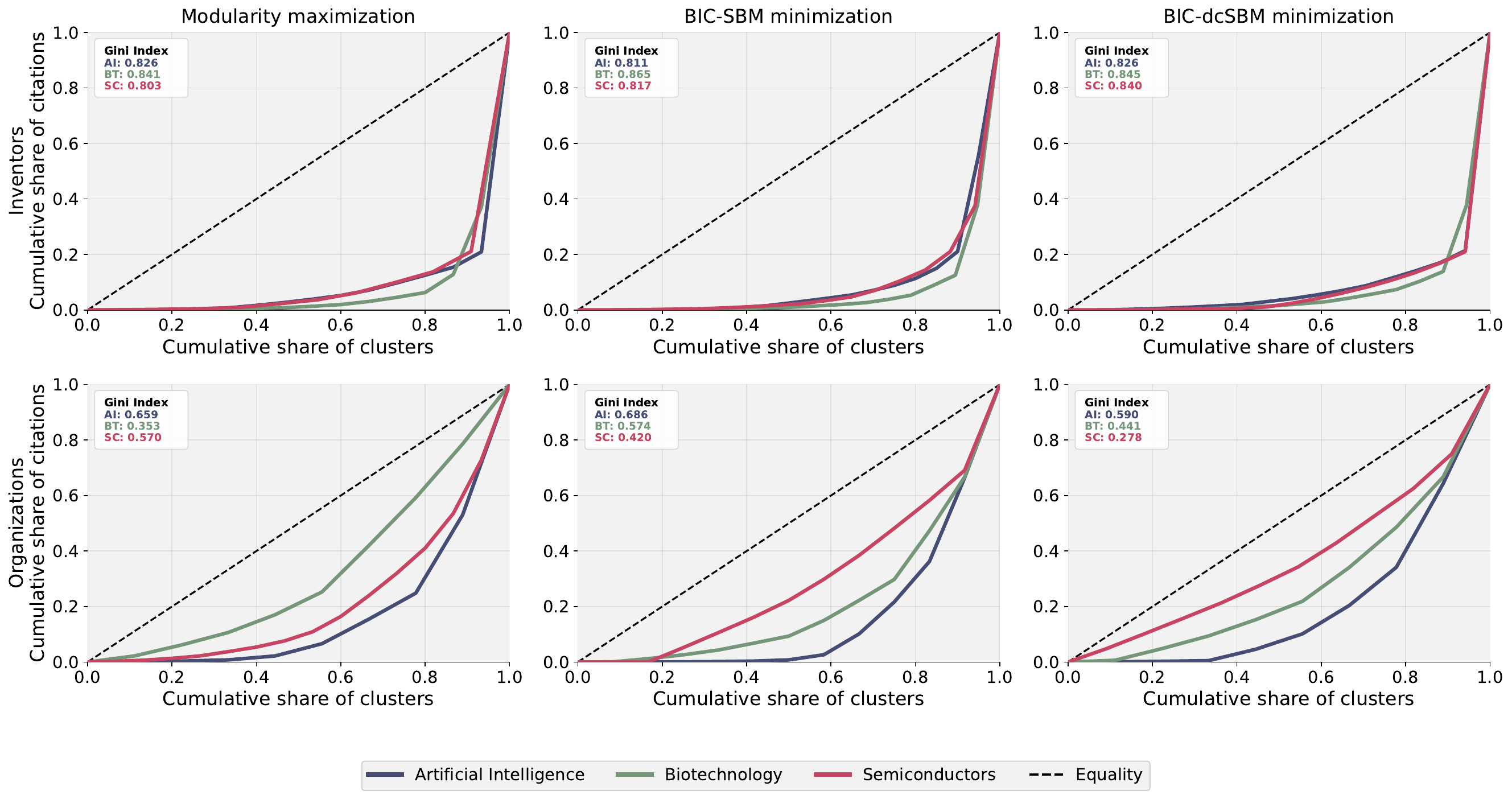}
\caption{\textbf{Inequality of the innovation impact.} Lorenz curves of forward citations for inventor and organization networks (only the top-$500$ actors have been considered). Inventor networks show a steadily high inequality across sectors, whereas organization networks are more diverse, with the AI sector exhibiting the highest level of inequality.}
\label{fig:lorenz}
\end{figure*}

The analysis of Lorenz curves and Gini indices (Fig.~\ref{fig:lorenz}) confirms that inequality in technological impact is pervasive, although its intensity varies across levels and sectors. Inventor networks display the strongest concentration of forward citations, indicating that technological influence is disproportionately located in a small number of co-inventor communities rather than being evenly distributed across the whole inventive population. This result reinforces the interpretation of inventor clusters as consequential units of innovation: cohesive teams do not only organize collaboration but also concentrate a large share of subsequent technological impact. Such a pattern is consistent with evidence that repeated collaborations reinforce network persistence and cumulative advantage among established inventors~\citep{Toth2021Repeated}, as well as with work interpreting forward citations as traces of follow-on inventive activity and cumulative innovation~\cite{galasso2015patents}. At the organization level, inequality is most pronounced in the AI sector, where few corporate-academic alliances attract a disproportionate share of forward citations; BT and SC, instead, display comparatively more balanced distributions. Overall, these patterns are consistent with cumulative-advantage mechanisms such as the Matthew effect~\cite{Merton1968}, although our static networks do not allow us to identify dynamic attachment processes directly.

\begin{figure*}[t!]
\centering
\includegraphics[width=\linewidth]{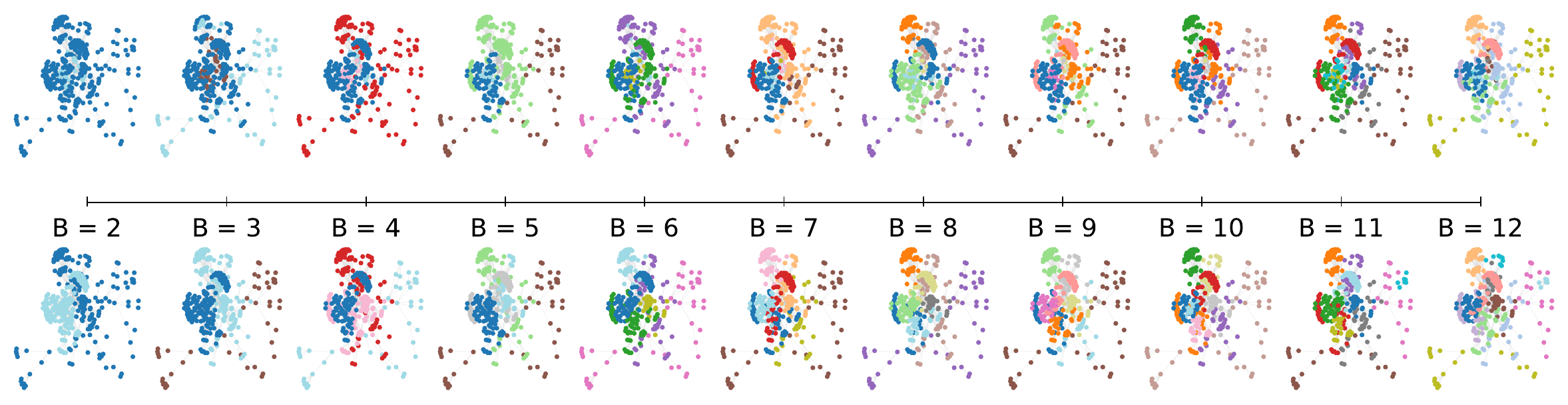}\\[2mm]
{\small (a)}\\[4mm]
\includegraphics[width=0.39\linewidth]{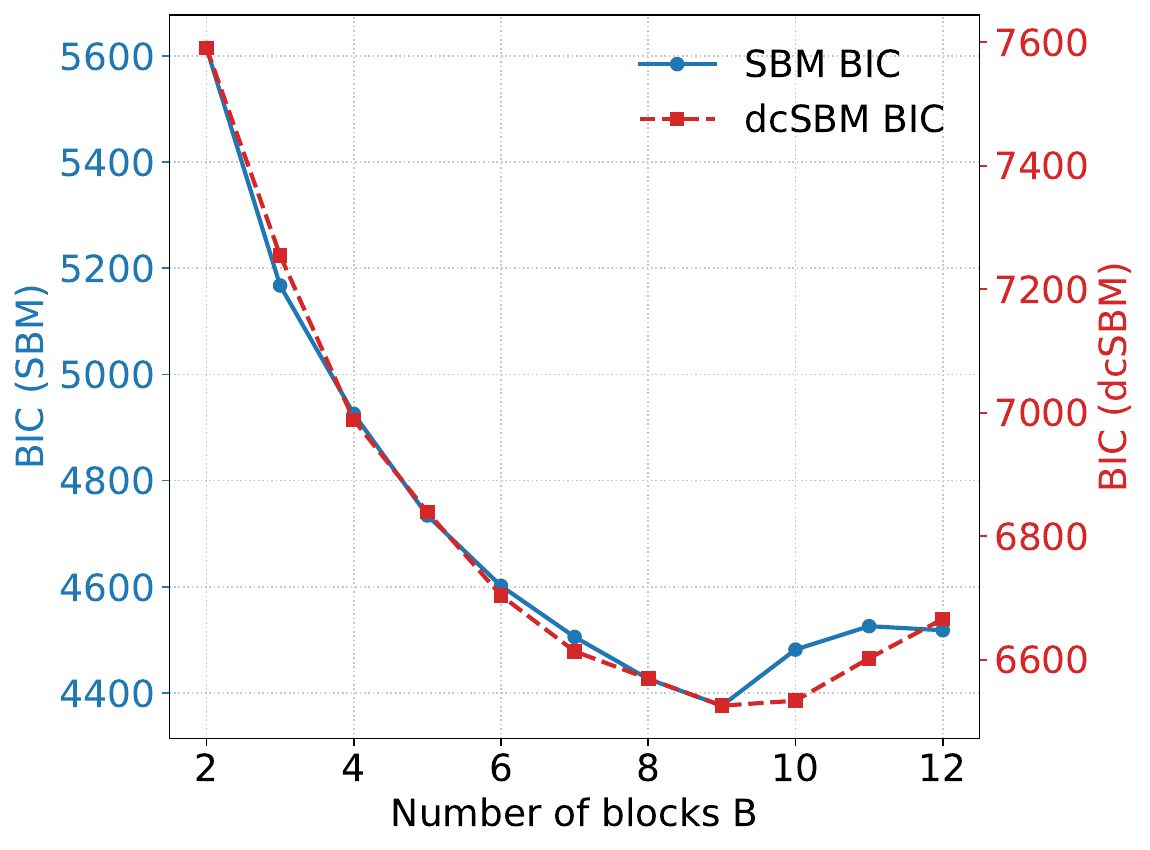}
\hfill
\includegraphics[width=0.295\linewidth]{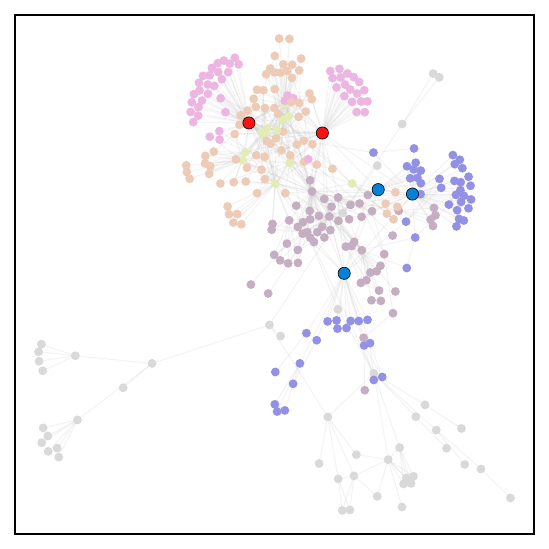}
\hfill
\includegraphics[width=0.295\linewidth]{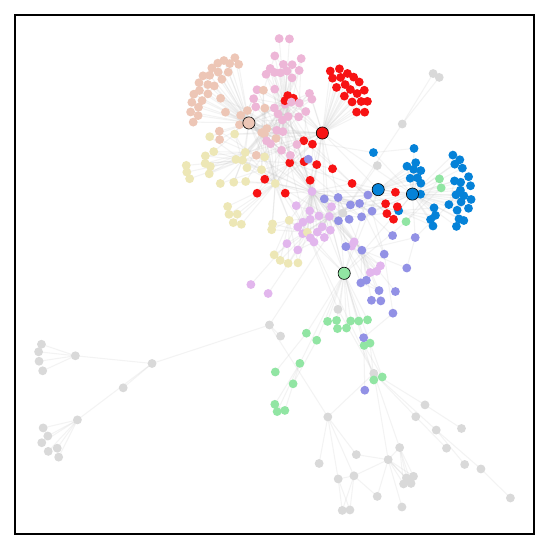}\\[2mm]
\parbox{0.39\linewidth}{\small (b)}
\hfill
\parbox{0.295\linewidth}{\small (c)}
\hfill
\parbox{0.295\linewidth}{\small (d)}
\caption{\textbf{BIC-based partition of the co-inventorship network in the sector of artificial intelligence.} Panel (a) depicts the BIC-SBM partition (top row) and the BIC-dcSBM partition (bottom row) on the same layout, thus making it possible to appreciate how the two models refine the same tree-like structure as the number of clusters, $B$, increases. Panel (b) reports the BIC values as a function of $B$: for both models, the minimum lies at $B=9$. Panels (c) and (d) illustrate this common optimum. In both cases, only part of the core-nodes (enlarged and contoured), as well as the periphery-nodes directly attached to them, are colored: under the BIC-SBM, they constitute small communities, the peripheries being assigned to separate blocks; under the BIC-dcSBM, instead, these nodes are put together. To reduce the computational burden, the analysis shown here has been carried out on the top-$300$ co-inventorship network.}
\label{fig:bic_evolution3}
\end{figure*}

Finally, let us notice that different algorithms for mesoscale structures detection return different results: while modularity maximization tends to fragment the network into more homogeneous clusters, thus distributing citations more evenly, BIC minimization is capable of individuating locally hierarchical structures on the cores of which impact is concentrated (Table~\ref{tab:communities}, Fig.~\ref{fig:bic_evolution3} and Fig.~\ref{fig:lorenz}).\\

\section{Conclusions}\label{sec:conclusions}

Taken together, our findings show that the architecture of innovation networks plays a central role in shaping how technological knowledge is created, coordinated and diffused across the global patent system. Across sectors, collaboration structures consistently reveal a dual pattern: at the inventor level, cohesive groups of individuals connect across organizational and technological boundaries, while at the organizational level, sparser and more hierarchical structures emerge, with a limited number of central actors coordinating wider ecosystems.

As of inventors, the clusters represent collaborations linking actors across organizational and technological boundaries. These groups do not necessarily represent enduring teams, mirroring, instead, the project-based character of inventive activity, where individuals from different firms cooperate on technologies of varying scope. Speaking of organizations, instead, the locally hierarchical patterns detected by our algorithms are especially pronounced in the AI sector, where national and corporate cores dominate the landscape of collaborations, while the BT and the SC sectors exhibit more regionally-anchored and supply chains-oriented structures.

We acknowledge some limitations of our study and pave the way for promising future work. We adopt a binary representation of collaboration networks, in line with a large literature on co-invention, inter-organizational collaboration and innovation networks, where the \textit{existence} of a collaborative tie is treated as the relevant unit of analysis~\citep{powell1996interorganizational,Balconi2004NetworksInventorsAcademia, BreschiCatalini2010}; moreover, the 2020-2024 window allows us to focus on recent and comparable collaboration patterns in three highly dynamic sectors. While this baseline is appropriate for our purpose and follows a long stream of previous work in innovation studies, several extensions can build on this representation: first, \textit{temporal} networks would allow one to study how mesoscale structures emerge, persist, merge or fragment over time and to assess whether citation concentration is driven by cumulative-advantage dynamics such as Matthew effects; second, \textit{weighted} networks would distinguish occasional from repeated collaborations by accounting for the intensity of co-invention or co-ownership ties; finally, \textit{multilayer} representations combining co-inventorship, co-ownership and citation networks would make it possible to model how interpersonal collaboration, institutional coordination and technological impact interact across levels.

In conclusion, this study shows that innovation networks are hierarchical systems where collaborations and technological influence are deeply intertwined: cohesive inventor teams sustain the creation of knowledge at a microscopic level, while organizational hierarchies coordinate and amplify that knowledge. Understanding the mesoscale architecture of collaboration networks is, therefore, crucial for understanding how they shape the distribution of inventive impact within and across technological domains.

\begin{table*}[t!]
\centering
\renewcommand{\arraystretch}{1.15}
\begin{tabularx}{\textwidth}{c|X|X|X}
\hline
\hline
\multicolumn{1}{c|}{\textbf{Sector}} & \multicolumn{1}{c|}{\textbf{Modularity maximization}} & \multicolumn{1}{c|}{\textbf{BIC-SBM minimization}} & \multicolumn{1}{c}{\textbf{BIC-dcSBM minimization}}\\
\hline
\hline
AI & \textbf{9 clusters.} Hub-driven, national innovation systems (China, Korea), dominated by large corporate-academic clusters such as Huawei, Samsung and KAIST & \textbf{12 clusters.} Role-based segmentation distinguishing the Guangdong digital, Korean industrial and State Grid-infrastructure ecosystems, plus a transnational academic-corporate hub & \textbf{9 clusters.} Degree correction merges subcommunities into cohesive, national macro-blocks (Guangdong, Korea, State Grid), revealing mirroring core-periphery structures\\
\hline
BT & \textbf{9 clusters.} Dense, regional and corporate clusters (Sinopec-CNPC-CNOOC, Sichuan, Zhejiang), shaped by geographic and industrial proximity & \textbf{14 clusters.} Finer, functional segmentation within corporate groups, distinguishing production, R\&D and regulatory divisions within the same industrial conglomerates & \textbf{9 clusters.} Coarse-grained macro-blocks such as the Sinopec industrial complex, the Sichuan translational biomedical hub and the Zhejiang R\&D cluster\\
\hline
SC & \textbf{15 clusters.} Hub-dominated clusters mirroring large ecosystems (IBM, Soitec-CEA-STMicro, TEL, Toyota) and mixed tool-materials groupings & \textbf{12 clusters.} Role-based blocks separating functionally different clusters (IBM, imec-Leuven, Toshiba/Kioxia, Toyota/Denso and TEL) & \textbf{11 clusters.} Degree-correction identifies macro-structures by unifying IBM with the GlobalFoundries federation, the two-tier TEL system and the R\&D cores of imec-Leuven and Soitec-CEA-STMicro\\
\hline
\hline
\end{tabularx}
\caption{\textbf{Comparative overview of the mesoscale structures detected by our algorithms on the organization networks.} Each entry of the table reports the number of detected clusters and a concise description of them.}
\label{tab:communities}
\end{table*}

\section{Materials and methods}\label{sec:methods}

\subsection{Network construction}

Here, we focus on the most recent five-year period (from January 1, 2020 - December 31, 2024) to capture contemporary patterns of technological collaboration while avoiding structural distortions due to older and less comparable data. We build three, sectorial collaboration networks - in the domains of \textit{artificial intelligence}, \textit{biotechnology} and \textit{semiconductors} - where nodes represent either \textit{patent inventors} or \textit{patent owners} and binary, undirected edges connect entities that co-appear on at least one patent. This representation focuses on the \textit{existence} of collaboration channels rather than on their \textit{intensity}, which is consistent with previous work on the matter~\cite{powell1996interorganizational,Balconi2004NetworksInventorsAcademia} and appropriate for comparing mesoscale structures across sectors and levels.

Sectorial delineation follows the classification scheme detailed in Appendix~\ref{AppA}. Patent-level data are retrieved from the \textit{ORBIS Intellectual Property} (ORBIS IP) database by Moody's~\cite{MoodyOrbisIP}, a comprehensive source linking global patent data to firm-level information, ownership structures and corporate networks\footnote{Although ORBIS and ORBIS IP are commercial datasets, requiring subscription-based access, their richness and data quality make them among the most widely-used sources for research in intellectual property, finance and economics~\cite{Gal2013Orbis}.}. ORBIS IP integrates over 138 million patents with data on approximately 2.4 million companies worldwide, including ownership timelines, litigation cases, M\&A information, patent valuations and technology classifications (IPC, CPC, USPC).

We focus on the most influential actors, restricting the analysis to the top-$500$ nodes ranked by forward citations; our choice is also consistent with the strengths of ORBIS, particularly well-suited to analyze large and high-performing firms, global statistics and frontier-level innovation dynamics~\cite{Bajgar2020Orbis}.

\subsection{Network analysis at the macroscale}

We characterize each collaboration network, the adjacency matrix of which will be indicated as $\mathbf{A}=\{a_{ij}\}_{i,j=1}^N$, as follows~\cite{newman2018networks}: upon indicating the total number of nodes with $N$ and the total number of edges with $L$, one can define the link density as

\begin{equation}
\rho=\frac{2L}{N(N-1)};
\end{equation}
a high value of link density indicates more pervasive interconnections and is often associated with mature, or well-integrated, technological fields.

The number of neighbors of node $i$ is captured by the degree

\begin{equation}
k_i=\sum_{j=1}^Na_{ij};
\end{equation}
from it, we can compute the average degree $\bar{k}=N^{-1}\sum_{i=1}^Nk_i$ and the degree standard deviation $\sigma_k=\sqrt{N^{-1}\sum_{i=1}^N\left(k_i-\bar{k}\right)^2}$, the ratio of which quantifies the relative dispersion of collaborations: values of the coefficient of variation $\text{CV}=\sigma_k/\bar{k}$ exceeding $1$ indicate that the first moment is not representative of the distribution or, equivalently, that such a distribution can be considered as representing a heterogeneous system. The concentration of connectivity can be further assessed by computing the Nakamoto index, i.e. the minimum number of nodes, ranked by degree, required to account for the $51\%$ of the total number of connections~\cite{srinivasan2017quantifying, Lin2022WeightedLightning}: smaller values indicate more unequal distributions dominated by few hubs.

The extent to which a collaboration network is cohesive can, instead, be quantified by computing the average clustering coefficient $\bar{c}=N^{-1}\sum_{i=1}^Nc_i$, where

\begin{align}
c_i=\frac{2t_i}{k_i(k_i-1)},
\end{align}
with $t_i=\sum_{j=1}^N\sum_{k(>j)}a_{ij}a_{jk}a_{ki}$ indicating the number of triangles incident to node $i$.

Finally, the degree distribution and the clustering coefficient distribution read $P(k)=N^{-1}\sum_{i=1}^N\mathbf{1}\{k_i=k\}$ and $P(c)=N^{-1}\sum_{i=1}^N\mathbf{1}\{c_i=c\}$. The degree distribution reveals the heterogeneity of collaborations, with heavy tails indicating the presence of few hubs alongside many, peripheral vertices; the clustering coefficient distribution quantifies the heterogeneity of local cohesion, thus complementing the aforementioned information.

To characterize the degrees, we fitted two candidate distributions on each network, i.e. a lognormal and a power-law. For each network, the lower cutoff was identified with the minimum degree, so that the full degree sequence was used in estimation. Parameters were estimated by maximizing the log-likelihood of the corresponding distribution~\cite{clauset2009power} and goodness-of-fit was assessed via the Kolmogorov-Smirnov test; lastly, lognormal and power-law distributions were compared via the Likelihood-Ratio Test.

\subsection{Network analysis at the mesoscale}

The problem of community detection has been extensively studied, leading to the development of numerous methods: among these, optimization-based approaches are, by far, the most popular ones.

\subsubsection{Maximization of the modularity}

The first, and most representative, method of the aforementioned class is the one prescribing to maximize modularity, defined as

\begin{align}
Q=\frac{1}{2L}\sum_{i=1}^N\sum_j\left(a_{ij}-\frac{k_ik_j}{2L}\right)\delta_{g_ig_j},
\end{align}
where $\delta_{g_ig_j}$ is the Kronecker-delta function, equaling $1$ if nodes $i$ and $j$ belong to the same group (i.e. $g_i=g_j$) and $0$ otherwise (i.e. $g_i\neq g_j$), and $p_{ij}=k_ik_j/2L$ is the probability that $i$ and $j$ are connected under the Chung-Lu model~\cite{chung2002connected,newman2006modularity}. Therefore, maximizing $Q$ identifies the partition characterized by the number of edges within groups exceeding, to the largest extent, what would be expected by chance. 

Despite its widespread use, modularity maximization is affected by both conceptual and practical flaws. As shown in~\cite{guimera2004modularity}, and later confirmed in~\cite{good2010performance}, a key limitation arises from the presence of a very rugged landscape: in other words, $Q$ often admits many, distinct partitions corresponding to nearly-identical, high scores, hence lacking a clearly-identifiable global maximum. Furthermore, in~\cite{guimera2004modularity} the authors demonstrate that even random graphs (i.e. without a modular structure) can display relatively high modularity scores, depending on their size and average degree.

A second, major issue is represented by the so-called \textit{resolution limit}~\cite{fortunato2007resolution}, whereby small, yet well-defined, communities may remain undetected, as the optimization procedure tends to merge them into larger groups. Such a problem is not specific to any network structure, as the size of the modules that $Q$ fails to identify depends on $L$ as well as on the number of intermodular connections.

Aware of the limitations of the modularity optimization, we have compared its performance with the one of genuinely inferential approaches~\cite{peixoto2014hierarchical,peixoto2017nonparametric,peixoto2019bayesian}.

\subsubsection{Minimization of the Bayesian Information Criterion}

BIC is an information criterion widely employed to compare statistical models~\cite{Burnham2002ModelSelection}: it embodies a trade-off between accuracy and complexity, penalizing models with too many parameters. More formally, the BIC of a probabilistic model with log-likelihood $\mathcal L$ is defined as

\begin{align}\label{eq:bic}
\text{BIC}=-2\mathcal L+\kappa\ln V,
\end{align}
where $\kappa$ denotes the number of parameters to be tuned and $V$ denotes the size of the system under consideration. Assuming that a given, real-world configuration is the output of a generative model probabilistic in nature, BIC minimization can be employed for mesoscale structures detection: in such a case, it would compare different partitions - not just those consisting of modules - and select the optimal one. In what follows, we will instantiate BIC with the Stochastic Block Model and its degree-corrected variant (see Appendix~\ref{AppB} for the calculations and Appendix~\ref{AppC} for the codes).\\

\noindent\textit{Stochastic Block Model (SBM).} Formally, it reads

\begin{align}
\mathscr{L}_\text{SBM}=&\prod_{r=1}^Bp_{rr}^{L_{rr}}(1-p_{rr})^{\binom{N_r}{2}-L_{rr}}\nonumber\\
&\cdot\prod_{r=1}^B\prod_{s(>r)}p_{rs}^{L_{rs}}(1-p_{rs})^{N_rN_s-L_{rs}},
\end{align}
where $N_r$ is the number of nodes constituting block $r$, $L_{rr}$ is the number of links within block $r$, $L_{rs}$ is the number of links between blocks $r$ and $s$, $\forall\:r<s$; the probability coefficients read $p_{rr}=2L_{rr}/N_r(N_r-1)$, $r=1\dots B$ and $p_{rs}=L_{rs}/N_rN_s$, $\forall\:r<s$ (see Appendix~\ref{AppB} for the derivation of the SBM within the framework of the ERGs).\\

\noindent\textit{Degree-corrected Stochastic Block Model (dcSBM).} Under the SBM, the nodes within the same block are treated as statistically equivalent. Real-world networks, however, are typically heterogeneous. In order to properly account for this, we have also considered the degree-corrected variant of the SBM, reading

\begin{align}
\mathscr{L}_\text{dcSBM}=\sum_{i=1}^N\sum_{j(>i)}[a_{ij}\ln p_{ij}+(1-a_{ij})\ln(1-p_{ij})],
\end{align}
where the generic probability coefficient is defined as

\begin{align}
    p_{ij}=\frac{x_ix_j\chi_{g_ig_j}}{1+x_ix_j\chi_{g_ig_j}}
\end{align}
where $x_i$ is the multiplier controlling for the degree of node $i$, $x_j$ is the multiplier controlling for the degree of node $j$ and $\chi_{g_ig_j}$ is the multiplier controlling for the link density between groups $g_i$ and $g_j$ (see Appendix~\ref{AppB} for the derivation of the dcSBM within the framework of the ERGs).\\

Our BIC-based approach was compared with the MDL-based one, implemented via \texttt{graph-tool} (see Appendix~\ref{AppD}), and other BIC-related ones (see Appendix~\ref{AppE}), aiming at enriching the `plain' definition provided in Eq.~\ref{eq:bic} with additional terms.

\subsubsection{Characterization of the partitions}

In order to describe the composition of the partition identified by each algorithm, we perform complementary analyses at the levels of individuals and organizations, further characterizing each cluster through a set of indicators that summarize its internal organization.

More specifically, we compute the cluster-specific average degree and degree standard deviation as well as the ratio between intra-cluster and inter-cluster connectivity (in symbols, IC/EC): while the first two quantities measure the average intensity of collaborations and the extent to which they are (un)evenly distributed across a cluster members, the IC/EC ratio is defined as the number of intra-cluster edges divided by the number of inter-cluster edges, thus capturing how sharply clusters are separated from the rest of the network: in words, high (low) values indicate less (more) integrated communities with the rest of the network.\\

\noindent\textit{Inventor-level.} At the level of inventors, we assess the diversity of each cluster along two, complementary dimensions: \textit{i)} the degree of \textit{organizational diversity}, measured by the number of distinct patent owners within each cluster, $D_{\text{own}}^{(r)}$, and distinguishing \textit{single-company} clusters ($D_{\text{own}}^{(r)}=1$) from \textit{multi-company} ones ($D_{\text{own}}^{(r)}>1$); \textit{ii)} the degree of \textit{technological diversity}, measured through the normalized Shannon entropy of IPC codes at the most granular level of classification, namely main groups~\cite{shannon1948mathematical}. For each cluster $r$, let $p_c^{(r)}$ denote the share of patents classified under IPC code $c$; the entropy is, then, defined as

\begin{align}
D_{\text{IPC}}^{(r)}=-N_\text{sector}^{-1}\sum_cp_c^{(r)}\ln p_c^{(r)},
\end{align}
where $N_\text{sector}$ is the total number of patents within a specific domain: Shannon entropy is particularly suitable in this context as it does not simply count the number of distinct IPC classes - which would only reflect variety - but also captures how evenly patents are distributed across those classes~\cite{zhang2017entropy}. It, therefore, increases when a cluster covers multiple technological domains with comparable intensity and decreases when patents are concentrated within a few IPC categories; besides, it represents a scale-independent measure of diversity that can be employed to carry out consistent comparisons across sectors and models. The combined use of the indicators above reveals whether a given cluster can be deemed as (primarily) representing intra-firm specialization, inter-firm collaborations, or multi-firm technological integration.\\

\noindent\textit{Organization-level.} At the level of organizations, we perform a qualitative examination of the clusters detected by each algorithm: for every sector and model, the largest clusters have been manually inspected to identify their institutional composition, the dominant firms and the scope (either national or regional). This information is retrieved from patent assignee names and cross-checked with publicly available corporate and institutional affiliations. The resulting summary is reported in Table~\ref{tab:communities}, providing a concise description of the structure captured by each algorithm (e.g. corporate clusters, national innovation systems, vertically-integrated industrial ecosystems).

\subsection{Lorenz curves of forward citations}

In order to quantify the heterogeneity of inventive impact across the mesoscale structure of each network, we compute the Lorenz curve~\cite{lorenz1905methods} and Gini coefficient~\cite{cowell2011measuring} for the corresponding distribution of patent forward citations. Given the model $M$ and the network level $\ell$ (either inventors or organizations), let $f_c^{(r)}$ denote the total number of forward citations received by all patents associated with cluster $r=1\dots k_M^\ell$. Upon ordering clusters by increasing value of $f_c^{(r)}$, the cumulative shares read

\begin{align}
X_j&=\frac{j}{k_M^\ell},\\
Y_j&=\frac{\sum_{r=1}^jf_c^{(r)}}{\sum_{r=1}^{k_M^\ell} f_c^{(r)}}
\end{align}
and plotting the pairs $(X_j,Y_j)$ yields the Lorenz curve, where $X_j$ is the cumulative fraction of partitions and $Y_j$ the cumulative fraction of total citations. With an abuse of notation, the corresponding Gini coefficient is computed as

\begin{align}
G=1-2\int_0^1Y(X)dX;
\end{align}
while a perfectly equal distribution corresponds to $G=0$, the value $G=1$ indicates that all citations are concentrated in a single cluster (see Also Appendix~\ref{AppD} for the calculation of the curves on the top-$300$ and top-$700$ representations of our system).

\subsection{Robustness checks}

For robustness, we have replicated the analysis described above on the configurations defined by the top-$300$ and the top-$700$ nodes of both inventor and organization networks: results are stable upon varying the threshold (see Appendix~\ref{AppF}).

\section{Acknowledgments}

This work has been supported by the projects: `RE-Net - Reconstructing economic networks: from physics to machine learning and back' - 2022MTBB22, Funded by the European Union Next Generation EU, PNRR Mission 4 Component 2 Investment 1.1, CUP: D53D23002330006; `C2T - From Crises to Theory: towards a science of resilience and recovery for economic and financial systems' - P2022E93B8, Funded by the European Union Next Generation EU, PNRR Mission 4 Component 2 Investment 1.1, CUP: D53D23019330001; `SoBigData RI PPP - SoBigData RI Preparatory Phase Project', funded by the European Union under the scheme HORIZON-INFRA-2021-DEV-02-01, preparatory phase of new ESFRI research infrastructure projects, G.A. 101079043; `FAIR - Future Artificial Intelligence Research' - Spoke 1 `Human-centered AI', funded by the European Commission under the Next Generation EU program, PNRR Mission 4 Component 2 Investment 3.1, G.A. PE00000013; SMaRT COnSTRUCT (CUP J53C24001460006), in the context of FAIR (PE0000013, CUP B53C22003630006) under the National Recovery and Resilience Plan (Mission 4, Component 2, Line of Investment 1.3) funded by the European Union - NextGenerationEU.

The authors are grateful to the participants of FAIR 2025, the workshop COMPASS, and NetSci 2026 for their valuable feedback.

\section{Code and Data availability}

The Python package named \texttt{DOMINO} (\textit{Detection Of Mesoscale structures via INfOrmation criteria}), implementing the algorithms described in the main text, is available on PyPI and at the URL \url{https://github.com/mattiamarzi/DOMINO}. The underlying patent-level data were obtained from Orbis Intellectual Property (Moody's), a proprietary database subject to licensing restrictions, and therefore cannot be publicly shared.

\section{Author contributions}

Study conception and design: AM, AV, TS. Literature review: AG, LE. Data collection: AM, AV, LE. Analysis and interpretation of results: AG, AM, AV, LE, MM, TS. Draft manuscript preparation: AG, LE, MM, TS. Draft manuscript revision: AG, AM, AV, LE, MM, TS. Python package preparation: MM.

\section{Competing interests}

The authors declare no competing interests.

\bibliography{biblio.bib}

\clearpage

\onecolumngrid

\appendix

\hypertarget{AppA}{}
\section{Classification of Artificial Intelligence, Biotechnology and Semiconductors patents}\label{AppA}

\setcounter{figure}{0}
\renewcommand\thefigure{A.\arabic{figure}}
\setcounter{table}{0}
\renewcommand\thetable{A.\Roman{table}}

To identify the patents belonging to artificial intelligence, biotechnology and semiconductor technologies, we adopted the following classification, ensuring that our sectorial delineation is consistent with prior work and internationally-recognized standards: AI patents were identified following the taxonomy of the \textit{WIPO Technology Trends 2019: Artificial Intelligence} report~\cite{WIPO2019AI}, mapping CPC codes across AI subdomains such as computer vision, natural language processing and control methods; BT patents were identified following the taxonomy of the \textit{Exploring the Global Landscape of Biotech Innovation} report by the European Commission Joint Research Center~\cite{Grassano2024Biotech}; SC patents were identified following the IPC class H01L (semiconductor devices) proposed in~\cite{Adams2013Semiconductors}. An overview of the codes used to identify patents is provided in Table~\ref{tab:cpc_codes}.

\begin{table}[ht!]
\centering
\begin{tabular}{p{3.4cm}|p{12cm}}
\hline
\hline
&\multicolumn{1}{c}{\textbf{Relevant CPC/IPC codes and categories}}\\
\hline
\hline
\textbf{Artificial intelligence} & \textbf{Computer vision and image processing:} G06T1/20, G06T2207/20081, G06T2207/20084, G06T3/4046, G06T9/002
\newline
\textbf{Augmented/mixed reality:} G02B27/01, G06T19/006, G06F3/011, G05B2219/32014
\newline
\textbf{Biometrics and character recognition:} G06K9/00006, G06K9/00221, G06K2009/00395, G06K9/00375, G06K9/00597, G06K9/00885, G06K9/00362, G06K9/00335, G06K9/00402, G06K9/00442, G06K2209/01, G06K9/00852
\newline
\textbf{Segmentation, tracking, scene understanding:} G06T7/10, G06T7/215, H04N5/147, G06K9/34, H04N1/40062, G06F3/012, G06K2017/0045, G06T7/246, G08B13/2402, G06K9/00624, G06T1/0014, G06K9/00798
\newline
\textbf{Control methods:} G05B13/0265, G05B13/027, G05B13/0275, G05B13/028, G05B13/0285, G05B13/029, G05B13/0295, G05D1/0088
\newline
\textbf{Natural language processing and knowledge representation:} G06F17/279, G06F17/2765, G06F17/2705, G06F17/28, G06F17/30669, G06F17/2755, G06F17/2881, G06F17/2282, G06F17/30401, G06F17/3043, G06F17/30654, G06F17/30663, G06F17/30666, G06F17/30731, G06F17/2785, G06N5/00
\newline
\textbf{Speech processing:} G10L13/00, G10L15/00, G10L17/00, G10L25/00, G10L99/00, G06F17/30784\\
\hline
\textbf{Biotechnology} & A01H1/00, A01H4/00, A01K67/00, A61K35/12, A61K38/00, A61K38/17, A61K39/00, A61K39/395, A61K48/00, C02F3/34, C07G11/00, C07G13/00, C07G15/00, C07K4/00, C07K14/00, C07K16/00, C07K16/28, C07K17/00, C07K19/00, C12M1/00, C12M1/34, C12N15/82, C12N15/113, C12P, C12Q, C12Q1/68, C40B10/00, C40B40/02, C40B40/06, C40B40/08, C40B50/06, G01N27/327, G01N33/50, G01N33/53, G01N33/54, G01N33/543, G01N33/55, G01N33/57, G01N33/574, G01N33/68, G01N33/74, G01N33/76, G01N33/78, G01N33/88, G01N33/92, G06F19/10\\
\hline
\textbf{Semiconductors} & H01L\\
\hline
\hline
\end{tabular}
\caption{\textbf{CPC/IPC codes used to identify patents in Artificial Intelligence, Biotechnology and Semiconductors.}
\label{tab:cpc_codes}}
\end{table}

\clearpage

\hypertarget{AppB}{}
\section{Stochastic block models as Exponential Random Graphs}\label{AppB}

\setcounter{figure}{0}
\renewcommand\thefigure{B.\arabic{figure}}
\setcounter{table}{0}
\renewcommand\thetable{B.\Roman{table}}

Both the SBM and the dcSBM can be derived within the framework of ERGs, following the approach introduced in~\cite{park2004statistical} and further developed in~\cite{Squartini2011a}. Such an approach looks for the probability distribution $P(\mathbf A)$ that maximizes Shannon entropy

\begin{align}
S=-\sum_{\mathbf A\in\mathbb A}P(\mathbf A)\ln P(\mathbf A),
\end{align}
where the sum runs over $\mathbb A$, i.e. the ensemble of all, possible binary, undirected graphs, under a set of constraints representing the expected values of certain properties. The formal solution to this problem reads

\begin{align}
P(\mathbf A)=\frac{e^{-H(\mathbf A)}}{Z}
\end{align}
where $H(\mathbf A)$ is the Hamiltonian, i.e. a linear combination of the constrained properties, each multiplied by the corresponding Lagrange multiplier, and $Z=\sum_{\mathbf A\in\mathbb A}e^{-H(\mathbf A)}$ is the partition function.

\subsection{Stochastic Block Model (SBM)}

In the framework of ERGs, the SBM is defined by the Hamiltonian

\begin{align}
H(\mathbf{A})&=\sum_{i=1}^N\sum_{j(>i)}\alpha_{g_ig_j}a_{ij}=\sum_{r=1}^B\sum_{s(\geq r)}\alpha_{rs}\sum_{i=1}^N\sum_{j(>i)}\delta_{g_ir}\delta_{g_{js}}a_{ij}=\sum_{r=1}^B\sum_{s(\geq r)}\alpha_{rs}L_{rs}(\mathbf{A}),
\end{align}
where $B$ is the total number of clusters and $g_i$ denotes the group membership of node $i$\footnote{Throughout the manuscript, whenever a sum runs over $r\leq s$, the product $\delta_{g_ir}\delta_{g_js}$ is understood as the shorthand for $\delta_{g_ir}\delta_{g_js}+\delta_{g_is}\delta_{g_jr}-\delta_{rs}\delta_{g_ir}\delta_{g_jr}$, so as to account for both node membership assignments.}. It induces the partition function reading

\begin{align}
Z=\sum_{\mathbf{A}\in\mathbb{A}}e^{-H(\mathbf{A})}=\sum_{\mathbf{A}\in\mathbb{A}}\prod_{i=1}^N\prod_{j(>i)}e^{-\alpha_{g_ig_j} a_{ij}}=\prod_{i=1}^N\prod_{j(>i)}\sum_{a_{ij}=0,1}e^{-\alpha_{g_ig_j} a_{ij}}=\prod_{i=1}^N\prod_{j(>i)}\left[1+e^{-\alpha_{g_ig_j}}\right];
\end{align}
as a consequence, we have

\begin{align}
P_\text{SBM}(\mathbf{A})&=\frac{\prod_{i=1}^N\prod_{j(>i)}e^{-\alpha_{g_ig_j}a_{ij}}}{\prod_{i=1}^N\prod_{j(>i)}[1+e^{-\alpha_{g_ig_j}}]}=\prod_{i=1}^N\prod_{j(>i)}\frac{x_{g_ig_j}^{a_{ij}}}{1+x_{g_ig_j}}=\prod_{i=1}^N\prod_{j(>i)}p_{g_ig_j}^{a_{ij}}(1-p_{g_ig_j})^{1-a_{ij}},
\end{align}
having posed $e^{-\alpha_{g_ig_j}}=x_{g_ig_j}$ and $p_{ij}=x_{g_ig_j}/(1+x_{g_ig_j})$. Let us notice that

\begin{align}
P_\text{SBM}(\mathbf{A})&=\prod_{i=1}^N\prod_{j(>i)}p_{g_ig_j}^{a_{ij}}(1-p_{g_ig_j})^{1-a_{ij}}\nonumber\\
&=\prod_{i=1}^N\prod_{j(>i)}\prod_{r=1}^B\prod_{s(\geq r)}\left[p_{rs}^{a_{ij}}(1-p_{rs})^{1-a_{ij}}\right]^{\delta_{g_ir}\delta_{g_js}}\nonumber\\
&=\prod_{r=1}^B\prod_{s(\geq r)}\prod_{i=1}^N\prod_{j(>i)}\left[p_{rs}^{a_{ij}\delta_{g_ir}\delta_{g_js}}(1-p_{rs})^{\delta_{g_ir}\delta_{g_js}(1-a_{ij})}\right]\nonumber\\
&=\prod_{r=1}^B\prod_{s(\geq r)}\left[p_{rs}^{\sum_{i=1}^N\sum_{j(>i)}\delta_{g_ir}\delta_{g_js}a_{ij}}(1-p_{rs})^{\sum_{i=1}^N\sum_{j(>i)}\delta_{g_ir}\delta_{g_js}(1-a_{ij})}\right]\nonumber\\
&=\prod_{r=1}^Bp_{rr}^{L_{rr}}(1-p_{rr})^{\binom{N_r}{2}-L_{rr}}\prod_{s(>r)}p_{rs}^{L_{rs}}(1-p_{rs})^{N_rN_s-L_{rs}},
\end{align}
where $N_r$ is the total number of nodes in the $r$-th block, $r=1\dots B$. The parameters that define the SBM can be estimated by maximizing the log-likelihood

\begin{align}
\mathscr{L}_\text{SBM}&=\ln P_\text{SBM}(\mathbf A)=\sum_{r=1}^B\left[L_{rr}(\mathbf{A})\ln x_{rr}-\binom{N_r}{2}\ln(1+x_{rr})\right]+\sum_{r=1}^B\sum_{s(>r)}[L_{rs}(\mathbf{A})\ln x_{rs}-N_rN_s\ln(1+x_{rs})]
\end{align}
with respect to $x_{rr}$ and $x_{rs}$. Upon doing so, we obtain the system of equations

\begin{align}
\frac{\partial\mathscr{L}_\text{SBM}}{\partial x_{rr}}&=\frac{L_{rr}(\mathbf A)}{x_{rr}}-\binom{N_r}{2}\left(\frac{1}{1+x_{rr}}\right),\\
\frac{\partial\mathscr{L}_\text{SBM}}{\partial x_{rs}}&=\frac{L_{rs}(\mathbf A)}{x_{rs}}-N_rN_s\left(\frac{1}{1+x_{rs}}\right)
\end{align}
and equating them to zero leads to the conditions

\begin{align}
p_{rr}&=\frac{2L_{rr}(\mathbf{A})}{N_r(N_r-1)},\quad r=1\dots B\\
p_{rs}&=\frac{L_{rs}(\mathbf{A})}{N_rN_s},\quad\forall\:r<s.
\end{align}

The expression of BIC, thus, reads

\begin{align}
\text{BIC}_\text{SBM}=-2\mathscr{L}_\text{SBM}+\kappa_\text{SBM}\ln V,
\end{align}
with $\kappa_\text{SBM}=B(B+1)/2$.

\subsection{Degree-corrected Stochastic Block Model (dcSBM)}

The dcSBM is, instead, defined by the Hamiltonian

\begin{align}
H(\mathbf A)&=\sum_{i=1}^N\sum_{j(>i)}\alpha_{ij}a_{ij}=\sum_{i=1}^N\sum_{j(>i)}(\alpha_i+\alpha_j+\alpha_{g_ig_j})a_{ij},
\end{align}
leading to the partition function

\begin{align}
Z&=\sum_{\mathbf A\in\mathbb A}e^{-H(\mathbf A)}=\sum_{\mathbf A\in\mathbb A}\prod_{i=1}^N\prod_{j(>i)}e^{-\left(\alpha_i+\alpha_j+\alpha_{g_ig_j}\right)a_{ij}}=\prod_{i=1}^N\prod_{j(>i)}\sum_{a_{ij}=0,1}e^{-\left(\alpha_i+\alpha_j+\alpha_{g_ig_j}\right)a_{ij}}=\prod_{i=1}^N\prod_{j(>i)}\left[1+e^{-\left(\alpha_i+\alpha_j+\alpha_{g_ig_j}\right)}\right],
\end{align}
further implying

\begin{align}
P_\text{dcSBM}(\mathbf A)&=\frac{\prod_{i=1}^N\prod_{j(>i)}e^{-\left(\alpha_i+\alpha_j+\alpha_{g_ig_j}\right)a_{ij}}}{\prod_{i=1}^N\prod_{j(>i)}\left[1+e^{-\left(\alpha_i+\alpha_j+\alpha_{g_ig_j}\right)}\right]}=\prod_{i=1}^N\prod_{j(>i)}\frac{(x_ix_j\chi_{g_ig_j})^{a_{ij}}}{1+x_ix_j\chi_{g_ig_j}},
\end{align}
having posed $x_i=e^{-\alpha_i}$ and $\chi_{g_ig_j}=e^{-\alpha_{g_ig_j}}$. The parameters that define the dcSBM can be estimated by maximizing the log-likelihood

\begin{align}
\mathscr{L}_\text{dcSBM}=\ln P_\text{dcSBM}(\mathbf A)&=\sum_{i=1}^N\sum_{j(>i)}[a_{ij}\ln(x_ix_j\chi_{g_ig_j})-\ln(1+x_ix_j\chi_{g_ig_j})]\nonumber\\
&=\sum_{i=1}^N\sum_{j(>i)}\sum_{r=1}^B\sum_{s(\geq r)}\delta_{g_ir}\delta_{g_js}[a_{ij}\ln(x_ix_j\chi_{rs})-\ln(1+x_ix_j\chi_{rs})]
\end{align}
with respect to $x_i$ and $\chi_{rs}$. Upon doing so, we obtain the system of equations

\begin{align}
\frac{\partial\mathscr{L}_\text{dcSBM}}{\partial x_i}&=\sum_{j(\neq i)}\sum_{r=1}^B\sum_{s(\geq r)}\delta_{g_ir}\delta_{g_js}\left(\frac{a_{ij}}{x_i}-\frac{x_j\chi_{rs}}{1+x_ix_j\chi_{rs}}\right),\\
\frac{\partial\mathscr{L}_\text{dcSBM}}{\partial\chi_{rs}}&=\sum_{i=1}^N\sum_{j(>i)}\delta_{g_ir}\delta_{g_js}\left(\frac{a_{ij}}{\chi_{rs}}-\frac{x_ix_j}{1+x_ix_j\chi_{rs}}\right)
\end{align}
and equating them to zero leads us to recover the conditions

\begin{align}
k_i(\mathbf A)&=\sum_{j(\neq i)}\sum_{r=1}^B\sum_{s(\geq r)}\delta_{g_ir}\delta_{g_js}\left(\frac{x_ix_j\chi_{rs}}{1+x_ix_j\chi_{rs}}\right),\quad i=1\dots N\\
L_{rs}(\mathbf A)&=\sum_{i=1}^N\sum_{j(>i)}\delta_{g_ir}\delta_{g_js}\left(\frac{x_ix_j\chi_{rs}}{1+x_ix_j\chi_{rs}}\right),\quad\forall\:r\leq s.
\end{align}

Since the resolution of the dcSBM is computationally demanding, here we have considered a `decoupled' approximation of it, defined by the following algorithm: first, we have solved the Undirected Binary Configuration Model (UBCM)~\cite{park2004statistical,Squartini2011a}, defined by the system of equations

\begin{align}
k_i(\mathbf A)&=\sum_{j(\neq i)}\left(\frac{x_i^\text{UBCM}x_j^\text{UBCM}}{1+x_i^\text{UBCM}x_j^\text{UBCM}}\right),\quad i=1\dots N
\end{align}
and, afterwards, we have solved the system of equations

\begin{align}
L_{rs}(\mathbf A)&=\sum_{i=1}^N\sum_{j(>i)}\delta_{g_ir}\delta_{g_js}\left(\frac{x_i^\text{UBCM}x_j^\text{UBCM}\chi_{rs}}{1+x_i^\text{UBCM}x_j^\text{UBCM}\chi_{rs}}\right),\quad\forall\:r\leq s
\end{align}
(with clear meaning of the symbols). The expression of BIC, thus, reads

\begin{align}
\text{BIC}_\text{dcSBM}=-2\mathscr{L}_\text{dcSBM}+\kappa_\text{dcSBM}\ln V,
\end{align}
with $\kappa_\text{dcSBM}=B(B+1)/2+N$.

\clearpage

\hypertarget{AppC}{}
\section{Codes for the implementation of the BIC-based community detection schemes}\label{AppC}

\setcounter{figure}{0}
\renewcommand\thefigure{C.\arabic{figure}}
\setcounter{table}{0}
\renewcommand\thetable{C.\Roman{table}}

The code operates on binary, undirected networks and implements a community detection framework resting upon the optimization of the Bayesian Information Criterion.

\begin{algorithm}[b!]
\caption{Leiden algorithm for the exploration of partitions}
\begin{algorithmic}
\item[\hspace{1.4pt}1: \textbf{function} \textit{Leiden}$(G,\text{quality},\text{initial})$]
\item[\hspace{1.4pt}2: \quad \textbf{if} an initial partition is provided \textbf{then}]
\item[\hspace{1.4pt}3: \quad\quad initialize each node to its community in the input partition]
\item[\hspace{1.4pt}4: \quad \textbf{else}]
\item[\hspace{1.4pt}5: \quad\quad assign each node to its own singleton community]
\item[\hspace{1.4pt}6: \quad \textbf{end if}]
\item[\hspace{1.4pt}7: \quad $\text{prev} \leftarrow \text{None}$]
\item[\hspace{1.4pt}8: \quad \textbf{while True do}]
\item[\hspace{1.4pt}9: \quad\quad randomly shuffle the nodes into a list $Q$]
\item[\hspace{1.4pt}10: \quad\quad \textbf{while} $Q \neq \emptyset$ \textbf{do}]
\item[\hspace{1.4pt}11: \quad\quad\quad pop first node $v$ from $Q$]
\item[\hspace{1.4pt}12: \quad\quad\quad let $C_v$ be the community of $v$]
\item[\hspace{1.4pt}13: \quad\quad\quad find communities of neighbors of $v$ and add an empty community as candidate]
\item[\hspace{1.4pt}14: \quad\quad\quad \textbf{for} each candidate community $C'$ \textbf{do}]
\item[\hspace{1.4pt}15: \quad\quad\quad\quad compute $\Delta Q =$ quality delta for moving $v$ to $C'$]
\item[\hspace{1.4pt}16: \quad\quad\quad \textbf{end for}]
\item[\hspace{1.4pt}17: \quad\quad\quad choose a community $C^{\ast}$ that gives the maximum positive $\Delta Q$ (if any)]
\item[\hspace{1.4pt}18: \quad\quad\quad \textbf{if} $\Delta Q > 0$ \textbf{then}]
\item[\hspace{1.4pt}19: \quad\quad\quad\quad move $v$ to $C^{\ast}$]
\item[\hspace{1.4pt}20: \quad\quad\quad\quad for each neighbor $u$ of $v$ that does not belong to $C^{\ast}$, add $u$ to $Q$ (if not already present)]
\item[\hspace{1.4pt}21: \quad\quad\quad \textbf{end if}]
\item[\hspace{1.4pt}22: \quad\quad \textbf{end while}]
\item[\hspace{1.4pt}23: \quad\quad \textbf{if} no node moved or all communities are singletons \textbf{then}]
\item[\hspace{1.4pt}24: \quad\quad\quad \textbf{return} current community assignment as flat partition]
\item[\hspace{1.4pt}25: \quad\quad \textbf{end if}]
\item[\hspace{1.4pt}26: \quad\quad $\text{prev} \leftarrow$ current partition]
\item[\hspace{1.4pt}27: \quad\quad start refinement step]
\item[\hspace{1.4pt}28: \quad\quad assign each node to its own singleton community in a refined partition]
\item[\hspace{1.4pt}29: \quad\quad \textbf{for} each community $C$ in previous partition \textbf{do}]
\item[\hspace{1.4pt}30: \quad\quad\quad let $T$ be the set of nodes in $C$]
\item[\hspace{1.4pt}31: \quad\quad\quad run a local Leiden pass on the induced subgraph on $T$, initialized from the current labels]
\item[\hspace{1.4pt}32: \quad\quad\quad update the refined partition within $T$ according to this local optimization]
\item[\hspace{1.4pt}33: \quad\quad \textbf{end for}]
\item[\hspace{1.4pt}34: \quad\quad aggregate nodes of the refined partition into supernodes]
\item[\hspace{1.4pt}35: \quad\quad build a new graph $G'$ where each node is a refined community]
\item[\hspace{1.4pt}36: \quad\quad lift the refined partition to $G'$ by assigning each supernode to a community label]
\item[\hspace{1.4pt}37: \quad\quad replace $G$ with $G'$ and continue]
\item[\hspace{1.4pt}38: \quad \textbf{end while}]
\item[\hspace{1.4pt}39: \textbf{end function}]
\end{algorithmic}
\label{alg:leiden_bic}
\end{algorithm}

Our mesoscale detection scheme is based on outer iterations that repeatedly rerun Leiden starting from the last available partition. A first run of Leiden is performed starting from either singleton communities or a user-supplied initialization; the resulting partition is, then, used as initial condition for a new Leiden run that uses the same BIC-based score function. After each run, the BIC of the resulting partition is evaluated and compared with the best value observed until then. The procedure is repeated until the partition becomes stable - meaning that two, consecutive outer runs return the same assignment - or the maximum number of outer iterations is reached. The algorithm returns the partition that achieves the lowest BIC across all runs.

The Leiden implementation follows the standard, three-step structure, consisting of \textit{i)} greedy node movement, \textit{ii)} refinement and \textit{iii)} aggregation. At the first step, nodes are visited in random order and moved to neighboring communities whenever this produces a positive gain in the quality score, i.e. a reduction of the BIC value. During the refinement step, each community from the previous iteration is further split by a local Leiden pass restricted to the nodes inside that community: this guarantees that each community in the refined partition is internally well-connected. At the aggregation step, communities are collapsed into supernodes and a coarse grained graph is built by summing the entries of the adjacency matrix over the corresponding node sets. The refined partition is lifted to this coarse graph, that becomes the input for the next iteration. The inner Leiden loop stops when no node moves or when a stopping criterion is reached.

Our framework exposes a single quality function interface, with specializations for the SBM or its degree-corrected variant. For what concerns the SBM, the probability of a link between nodes $i$ and $j$ depends only on their community labels. The model parameters are block-wise probabilities, estimated from the observed adjacency matrix and the candidate partition by counting the (missing) links between communities. The corresponding log-likelihood and BIC values are computed once per partition and used to define the quality score used by Leiden.

\begin{algorithm}[h!]
\caption{Community detection by minimizing BIC-SBM}
\begin{algorithmic}
\item[\hspace{1.4pt}1:] \textbf{function} \textit{iterative\_leiden\_SBM}$(G, \mathbf A, \dots)$
\item[\hspace{1.4pt}2:] \quad $qf \leftarrow \text{SBM\_BIC\_Quality}(\mathbf A)$
\item[\hspace{1.4pt}3:] \quad $\text{part} \leftarrow \text{Leiden}(G, qf, \text{initial = None})$
\item[\hspace{1.4pt}4:] \quad $\text{best\_part} \leftarrow \text{part}$, $\text{best\_bic} \leftarrow \text{BIC\_SBM}(\mathbf A,\text{part})$
\item[\hspace{1.4pt}5:] \quad \textbf{for} $i = 1 \dots \text{max\_outer}$ \textbf{do}]
\item[\hspace{1.4pt}6:] \quad\quad $\text{new\_part} \leftarrow \text{Leiden}(G, qf, \text{initial = best\_part})$
\item[\hspace{1.4pt}7:] \quad\quad \textbf{if} $\text{new\_part} = \text{best\_part}$ \textbf{then break}]
\item[\hspace{1.4pt}8:] \quad\quad $\text{new\_bic} \leftarrow \text{BIC\_SBM}(\mathbf A,\text{new\_part})$
\item[\hspace{1.4pt}9:] \quad\quad \textbf{if} $\text{new\_bic} < \text{best\_bic}$ \textbf{then}]
\item[\hspace{1.4pt}10:] \quad\quad\quad $\text{best\_part} \leftarrow \text{new\_part}$, $\text{best\_bic} \leftarrow \text{new\_bic}$
\item[\hspace{1.4pt}11:] \quad\quad \textbf{end if}]
\item[\hspace{1.4pt}12:] \quad \textbf{end for}]
\item[\hspace{1.4pt}13:] \quad \textbf{return} $\text{best\_part}, \text{best\_bic}$]
\item[\hspace{1.4pt}14:] \textbf{end function}]
\end{algorithmic}
\end{algorithm}

The dcSBM introduces node-specific degree parameters and block-specific coefficients. Since a direct calculation of all parameters at every step would be computationally expensive and prone to convergence issues, our implementation follows a two-step approach. First, the UBCM is solved through a root finding routine: this provides an approximate set of degree parameters that is kept fixed throughout all outer Leiden iterations. Given these parameters, the code runs Leiden using a dcSBM-based quality score in which only the block-specific coefficients are adjusted. The outer loop described above is, then, applied exactly as for the SBM: each run of Leiden starts from the current best partition and the resulting BIC-dcSBM compared with the best value observed until then. After convergence of the outer loop, a final fit of the dcSBM is performed on the selected partition, by jointly updating the degree and block-specific parameters. This last step is solely used to calculate the correct value of the BIC-dcSBM associated with the final partition - which is the value reported in the analysis.

\begin{algorithm}[h!]
\caption{Community detection by minimizing BIC-dcSBM}
\begin{algorithmic}
\item[\hspace{1.4pt}1:] \textbf{function} \textit{iterative\_leiden\_dcSBM}$(G, \mathbf A, \dots)$
\item[\hspace{1.4pt}2:] \quad estimate initial degree parameters $\mathbf x$ by calling \textit{solve\_UBCM\_iterative} on the observed degree sequence]
\item[\hspace{1.4pt}3:] \quad define dcSBM BIC quality function $qf$ using $\mathbf A$ and fixed $\mathbf x$]
\item[\hspace{1.4pt}4:] \quad $\text{part} \leftarrow \text{Leiden}(G, qf, \text{initial = None})$
\item[\hspace{1.4pt}5:] \quad $\text{best\_part} \leftarrow \text{part}$, $\text{best\_bic} \leftarrow \text{BIC\_dcSBM}(\mathbf A,\text{part},\mathbf x)$]
\item[\hspace{1.4pt}6:] \quad \textbf{for} $i = 1 \dots \text{max\_outer}$ \textbf{do}]
\item[\hspace{1.4pt}7:] \quad\quad $\text{new\_part} \leftarrow \text{Leiden}(G, qf, \text{initial = best\_part})$
\item[\hspace{1.4pt}8:] \quad\quad \textbf{if} $\text{new\_part} = \text{best\_part}$ \textbf{then break}]
\item[\hspace{1.4pt}9:] \quad\quad $\text{new\_bic} \leftarrow \text{BIC\_dcSBM}(\mathbf A,\text{new\_part},\mathbf x)$]
\item[\hspace{1.4pt}10:] \quad\quad \textbf{if} $\text{new\_bic} < \text{best\_bic}$ \textbf{then}]
\item[\hspace{1.4pt}11:] \quad\quad\quad $\text{best\_part} \leftarrow \text{new\_part}$, $\text{best\_bic} \leftarrow \text{new\_bic}$
\item[\hspace{1.4pt}12:] \quad\quad \textbf{end if}]
\item[\hspace{1.4pt}13:] \quad \textbf{end for}]
\item[\hspace{1.4pt}14:] \quad compute block counts $\mathbf L_{\text{obs}}$ from $\text{best\_part}$]
\item[\hspace{1.4pt}15:] \quad obtain refined $(\mathbf x_{\text{final}},\boldsymbol\chi_{\text{final}})$ by calling \textit{solve\_dcSBM\_iterative} on $(\mathbf k,\text{best\_part},\mathbf L_{\text{obs}},\mathbf x)$]
\item[\hspace{1.4pt}16:] \quad $\text{final\_bic} \leftarrow \text{BIC\_dcSBM}(\mathbf A,\text{best\_part},\mathbf x_{\text{final}},\boldsymbol\chi_{\text{final}})$]
\item[\hspace{1.4pt}17:] \quad \textbf{return} $\text{best\_part}, \text{final\_bic}$]
\item[\hspace{1.4pt}18:] \textbf{end function}]
\end{algorithmic}
\end{algorithm}

The implementation is designed to be reproducible. All sources of randomness, including the random order in which nodes are visited and the initial partition when Leiden is seeded from random labels, are controlled by explicit seeds that can be passed to the high-level interface. Using the same seed, data, and hyperparameters produces the same sequence of partitions and the same BIC trajectory; at the same time, the code allows the user to initialize Leiden with a user-supplied partition, a modularity-based partition, or with the partition constituted by singletons. When a modularity-based initialization is used, an external modularity-maximization routine is run once to obtain a starting partition: this partition is, then, refined by the BIC-based outer iterations; when the partition constituted by singletons is used, each node starts as a community on its own and the method refines such an assignment from scratch.

\begin{algorithm}[h!]
\caption{Iterative resolution of the UBCM}
\begin{algorithmic}
\item[\hspace{1.4pt}1:] \textbf{function} \textit{solve\_UBCM\_iterative}$(\mathbf k, \mathbf x_0, \dots)$
\item[\hspace{1.4pt}2:] \quad $\mathbf x \leftarrow \mathbf x_0$, $n \leftarrow \lVert \text{residuals\_UBCM}(\mathbf x, \mathbf k) \rVert$
\item[\hspace{1.4pt}3:] \quad store $(\mathbf x, n)$ as current best solution, initialise patience counter]
\item[\hspace{1.4pt}4:] \quad \textbf{for} $i = 1 \dots \text{max\_iter}$ \textbf{do}]
\item[\hspace{1.4pt}5:] \quad\quad $\mathbf x' \leftarrow \text{root}(\text{residuals\_UBCM}, \mathbf x)$
\item[\hspace{1.4pt}6:] \quad\quad $n' \leftarrow \lVert \text{residuals\_UBCM}(\mathbf x', \mathbf k) \rVert$
\item[\hspace{1.4pt}7:] \quad\quad \textbf{if} $n' < n$ \textbf{then}]
\item[\hspace{1.4pt}8:] \quad\quad\quad update best solution to $(\mathbf x', n')$, reset patience counter]
\item[\hspace{1.4pt}9:] \quad\quad\quad $\mathbf x \leftarrow \mathbf x'$, $n \leftarrow n'$
\item[\hspace{1.4pt}10:] \quad\quad \textbf{else}]
\item[\hspace{1.4pt}11:] \quad\quad\quad decrease patience counter]
\item[\hspace{1.4pt}12:] \quad\quad \textbf{end if}]
\item[\hspace{1.4pt}13:] \quad\quad \textbf{if} $n' < \text{tol}$ or patience counter expired \textbf{then break}]
\item[\hspace{1.4pt}14:] \quad \textbf{end for}]
\item[\hspace{1.4pt}15:] \quad \textbf{return} best pair $(\mathbf x, n)$]
\item[\hspace{1.4pt}16:] \textbf{end function}]
\end{algorithmic}
\end{algorithm}

\begin{algorithm}[h!]
\caption{Iterative resolution of the dcSBM}
\begin{algorithmic}
\item[\hspace{1.4pt}1:] \textbf{function} \textit{solve\_dcSBM\_iterative}$(\mathbf k, \mathbf c, \mathbf L_{\text{obs}}, \mathbf u_0, \dots)$
\item[\hspace{1.4pt}2:] \quad $(\text{pairs}, \text{idx}) \leftarrow \text{make\_block\_struct}(\mathbf c)$
\item[\hspace{1.4pt}3:] \quad define $f(\log \mathbf u) = \text{residuals\_numba}(\log \mathbf u, \mathbf k, \mathbf c, \mathbf L_{\text{obs}}, \text{pairs}, \text{idx})$
\item[\hspace{1.4pt}4:] \quad initialise $\log \mathbf u \leftarrow \log(\mathbf u_0)$, and compute $n \leftarrow \lVert f(\log \mathbf u) \rVert$
\item[\hspace{1.4pt}5:] \quad store $(\log \mathbf u, n)$ as current best solution, initialise patience counter]
\item[\hspace{1.4pt}6:] \quad \textbf{for} $i = 1 \dots \text{max\_iter}$ \textbf{do}]
\item[\hspace{1.4pt}7:] \quad\quad compute an updated vector $\log \mathbf u'$ using a nonlinear solver applied to $f$
\item[\hspace{1.4pt}8:] \quad\quad $n' \leftarrow \lVert f(\log \mathbf u') \rVert$
\item[\hspace{1.4pt}9:] \quad\quad \textbf{if} $n' < n$ \textbf{then}]
\item[\hspace{1.4pt}10:] \quad\quad\quad update best solution to $(\log \mathbf u', n')$, reset patience counter]
\item[\hspace{1.4pt}11:] \quad\quad\quad $\log \mathbf u \leftarrow \log \mathbf u'$, $n \leftarrow n'$
\item[\hspace{1.4pt}12:] \quad\quad \textbf{else}]
\item[\hspace{1.4pt}13:] \quad\quad\quad decrease patience counter]
\item[\hspace{1.4pt}14:] \quad\quad \textbf{end if}]
\item[\hspace{1.4pt}15:] \quad\quad \textbf{if} $n' < \text{tol}$ or patience counter expired \textbf{then break}]
\item[\hspace{1.4pt}16:] \quad \textbf{end for}]
\item[\hspace{1.4pt}17:] \quad \textbf{return} $(\exp(\log \mathbf u_{\text{best}}), n_{\text{best}})$]
\item[\hspace{1.4pt}18:] \textbf{end function}]
\end{algorithmic}
\end{algorithm}

\clearpage

\hypertarget{AppD}{}
\section{Comparison with \texttt{graph-tool}}\label{AppD}

\setcounter{figure}{0}
\renewcommand\thefigure{D.\arabic{figure}}
\setcounter{table}{0}
\renewcommand\thetable{D.\Roman{table}}

To assess the robustness of the partitions obtained with the BIC-based inference procedure, we have compared them with the partitions inferred by the procedure introduced in~\cite{peixoto2017nonparametric} and implemented via \texttt{graph-tool}. Such a comparison has been carried out for the binary, undirected networks with $N=500$, separately for the SBM and the dcSBM. For each network, we report the number of clusters inferred by the BIC-based procedure and the MDL-based one, together with three measures of similarity, i.e. the Normalized Mutual Information (NMI), the Rand Index (RI) and the Jaccard Index (JI): values closer to $1$ indicate a stronger agreement between the compared partitions (see Appendix~\ref{AppF}).

\begin{figure}[t!]
\centering
\includegraphics[width=\textwidth]{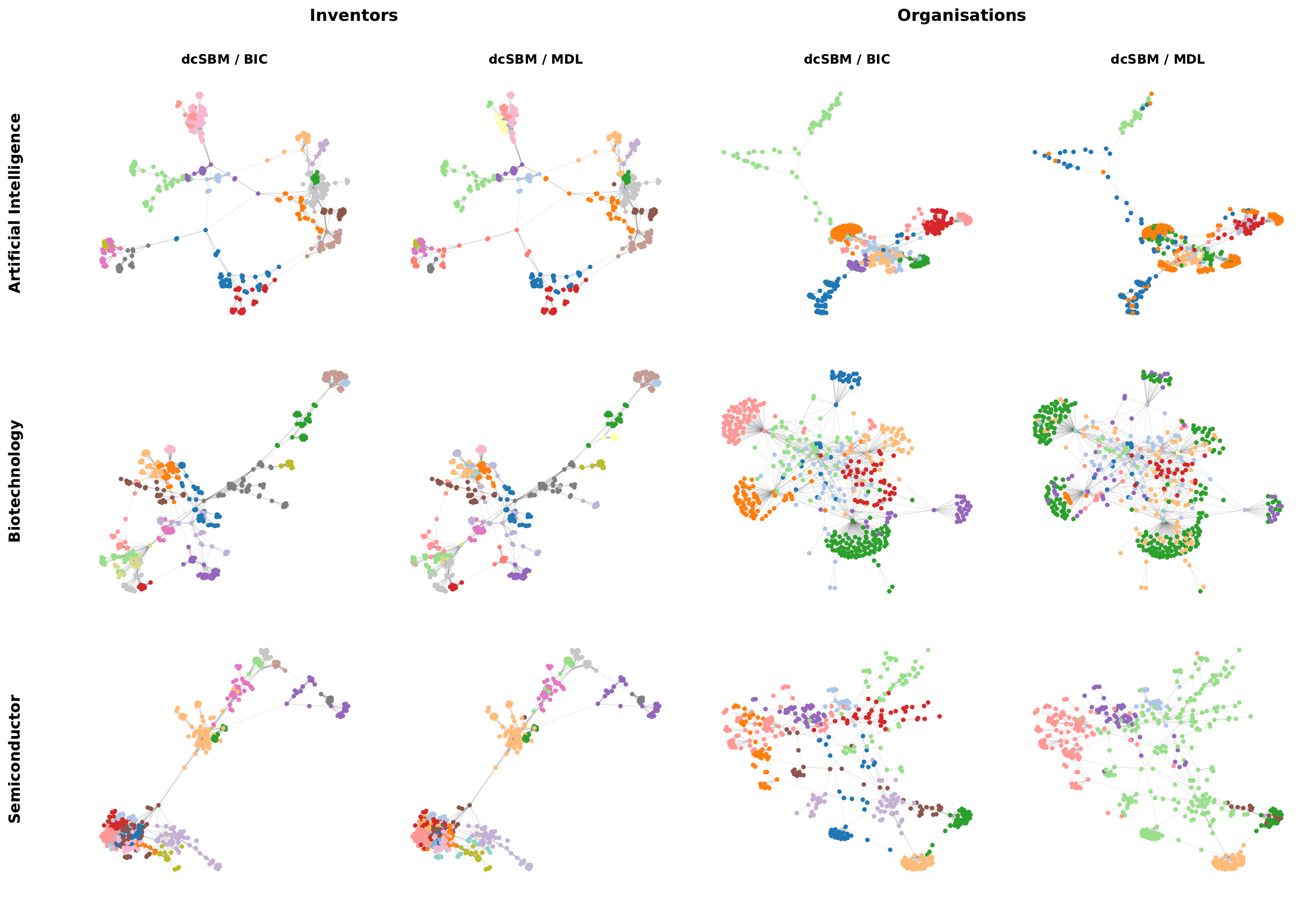}
\caption{\textbf{Comparison between the partitions induced by BIC-dcSBM and MDL.} While rows correspond to sectors, columns compare the partitions of inventor and organization networks detected by the BIC-SBM (left) and MDL. Node colors indicate the detected clusters (colors are aligned whereas the clusters highly overlap). The comparison reveals a substantial agreement between the two algorithms.}
\label{fig:dcsbm_bic_mdl_comparison}
\end{figure}

As shown in Figs.~\ref{fig:sbm_bic_mdl_comparison} and~\ref{fig:dcsbm_bic_mdl_comparison} and reported in Table~\ref{tab:bic_mdl_microcanonical_sbm}, the BIC-based and MDL-based approaches lead to broadly consistent partitions, especially for what concerns inventor networks: this is particularly evident upon looking at the values of the RI, accounting for the number of true positives and true negatives, i.e. the number of pairs of nodes that should and should not be gathered; still the MDL-based procedure individuates more clusters than the BIC-based one and the JI is sensitive to such differences - although minimal, typically concerning few groups of peripheral nodes.

However, we suspect that \texttt{graph-tool} may perform only sub-optimally on certain organization networks, as the degree-corrected variant of the SBM implemented there happens to put the `leaves' of a number of distinct trees together.

\begin{figure}[t!]
\centering
\includegraphics[width=\textwidth]{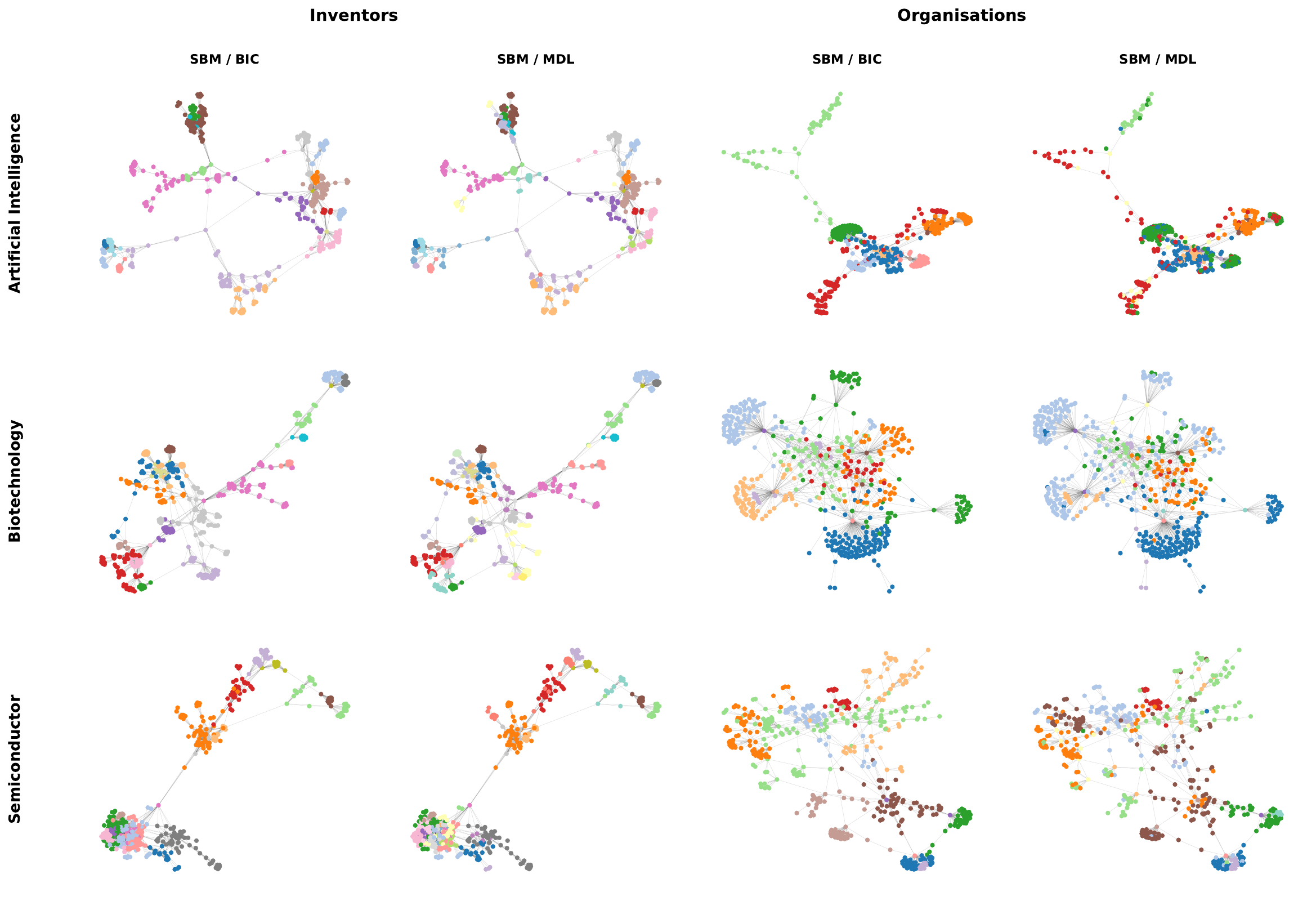}
\caption{\textbf{Comparison between the partitions induced by BIC-SBM and MDL.} While rows correspond to sectors, columns compare the partitions of inventor and organization networks detected by the BIC-SBM (left) and MDL. Node colors indicate the detected clusters (colors are aligned whereas the clusters highly overlap). The comparison reveals a substantial agreement between the two algorithms.}
\label{fig:sbm_bic_mdl_comparison}
\end{figure}

\clearpage

\begin{table}[b!]
\centering
\small
\begin{tabular}{c|c|c|c|c|c|c|c}
\hline
\hline
\textbf{Model} & \textbf{Level} & \textbf{Sector} & \textbf{\# BIC clusters} & \textbf{\# MDL clusters} & \textbf{NMI} & \textbf{RI} & \textbf{JI} \\
\hline
\hline
\multirow{6}{*}{SBM}
& \multirow{3}{*}{Inventors}
&   AI & 20 & 27 & 0.867 & 0.963 & 0.545 \\
& & BT & 19 & 31 & 0.881 & 0.966 & 0.539 \\
& & SC & 17 & 27 & 0.809 & 0.954 & 0.457 \\
\cline{2-8}
& \multirow{3}{*}{Organizations}
&   AI & 12 & 11 & 0.488 & 0.808 & 0.262 \\
& & BT & 14 & 13 & 0.453 & 0.769 & 0.281 \\
& & SC & 12 & 15 & 0.533 & 0.834 & 0.230 \\
\hline
\multirow{6}{*}{dcSBM}
& \multirow{3}{*}{Inventors}
&   AI & 17 & 22 & 0.890 & 0.975 & 0.665 \\
& & BT & 18 & 22 & 0.888 & 0.974 & 0.620 \\
& & SC & 19 & 22 & 0.827 & 0.963 & 0.580 \\
\cline{2-8}
& \multirow{3}{*}{Organizations}
&   AI & 9  & 12 & 0.456 & 0.786 & 0.201 \\
& & BT & 9  & 12 & 0.321 & 0.718 & 0.170 \\
& & SC & 11 &  7 & 0.627 & 0.793 & 0.269 \\
\hline
\hline
\end{tabular}
\caption{\textbf{Comparison between the partitions induced by BIC and MDL.} For each level, sector and model, the table compares the number of clusters inferred by the BIC-based procedure with the number of clusters inferred by the MDL-based procedure introduced in~\cite{peixoto2017nonparametric} and implemented via \texttt{graph-tool}. The similarity of partitions is quantified via the Normalized Mutual Information (NMI), the Rand Index (RI) and the Jaccard Index (JI): values closer to $1$ indicate a stronger agreement between the compared partitions.}
\label{tab:bic_mdl_microcanonical_sbm}
\end{table}

\clearpage

\hypertarget{AppE}{}
\section{Information criteria}\label{AppE}

Following~\cite{gallo2024assessing}, one can adopt an `agnostic' attitude and explore the mesoscale organisation of a network without aligning with any, specific, conceptual framework: a principled approach to achieve such a goal is that of considering the \textit{marginal likelihood}, reading

\begin{align}
P(\mathbf{G}|M)&=\int P(\mathbf{G},\bm{\theta}|M)d\bm{\theta}=\int P(\mathbf{G}|\bm{\theta},M)\pi(\bm{\theta}|M)d\bm{\theta};
\end{align}
several possibilities exist to make the expression above (also named \textit{evidence}), or its logarithm, explicit.

\subsection{Instantiating the marginal likelihood with the conjugate prior}

A first choice concerns the prior. Hereby, we are going to substitute $\pi(\bm{\theta}|M)$ with the conjugate prior. For what concerns the SBM, $\bm{\theta}=\bm{p}$ and $\pi(\bm{p}|\text{SBM})=\prod_{r=1}^B\prod_{s(\geq r)}p_{rs}^{\gamma_{rs}-1}(1-p_{rs})^{\delta_{rs}-1}/\text{B}(\gamma_{rs},\delta_{rs})$ which, in turn, leads to the expression

\begin{align}
P(\mathbf{G}|\text{SBM})&=\int P(\mathbf{G}|\bm{\theta},\text{SBM})\pi(\bm{\theta}|\text{SBM})d\bm{\theta}\nonumber\\
&=\prod_{r=1}^B\prod_{s(\geq r)}\int_0^1 p_{rs}^{L_{rs}}(1-p_{rs})^{V_{rs}-L_{rs}}\cdot\frac{p_{rs}^{\gamma_{rs}-1}(1-p_{rs})^{\delta_{rs}-1}}{\text{B}(\gamma_{rs},\delta_{rs})}dp_{rs}\nonumber\\
&=\prod_{r=1}^B\prod_{s(\geq r)}\int_0^1\frac{p_{rs}^{L_{rs}+\gamma_{rs}-1}(1-p_{rs})^{V_{rs}-L_{rs}+\delta_{rs}-1}}{\text{B}(\gamma_{rs},\delta_{rs})}dp_{rs}\nonumber\\
&=\prod_{r=1}^B\prod_{s(\geq r)}\frac{\text{B}(L_{rs}+\gamma_{rs},V_{rs}-L_{rs}+\delta_{rs})}{\text{B}(\gamma_{rs},\delta_{rs})},
\end{align}
which can be made explicit upon instantiating the hyperparameters as follows

\begin{align}
P(\mathbf{G}|\text{SBM})=\prod_{r=1}^B\prod_{s(\geq r)}\frac{\text{B}(L_{rs}+1,V_{rs}-L_{rs}+1)}{\text{B}(1,1)}.
\end{align}

For what concerns the dcSBM, instead, $\bm{\theta}=\bm{\chi}$ and $\pi(\boldsymbol{\chi}|\text{dcSBM})=\prod_{r=1}^B\prod_{s(\geq r)}\beta_{rs}^{\alpha_{rs}}\chi_{rs}^{\alpha_{rs}-1}e^{-\beta_{rs}\chi_{rs}}/\Gamma(\alpha_{rs})$ which, in turn, leads to the expression

\begin{align}
P(\mathbf{G}|\text{dcSBM})&=\int P(\mathbf{G}|\bm{\theta},\text{dcSBM})\pi(\bm{\theta}|M_\text{dcSBM})d\bm{\theta}\nonumber\\
&=\int\prod_{i=1}^N\prod_{j(>i)} p_{ij}^{a_{ij}}(1-p_{ij})^{1-a_{ij}}\cdot\pi(\bm{\theta}|M_\text{dcSBM})d\bm{\theta}\nonumber\\
&=\int\prod_{i=1}^N\prod_{j(>i)} \left(\frac{p_{ij}}{1-p_{ij}}\right)^{a_{ij}}(1-p_{ij})\cdot\pi(\bm{\theta}|M_\text{dcSBM})d\bm{\theta}\nonumber\\
&\simeq\int\prod_{i=1}^N\prod_{j(>i)} p_{ij}^{a_{ij}}e^{-p_{ij}}\cdot\pi(\bm{\theta}|M_\text{dcSBM})d\bm{\theta}\nonumber\\
&=\int\prod_{i=1}^N\prod_{j(>i)} \frac{p_{ij}^{a_{ij}}e^{-p_{ij}}}{a_{ij}!}\cdot\pi(\bm{\theta}|M_\text{dcSBM})d\bm{\theta}\nonumber\\
&=\int\prod_{i=1}^N\prod_{j(>i)} (x_ix_j\chi_{g_ig_j})^{a_{ij}}e^{-x_ix_j\chi_{g_ig_j}}\cdot\pi(\bm{\theta}|M_\text{dcSBM})d\bm{\theta},
\end{align}
where we have approximated each Bernoulli distribution with a Poisson distribution - since $a_{ij}\in\{0,1\}$, each factor $a_{ij}!$ is identically $1$, henceforth omitted. Consistently with the `decoupled' approximation introduced in Appendix~\ref{AppB} and implemented in Appendix~\ref{AppC}, we consider the node-specific Lagrange multipliers as determined by the UBCM and treat the corresponding fitnesses as `quenched' quantities: as a consequence, we can design priors for the block-specific parameters only. Upon rewriting the expression above as

\begin{align}
P(\mathbf{G}|\text{dcSBM})&=\int\prod_{i=1}^N\prod_{j(>i)}(x_ix_j)^{a_{ij}}e^{-\sum_{i=1}^N\sum_{j(>i)}x_ix_j\chi_{g_ig_j}}\prod_{r=1}^B\prod_{s(\geq r)}\chi_{rs}^{L_{rs}}\cdot\pi(\bm{\theta}|M_\text{dcSBM})d\bm{\theta}
\end{align}
and defining $S_{rs}=\sum_{i=1}^N\sum_{j(>i)}\delta_{g_ir}\delta_{g_js}x_ix_j$, the conjugate prior, in this case, leads to

\begin{align}
P(\mathbf{G}|\text{dcSBM})&=\int\prod_{i=1}^N\prod_{j(>i)}(x_ix_j)^{a_{ij}}\prod_{r=1}^B\prod_{s(\geq r)}\chi_{rs}^{L_{rs}}e^{-S_{rs}\chi_{rs}}\cdot\pi(\bm{\theta}|M_\text{dcSBM})d\bm{\theta}\nonumber\\
&=\prod_{i=1}^N\prod_{j(>i)}(x_ix_j)^{a_{ij}}\int_0^{+\infty}\prod_{r=1}^B\prod_{s(\geq r)}\chi_{rs}^{L_{rs}}e^{-S_{rs}\chi_{rs}}\cdot\frac{\beta_{rs}^{\alpha_{rs}}}{\Gamma(\alpha_{rs})}\chi_{rs}^{\alpha_{rs}-1}e^{-\beta_{rs}\chi_{rs}}d\chi_{rs}\nonumber\\
&=\prod_{i=1}^N\prod_{j(>i)}(x_ix_j)^{a_{ij}}\int_0^{+\infty}\prod_{r=1}^B\prod_{s(\geq r)}\frac{\beta_{rs}^{\alpha_{rs}}}{\Gamma(\alpha_{rs})}\chi_{rs}^{L_{rs}+\alpha_{rs}-1}e^{-(S_{rs}+\beta_{rs})\chi_{rs}}d\chi_{rs}\nonumber\\
&=\prod_{i=1}^N\prod_{j(>i)}(x_ix_j)^{a_{ij}}\prod_{r=1}^B\prod_{s(\geq r)}\frac{\beta_{rs}^{\alpha_{rs}}}{\Gamma(\alpha_{rs})}\int_0^{+\infty}\chi_{rs}^{L_{rs}+\alpha_{rs}-1}e^{-(S_{rs}+\beta_{rs})\chi_{rs}}d\chi_{rs}\nonumber\\
&=\prod_{i=1}^N x_i^{k_i}\prod_{r=1}^B\prod_{s(\geq r)}\frac{\beta_{rs}^{\alpha_{rs}}\Gamma(L_{rs}+\alpha_{rs})}{\Gamma(\alpha_{rs})(S_{rs}+\beta_{rs})^{L_{rs}+\alpha_{rs}}}
\end{align}
(where we have employed the relationship $\int_0^{+\infty}u^{a-1}e^{-bu}du=\Gamma(a)/b^a$). 

Unlike the SBM probabilities, the coefficients $\bm{\chi}$ are unbounded, so a uniform prior cannot be defined. We, therefore, adopt the same Gamma prior for all block pairs, by posing $\alpha_{rs}=\beta_{rs}=1$, i.e. $\pi(\chi_{rs}|\text{dcSBM})=e^{-\chi_{rs}}$ - a choice that allows the evidence to be normalized without favoring any particular block pair: under the Poisson approximation, in fact, the expected values $\chi_{rs}=1$, $\forall\:r\leq s$ are the `reference' coefficients associated with the UBCM alone.

\subsection{Instantiating the marginal likelihood with the Jeffreys prior}

Hereby, we are going to substitute $\pi(\bm{\theta}|M)$ with the Jeffreys prior, i.e. to pose

\begin{align}
\pi(\bm{\theta}|M)=\frac{\sqrt{|\mathscr{H}(\bm{\theta})|}}{\int\sqrt{|\mathscr{H}(\bm{\theta})|}d\bm{\theta}},
\end{align}
where $|\mathscr{H}(\bm{\theta})|$ indicates the determinant of the Hessian matrix of the log-likelihood function (also named \textit{Fisher Information Matrix}): in general,

\begin{align}
P(\mathbf{G}|M)&=\int e^{\ln P(\mathbf{G}|\bm{\theta},M)}\pi(\bm{\theta}|M)d\bm{\theta}\nonumber\\
&=\int e^{\mathscr{L}(\bm{\theta})}\pi(\bm{\theta}|M)d\bm{\theta}\nonumber\\
&\simeq\int e^{\mathscr{L}(\hat{\bm{\theta}})-\frac{(\bm{\theta}-\hat{\bm{\theta}})\mathbf{H}(\hat{\bm{\theta}})(\bm{\theta}-\hat{\bm{\theta}})^T}{2}}\pi(\bm{\theta}|M)d\bm{\theta}\nonumber\\
&=e^{\mathscr{L}(\hat{\bm{\theta}})}\int e^{-V\frac{(\bm{\theta}-\hat{\bm{\theta}})\mathscr{H}(\hat{\bm{\theta}})(\bm{\theta}-\hat{\bm{\theta}})^T}{2}}\pi(\bm{\theta}|M)d\bm{\theta}\nonumber\\
&=e^{\mathscr{L}(\hat{\bm{\theta}})}\int e^{-V\frac{(\bm{\theta}-\hat{\bm{\theta}})\mathscr{H}(\hat{\bm{\theta}})(\bm{\theta}-\hat{\bm{\theta}})^T}{2}}\cdot\frac{\sqrt{|\mathscr{H}(\bm{\theta})|}}{\int\sqrt{|\mathscr{H}(\bm{\theta})|}d\bm{\theta}}d\bm{\theta}\nonumber\\
&=\frac{e^{\mathscr{L}(\hat{\bm{\theta}})}}{\int\sqrt{|\mathscr{H}(\bm{\theta})|}d\bm{\theta}}\int e^{-V\frac{(\bm{\theta}-\hat{\bm{\theta}})\mathscr{H}(\hat{\bm{\theta}})(\bm{\theta}-\hat{\bm{\theta}})^T}{2}}\sqrt{|\mathscr{H}(\bm{\theta})|}d\bm{\theta}\nonumber\\
&\simeq\frac{e^{\mathscr{L}(\hat{\bm{\theta}})}}{\int\sqrt{|\mathscr{H}(\bm{\theta})|}d\bm{\theta}}\sqrt{\frac{(2\pi)^\kappa}{V^\kappa|\mathscr{H}(\hat{\bm{\theta}})|}}\sqrt{|\mathscr{H}(\hat{\bm{\theta}})|}\nonumber\\
&\simeq\frac{e^{\mathscr{L}(\hat{\bm{\theta}})}}{\int\sqrt{|\mathscr{H}(\bm{\theta})|}d\bm{\theta}}\sqrt{\left(\frac{2\pi}{V}\right)^\kappa}.
\end{align}

Maximizing the log-evidence is, thus, equivalent at minimizing the variant of the Bayesian Information Criterion (BIC) reading

\begin{equation}\label{eq:jeffreys-bic}
\text{BIC}=-2\mathscr{L}(\hat{\bm{\theta}})+2\ln\left[\int\sqrt{|\mathscr{H}(\bm{\theta})|}d\bm{\theta}\right]+\kappa\ln\left(\frac{V}{2\pi}\right)
\end{equation}
(while the first addendum proxies the \textit{accuracy} of a model with its log-likelihood, $\mathscr{L}(\hat{\bm{\theta}})$, the second and third ones proxy the \textit{complexity} of a model with the number of its parameters, $\kappa$, and the integral of the square root of the determinant of the Fisher Information Matrix, $|\mathscr{H}(\bm{\theta})|$; lastly, $V=N(N-1)/2$ accounts for the system dimensions).

\subsubsection{Stochastic Block Model (SBM)}

For what concerns the SBM,

\begin{align}
\mathscr{L}_\text{SBM}
&=\ln\left[\prod_{r=1}^B\prod_{s(\geq r)}p_{rs}^{L_{rs}}(1-p_{rs})^{V_{rs}-L_{rs}}\right]\nonumber\\
&=\sum_{r=1}^B\sum_{s(\geq r)}[L_{rs}\ln p_{rs}+(V_{rs}-L_{rs})\ln (1-p_{rs})]\nonumber\\
&=\sum_{r=1}^B\sum_{s(\geq r)}\left[L_{rs}\ln e^{-\alpha_{rs}}-V_{rs}\ln (1+e^{-\alpha_{rs}})\right]\nonumber\\
&=\sum_{r=1}^B\sum_{s(\geq r)}\left[-L_{rs}\alpha_{rs}-V_{rs}\ln (1+e^{-\alpha_{rs}})\right];
\end{align}
as a consequence,

\begin{align}
\frac{\partial\mathscr{L}_\text{SBM}}{\partial\alpha_{rs}}&=-L_{rs}+V_{rs}\frac{e^{-\alpha_{rs}}}{1+e^{-\alpha_{rs}}}=-L_{rs}+V_{rs}p_{rs},\quad\forall\:r\leq s.
\end{align}

Let us, now, evaluate the determinant of the Fisher Information Matrix. To this aim, let us consider that

\begin{align}
\frac{\partial^2\mathscr{L}_\text{SBM}}{\partial\alpha_{rs}^2}&=-V_{rs}\frac{e^{-\alpha_{rs}}}{(1+e^{-\alpha_{rs}})^2}=-V_{rs}p_{rs}(1-p_{rs}),\quad\forall\:r\leq s;
\end{align}
since different block are independent, the Fisher Information Matrix is diagonal and its determinant coincides with the product of the individual elements: upon normalizing by $V=N(N-1)/2$, we find that

\begin{equation}
|\mathscr{H}_\text{SBM}|=\prod_{r=1}^B\prod_{s(\geq r)}\frac{V_{rs}p_{rs}(1-p_{rs})}{V}=\prod_{r=1}^B\prod_{s(\geq r)}\frac{\text{Var}[L_{rs}]}{V}.
\end{equation}

The last term of BIC, thus, becomes 

\begin{align}
2\ln\left[\int\sqrt{|\mathscr{H}(\bm{\theta})|}d\bm{\theta}\right]&=2\ln\left[\int\sqrt{\prod_{r=1}^B\prod_{s(\geq r)}\frac{V_{rs}}{V}\frac{e^{-\alpha_{rs}}}{(1+e^{-\alpha_{rs}})^2}}d\bm{\alpha}\right]\nonumber\\
&=2\ln\left[\int\prod_{r=1}^B\prod_{s(\geq r)}\sqrt{\frac{V_{rs}}{V}\frac{e^{-\alpha_{rs}}}{(1+e^{-\alpha_{rs}})^2}}d\bm{\alpha}\right]\nonumber\\
&=2\ln\left[\prod_{r=1}^B\prod_{s(\geq r)}\sqrt{\frac{V_{rs}}{V}}\int_{-\infty}^{+\infty}\left(\frac{e^{-\alpha_{rs}/2}}{1+e^{-\alpha_{rs}}}\right)d\alpha_{rs}\right]\nonumber\\
&=2\ln\left[\prod_{r=1}^B\prod_{s(\geq r)}\sqrt{\frac{V_{rs}}{V}}\pi\right]\nonumber\\
&=\sum_{r=1}^B\sum_{s(\geq r)}\ln V_{rs}+2\sum_{r=1}^B\sum_{s(\geq r)}\ln\pi-\sum_{r=1}^B\sum_{s(\geq r)}\ln V,
\end{align}
causing its full expression to read

\begin{equation}\label{eq:bic1}
\text{BIC}_\text{SBM}=-2\mathscr{L}(\hat{\bm{\theta}})+\sum_{r=1}^B\sum_{s(\geq r)}\ln V_{rs}+\frac{B(B+1)}{2}\ln\left(\frac{\pi}{2}\right).
\end{equation}

\subsubsection{Degree-corrected Stochastic Block Model (dcSBM)}

For what concerns the dcSBM,

\begin{align}
\mathscr{L}_\text{dcSBM}
&=\ln\left[\prod_{i=1}^N\prod_{j(>i)}p_{ij}^{a_{ij}}(1-p_{ij})^{1-a_{ij}}\right]\nonumber\\
&=\sum_{i=1}^N\sum_{j(>i)}[a_{ij}\ln p_{ij}+(1-a_{ij})\ln(1-p_{ij})]\nonumber\\
&=\sum_{i=1}^N\sum_{j(>i)}[a_{ij}\ln(x_ix_j\chi_{g_ig_j})-\ln(1+x_ix_j\chi_{g_ig_j})]\nonumber\\
&=\sum_{i=1}^N\sum_{j(>i)}a_{ij}\ln(x_ix_j)+\sum_{i=1}^N\sum_{j(>i)}a_{ij}\ln\chi_{g_ig_j}-\sum_{i=1}^N\sum_{j(>i)}\ln(1+x_ix_j\chi_{g_ig_j})\nonumber\\
&=\sum_{i=1}^N\sum_{j(>i)}a_{ij}\ln(x_ix_j)+\sum_{i=1}^N\sum_{j(>i)}\sum_{r=1}^B\sum_{s(\geq r)}\delta_{g_ir}\delta_{g_js}[a_{ij}\ln\chi_{rs}-\ln(1+x_ix_j\chi_{rs})]\nonumber\\
&=-\sum_{i=1}^N\sum_{j(>i)}a_{ij}(\alpha_i+\alpha_j)-\sum_{i=1}^N\sum_{j(>i)}\sum_{r=1}^B\sum_{s(\geq r)}\delta_{g_ir}\delta_{g_js}\left[a_{ij}\alpha_{rs}+\ln\left(1+e^{-(\alpha_i+\alpha_j+\alpha_{rs})}\right)\right]\nonumber\\
&=-\sum_{i=1}^N\sum_{j(>i)}\sum_{r=1}^B\sum_{s(\geq r)}\delta_{g_ir}\delta_{g_js}\left[a_{ij}(\alpha_i+\alpha_j+\alpha_{rs})+\ln\left(1+e^{-(\alpha_i+\alpha_j+\alpha_{rs})}\right)\right];
\end{align}
as a consequence,

\begin{align}
\frac{\partial\mathscr{L}_\text{dcSBM}}{\partial\alpha_i}&=-k_i+\sum_{j(\neq i)}\sum_{r=1}^B\sum_{s(\geq r)}\delta_{g_ir}\delta_{g_js}\frac{e^{-(\alpha_i+\alpha_j+\alpha_{rs})}}{1+e^{-(\alpha_i+\alpha_j+\alpha_{rs})}}=-k_i+\sum_{j(\neq i)}\sum_{r=1}^B\sum_{s(\geq r)}\delta_{g_ir}\delta_{g_js}p_{ij},\quad i=1\dots N\\
\frac{\partial\mathscr{L}_\text{dcSBM}}{\partial\alpha_{rs}}&=-L_{rs}+\sum_{i=1}^N\sum_{j(>i)}\delta_{g_ir}\delta_{g_js}\frac{e^{-(\alpha_i+\alpha_j+\alpha_{rs})}}{1+e^{-(\alpha_i+\alpha_j+\alpha_{rs})}}=-L_{rs}+\sum_{i=1}^N\sum_{j(>i)}\delta_{g_ir}\delta_{g_js}p_{ij},\quad\forall\:r\leq s;
\end{align}
let us, now, evaluate the determinant of the Fisher Information Matrix: to this aim, let us consider that its non-null entries read

\begin{align}
\frac{\partial^2\mathscr{L}_\text{dcSBM}}{\partial\alpha_i^2}&=-\sum_{j(\neq i)}\sum_{r=1}^B\sum_{s(\geq r)}\delta_{g_ir}\delta_{g_js}\frac{e^{-(\alpha_i+\alpha_j+\alpha_{rs})}}{[1+e^{-(\alpha_i+\alpha_j+\alpha_{rs})}]^2}=-\sum_{j(\neq i)}\sum_{r=1}^B\sum_{s(\geq r)}\delta_{g_ir}\delta_{g_js}p_{ij}(1-p_{ij}),\quad i=1\dots N\\
\frac{\partial^2\mathscr{L}_\text{dcSBM}}{\partial\alpha_i\partial\alpha_j}&=-\sum_{r=1}^B\sum_{s(\geq r)}\delta_{g_ir}\delta_{g_js}\frac{e^{-(\alpha_i+\alpha_j+\alpha_{rs})}}{[1+e^{-(\alpha_i+\alpha_j+\alpha_{rs})}]^2}=-\sum_{r=1}^B\sum_{s(\geq r)}\delta_{g_ir}\delta_{g_js}p_{ij}(1-p_{ij}),\quad\forall\:i<j\\
\frac{\partial^2\mathscr{L}_\text{dcSBM}}{\partial\alpha_i\partial\alpha_{rs}}&=-\sum_{j(\neq i)}\delta_{g_ir}\delta_{g_js}\frac{e^{-(\alpha_i+\alpha_j+\alpha_{rs})}}{[1+e^{-(\alpha_i+\alpha_j+\alpha_{rs})}]^2}=-\sum_{j(\neq i)}\delta_{g_ir}\delta_{g_js}p_{ij}(1-p_{ij}),\quad i=1\dots N,\:\forall\:r\leq s\\
\frac{\partial^2\mathscr{L}_\text{dcSBM}}{\partial\alpha_{rs}^2}&=-\sum_{i=1}^N\sum_{j(>i)}\delta_{g_ir}\delta_{g_js}\frac{e^{-(\alpha_i+\alpha_j+\alpha_{rs})}}{[1+e^{-(\alpha_i+\alpha_j+\alpha_{rs})}]^2}=-\sum_{i=1}^N\sum_{j(>i)}\delta_{g_ir}\delta_{g_js}p_{ij}(1-p_{ij}),\quad\forall\:r\leq s.
\end{align}

The Fisher Information Matrix is, however, singular: for any set of positive constants $\{c_r\}_{r=1}^B$, in fact, the transformations\footnote{This is the logistic counterpart of the block-wise scaling ambiguity established for the multiplicative dcSBM in~\cite{ParkZhaoHao2025}.} defined by $x_i\rightarrow c_{g_i}x_i$ and $\chi_{rs}\rightarrow \chi_{rs}/c_rc_s$ leaves each coefficient $p_{ij}$ unaltered. For each non-empty block, this invariance identifies one direction along which the likelihood remains constant; since the $B$ transformations are linearly independent (their node-specific components have disjoint non-empty supports), the Fisher Information Matrix has, at least, $B$ null eigenvalues and $|\mathscr H_\text{dcSBM}|=0$. Such a result is consistent with the general equivalence, under regularity conditions, between the local identifiability of a model and the non-singularity of its Fisher Information Matrix\footnote{As a numerical check, we considered the standard BIC-dcSBM partitions of the AI co-inventorship, SC co-inventorship, and SC co-ownership networks: the number of null eigenvalues of the Fisher Information Matrix matched exactly the corresponding number of groups.}~\cite{Rothenberg1971}.

Consistently with the `decoupled' approximation introduced in Appendix~\ref{AppB} and implemented in Appendix~\ref{AppC}, we consider the node-specific Lagrange multipliers as determined by the UBCM and treat the corresponding fitnesses as `quenched' quantities during optimization: excluding the derivatives with respect to $\alpha_i$ leads to a diagonal Fisher Information Matrix, the determinant of which coincides with the product of the individual elements, i.e.

\begin{align}
\frac{\partial^2\mathscr{L}_\text{dcSBM}}{\partial\alpha_{rs}^2}&=-\sum_{i=1}^N\sum_{j(>i)}\delta_{g_ir}\delta_{g_js}\frac{x_ix_je^{-\alpha_{rs}}}{(1+x_ix_je^{-\alpha_{rs}})^2}=-\sum_{i=1}^N\sum_{j(>i)}\delta_{g_ir}\delta_{g_js}p_{ij}(1-p_{ij}),\quad\forall\:r\leq s;
\end{align}
upon normalizing by $V=N(N-1)/2$, we find that

\begin{equation}
|\mathscr{H}_\text{dcSBM}|=\prod_{r=1}^B\prod_{s(\geq r)}\frac{\sum_{i=1}^N\sum_{j(>i)}\delta_{g_ir}\delta_{g_js}p_{ij}(1-p_{ij})}{V}=\prod_{r=1}^B\prod_{s(\geq r)}\frac{\text{Var}[L_{rs}]}{V}
\end{equation}
and the last term of BIC becomes 

\begin{align}
2\ln\left[\int\sqrt{|\mathscr{H}(\bm{\theta})|}d\bm{\theta}\right]&=2\ln\left[\int\sqrt{\prod_{r=1}^B\prod_{s(\geq r)}\frac{1}{V}\sum_{i=1}^N\sum_{j(>i)}\delta_{g_ir}\delta_{g_js}\frac{x_ix_je^{-\alpha_{rs}}}{(1+x_ix_je^{-\alpha_{rs}})^2}}d\bm{\alpha}\right]\nonumber\\
&=2\ln\left[\int\prod_{r=1}^B\prod_{s(\geq r)}\sqrt{\frac{1}{V}\sum_{i=1}^N\sum_{j(>i)}\delta_{g_ir}\delta_{g_js}\frac{x_ix_je^{-\alpha_{rs}}}{(1+x_ix_je^{-\alpha_{rs}})^2}}d\bm{\alpha}\right]\nonumber\\
&=2\ln\left[\prod_{r=1}^B\prod_{s(\geq r)}\sqrt{\frac{1}{V}}\int\sqrt{\sum_{i=1}^N\sum_{j(>i)}\delta_{g_ir}\delta_{g_js}\frac{x_ix_je^{-\alpha_{rs}}}{(1+x_ix_je^{-\alpha_{rs}})^2}}d\alpha_{rs}\right].
\end{align}

Let us, now, try to simplify the expression above. To this aim, let us consider the prescription characterizing the \textit{fitness model}~\cite{cimini2021reconstructing} and write

\begin{align}
\int\sqrt{|\mathscr{H}(\bm{\theta})|}d\bm{\theta}&\simeq\prod_{r=1}^B\prod_{s(\geq r)}\sqrt{\frac{V_{rs}}{V}}\int\left[\int\int\frac{xye^{-\alpha_{rs}}}{(1+xye^{-\alpha_{rs}})^2}\rho(x)\rho(y)dxdy\right]^{1/2}d\alpha_{rs},
\end{align}
causing its full expression to approximately read

\begin{equation}\label{eq:bic2}
\text{BIC}_\text{dcSBM}\simeq-2\mathscr{L}(\hat{\bm{\theta}})+\sum_{r=1}^B\sum_{s(\geq r)}\ln V_{rs}+\frac{B(B+1)}{2}\ln C+N\ln\left(\frac{V}{2\pi}\right).
\end{equation}
(after having noticed that the integral after the volume term reduces to a number, $C$, affecting all partitions in the same way and having posed $\ln C=1$). Although the determinant above is conditional on the `quenched' fitnesses, the term $N\ln V$ is retained since the $\{x_i\}_{i=1}^N$ remain parameters of the dcSBM - besides, they are jointly updated with the block-specific ones once the final partition is obtained (see Appendix~\ref{AppC}).

For numerical evaluation, the sums involving $\ln V_{rs}$, in Eqs.~\eqref{eq:bic1} and~\eqref{eq:bic2}, are let run solely over the block pairs with $V_{rs}>0$.

\subsection{Bayesian Information Criterion (BIC) and Information matrix-based BIC (IBIC)}

Let us, now, provide a different approximation of the log-evidence, i.e.

\begin{align}
P(\mathbf{G}|M)&=\int e^{\ln P(\mathbf{G}|\bm{\theta},M)}\pi(\bm{\theta}|M)d\bm{\theta}\nonumber\\
&=\int e^{\mathscr{L}(\bm{\theta})}\pi(\bm{\theta}|M)d\bm{\theta}\nonumber\\
&\simeq\int e^{\mathscr{L}(\hat{\bm{\theta}})-\frac{(\bm{\theta}-\hat{\bm{\theta}})\mathbf{H}(\hat{\bm{\theta}})(\bm{\theta}-\hat{\bm{\theta}})^T}{2}}\pi(\bm{\theta}|M)d\bm{\theta}\nonumber\\
&=e^{\mathscr{L}(\hat{\bm{\theta}})}\int e^{-V\frac{(\bm{\theta}-\hat{\bm{\theta}})\mathscr{H}(\hat{\bm{\theta}})(\bm{\theta}-\hat{\bm{\theta}})^T}{2}}\pi(\bm{\theta}|M)d\bm{\theta}\nonumber\\
&\simeq e^{\mathscr{L}(\hat{\bm{\theta}})}\sqrt{\frac{(2\pi)^\kappa}{V^\kappa|\mathscr{H}(\hat{\bm{\theta}})|}}\pi(\hat{\bm{\theta}}|M),
\end{align}
where the Laplace's approximation has been employed. Maximizing the log-evidence is, thus, equivalent at minimizing the variant of the Bayesian Information Criterion
(BIC) reading

\begin{equation}\label{eq:ibic}
\text{BIC}=-2\mathscr{L}(\hat{\bm{\theta}})+\ln|\mathscr{H}(\hat{\bm{\theta}})|+\kappa\ln\left(\frac{V}{2\pi}\right).
\end{equation}
(with clear meaning of the symbols).

\subsubsection{Stochastic Block Model (SBM)}

Since the determinant of the Fisher Information Matrix reads

\begin{equation}
|\mathscr{H}_\text{SBM}|=\prod_{r=1}^B\prod_{s(\geq r)}\frac{V_{rs}p_{rs}(1-p_{rs})}{V}=\prod_{r=1}^B\prod_{s(\geq r)}\frac{\text{Var}[L_{rs}]}{V},
\end{equation}
the full BIC expression becomes

\begin{align}
\text{BIC}_\text{SBM}=-2\mathscr{L}(\hat{\bm{\theta}})+\sum_{r=1}^B\sum_{s(\geq r)}\ln\left[\frac{\text{Var}[L_{rs}]}{2\pi}\right]=-2\mathscr{L}(\hat{\bm{\theta}})+\sum_{r=1}^B\sum_{s(\geq r)}\ln\left[\frac{V_{rs}p_{rs}(1-p_{rs})}{2\pi}\right]
\end{align}
(for numerical evaluation, the proper Fisher contribution is retained only for block pairs with $V_{rs}>0$ and $0<L_{rs}<V_{rs}$, otherwise it is replaced by the corresponding `plain' BIC contribution; block pairs with $V_{rs}=0$ are, instead, omitted). This expression can be upper bounded as follows

\begin{align}
\text{BIC}_\text{SBM}=-2\mathscr{L}(\hat{\bm{\theta}})+\sum_{r=1}^B\sum_{s(\geq r)}\ln V_{rs}+\sum_{r=1}^B\sum_{s(\geq r)}\ln\left[\frac{p_{rs}(1-p_{rs})}{2\pi}\right]\leq-2\mathscr{L}(\hat{\bm{\theta}})+\sum_{r=1}^B\sum_{s(\geq r)}\ln V_{rs}-\frac{B(B+1)}{2}\ln 8\pi,
\end{align}
the last term being characterized by a `minus' instead of a `plus': such an `under-correction' causes the criterion above to over-fragment the benchmarks considered here.\\

Ignoring the contribution coming from the Fisher Information Matrix leads one to recover the traditional expression of BIC, reading

\begin{align}
\text{BIC}_\text{SBM}=-2\mathscr{L}_\text{SBM}+\kappa_\text{SBM}\ln V,
\end{align}
with $\kappa_\text{SBM}=B(B+1)/2$.

\subsubsection{Degree-corrected Stochastic Block Model (dcSBM)}

Under the `quenched' approximation considered above, the determinant of the Fisher Information Matrix reads

\begin{equation}
|\mathscr{H}_\text{dcSBM}|=\prod_{r=1}^B\prod_{s(\geq r)}\frac{\sum_{i=1}^N\sum_{j(>i)}\delta_{g_ir}\delta_{g_js}p_{ij}(1-p_{ij})}{V}=\prod_{r=1}^B\prod_{s(\geq r)}\frac{\text{Var}[L_{rs}]}{V},
\end{equation}
the full BIC expression becomes

\begin{align}
\text{BIC}_\text{dcSBM}&=-2\mathscr{L}(\hat{\bm{\theta}})+\sum_{r=1}^B\sum_{s(\geq r)}\ln\left[\frac{\text{Var}[L_{rs}]}{2\pi}\right]+N\ln\left(\frac{V}{2\pi}\right)\nonumber\\
&=-2\mathscr{L}(\hat{\bm{\theta}})+\sum_{r=1}^B\sum_{s(\geq r)}\ln\left[\frac{\sum_{i=1}^N\sum_{j(>i)}\delta_{g_ir}\delta_{g_js}p_{ij}(1-p_{ij})}{2\pi}\right]+N\ln\left(\frac{V}{2\pi}\right)
\end{align}
(for numerical evaluation, the proper Fisher contribution is retained only for block pairs with $V_{rs}>0$ and $0<L_{rs}<V_{rs}$, otherwise it is replaced by the corresponding `plain' BIC contribution; block pairs with $V_{rs}=0$ are, instead, omitted). This expression can be upper bounded by terms analogous to the ones of the SBM, leading to over-fragment the corresponding benchmarks.\\

Ignoring the contribution coming from the Fisher Information Matrix one recovers the traditional expression of BIC, reading

\begin{align}
\text{BIC}_\text{dcSBM}=-2\mathscr{L}_\text{dcSBM}+\kappa_\text{dcSBM}\ln V,
\end{align}
with $\kappa_\text{dcSBM}=B(B+1)/2+N$.\\

We have also assessed the robustness of the partitions obtained by the BIC variants induced by the conjugate and Jeffreys priors: as Table~\ref{tab:bic_corrections} shows, they lead to broadly consistent partitions with the ones obtained by the `plain' BIC.

\begin{table*}[b!]
\centering
\small
\begin{tabular}{c|c|c|c|c|c|c|c|c}
\hline
\hline
\textbf{Sector} & \textbf{Level} & \textbf{Model} & \textbf{Criterion} & $B_\text{BIC}$ & $B_\text{prior}$ & \textbf{NMI} & \textbf{RI} & \textbf{JI} \\
\hline
\hline
\multirow{8}{*}{SC}
& \multirow{4}{*}{Inventors}
& \multirow{2}{*}{SBM}
& Conjugate & 17 & 21 & 0.9285 & 0.9823 & 0.7796 \\
& & & Jeffreys & 17 & 23 & 0.9540 & 0.9908 & 0.8833 \\
\cline{3-9}
& & \multirow{2}{*}{dcSBM}
& Conjugate & 17 & 25 & 0.8568 & 0.9660 & 0.5866 \\
& & & Jeffreys & 17 & 22 & 0.8939 & 0.9740 & 0.6787 \\
\cline{2-9}
& \multirow{4}{*}{Organizations}
& \multirow{2}{*}{SBM}
& Conjugate & 12 & 15 & 0.9232 & 0.9749 & 0.8064 \\
& & & Jeffreys & 12 & 19 & 0.9559 & 0.9877 & 0.9005 \\
\cline{3-9}
& & \multirow{2}{*}{dcSBM}
& Conjugate & 11 & 15 & 0.8243 & 0.9484 & 0.5228 \\
& & & Jeffreys & 11 & 15 & 0.8687 & 0.9630 & 0.6604 \\
\hline
\hline
\end{tabular}
\caption{\textbf{Comparison between the partitions induced by the variants of BIC considered here.} For the top-$500$ SC inventor and organization networks, the table compares the number of clusters inferred by the `plain' BIC with the number of clusters inferred by using the BIC variants induced by the conjugate and Jeffreys priors.}
\label{tab:bic_corrections}
\end{table*}

\clearpage

\hypertarget{AppF}{}
\section{Robustness checks}\label{AppF}

\setcounter{figure}{0}
\renewcommand\thefigure{F.\arabic{figure}}
\setcounter{table}{0}
\renewcommand\thetable{F.\Roman{table}}

To test the robustness of our results, we have replicated our analysis on the representations of our system induced by the top-$300$ and top-$700$ most connected nodes and compared them with the one carried out on the main sample.

\begin{figure*}[t!]
\centering
\includegraphics[width=\linewidth]{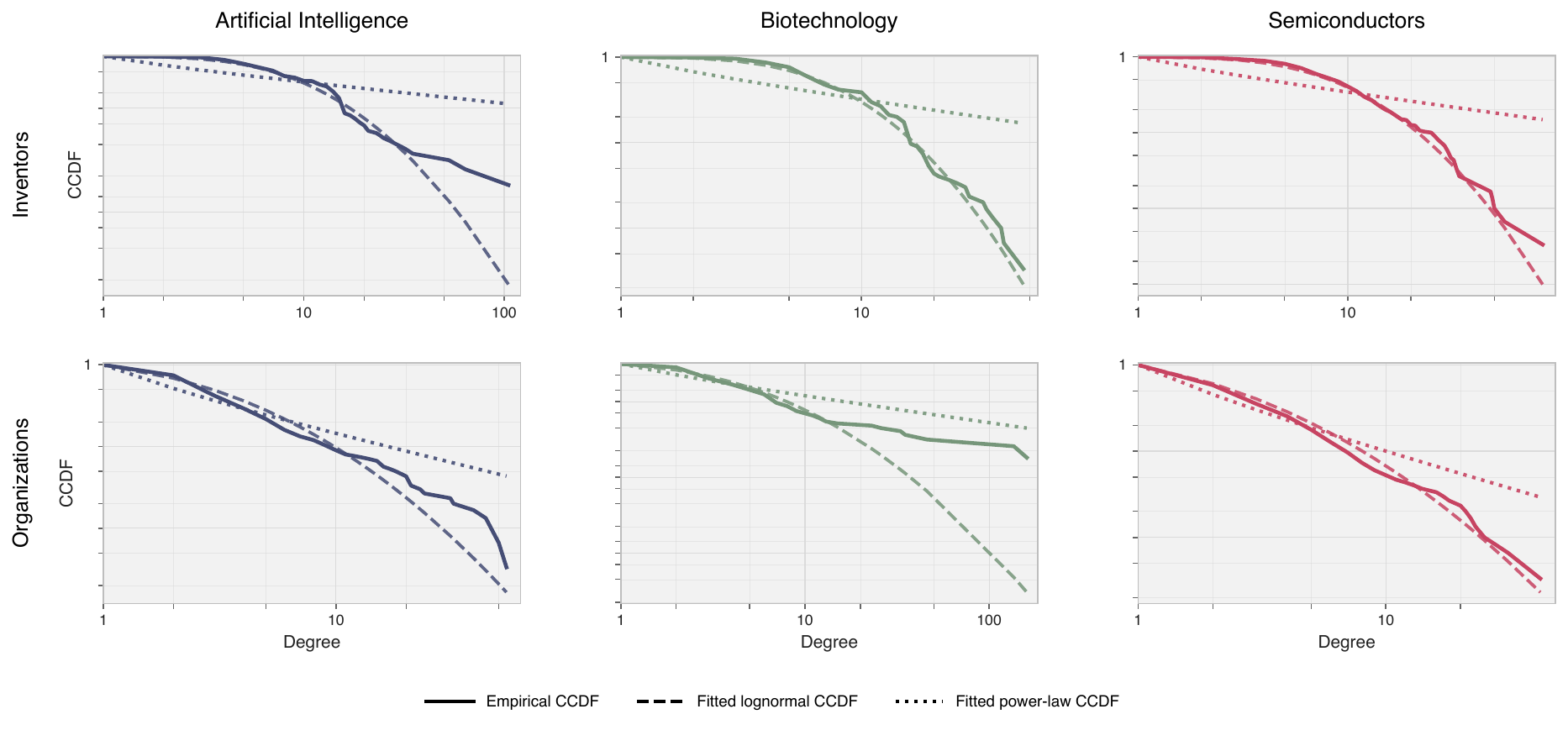}\\
\includegraphics[width=\linewidth]{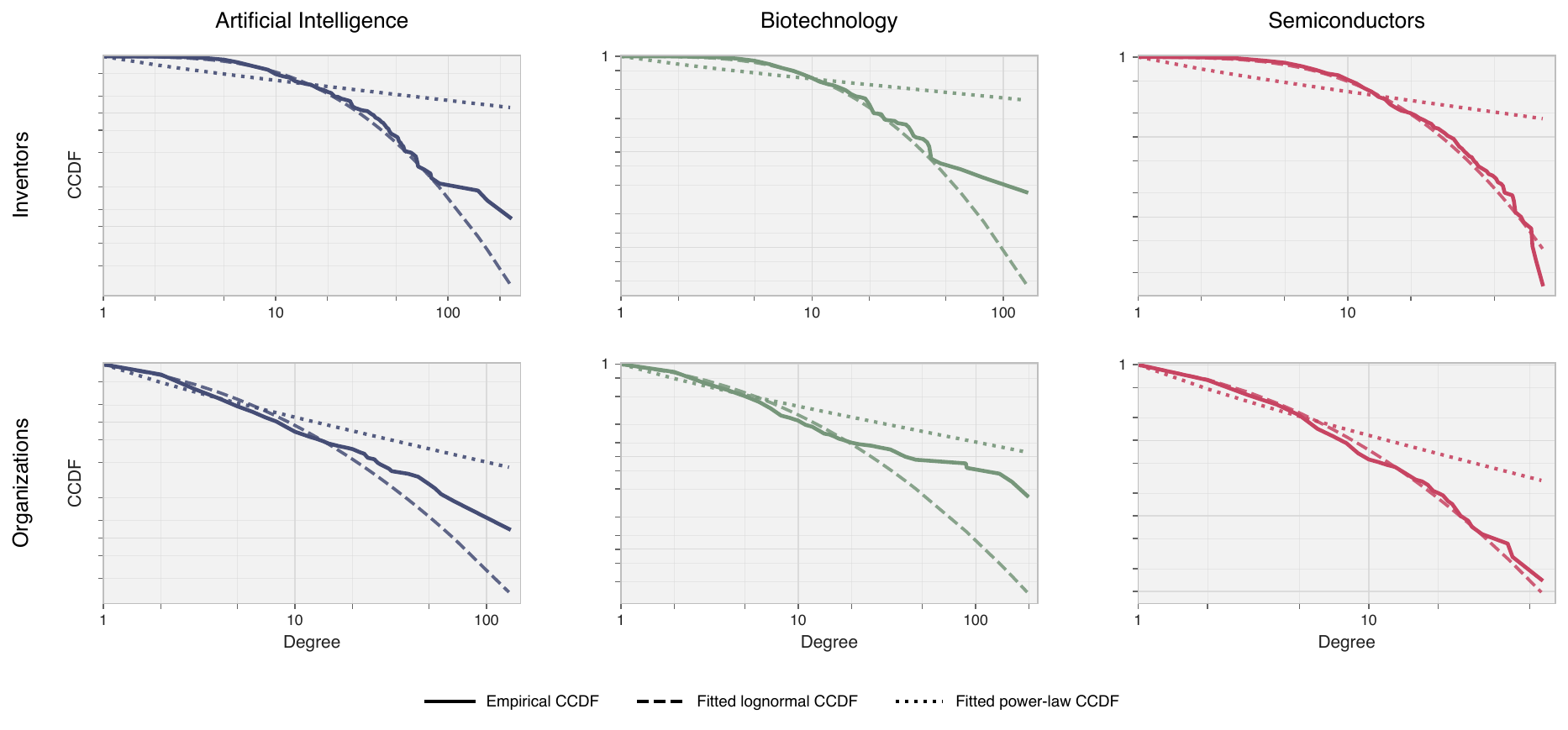}
\caption{\textbf{Node connectivity of innovation networks across sizes.} Complementary cumulative degree distributions (CCDDs) of inventor and organization networks in the three, strategic sectors of artificial intelligence, biotechnology and semiconductors (top: only the top-$300$ actors have been considered; bottom: only the top-$700$ actors have been considered). Solid lines show empirical CCDDs, while dashed and dotted lines show the fitted lognormal and power-law CCDDs, respectively. Across all levels and sectors, the patterns are consistent with the ones shown in the main text.}
\label{fig:robustness1}
\end{figure*}

\begin{figure*}[t!]
\centering
\includegraphics[width=\linewidth]{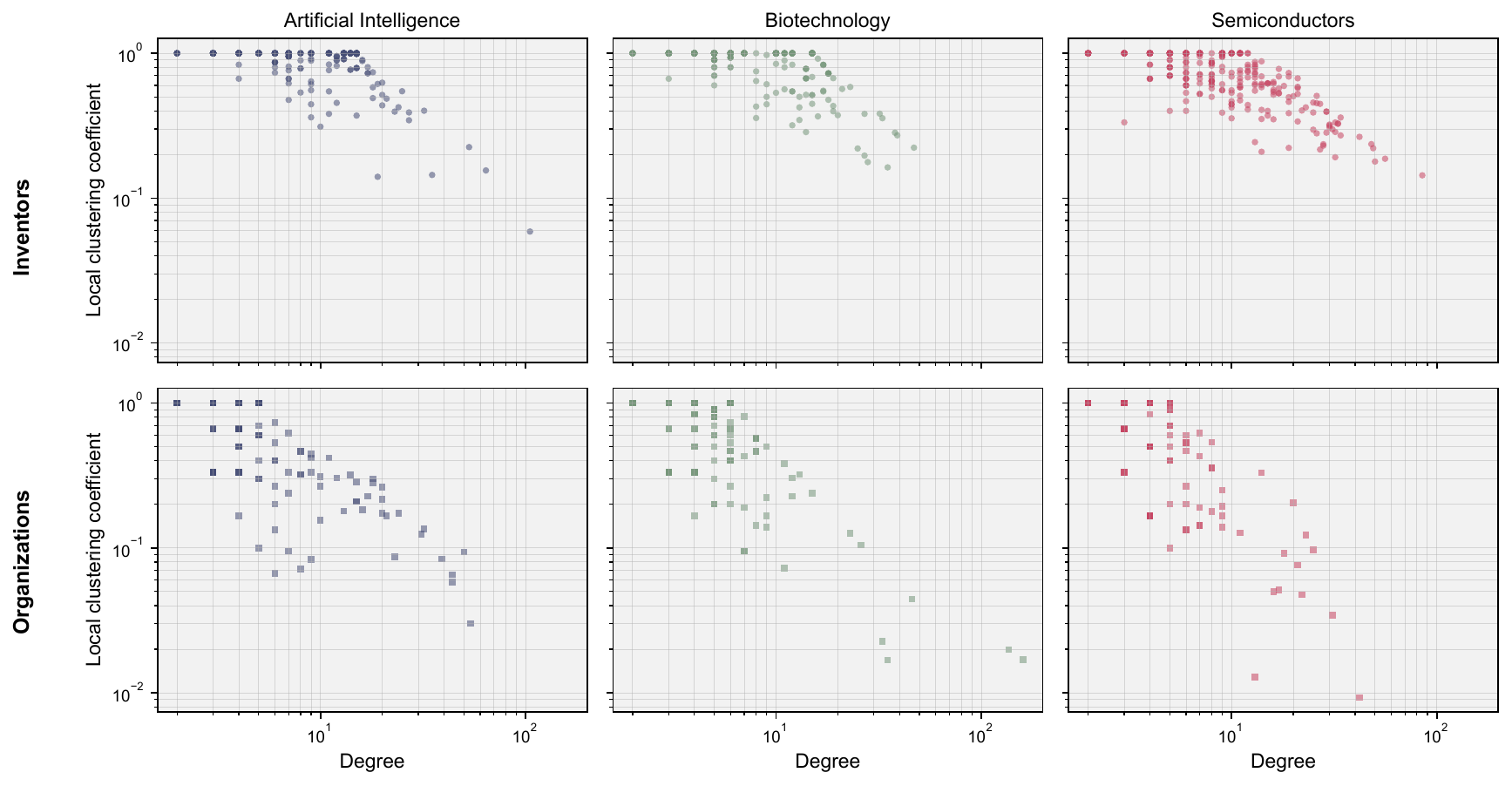}
\includegraphics[width=\linewidth]{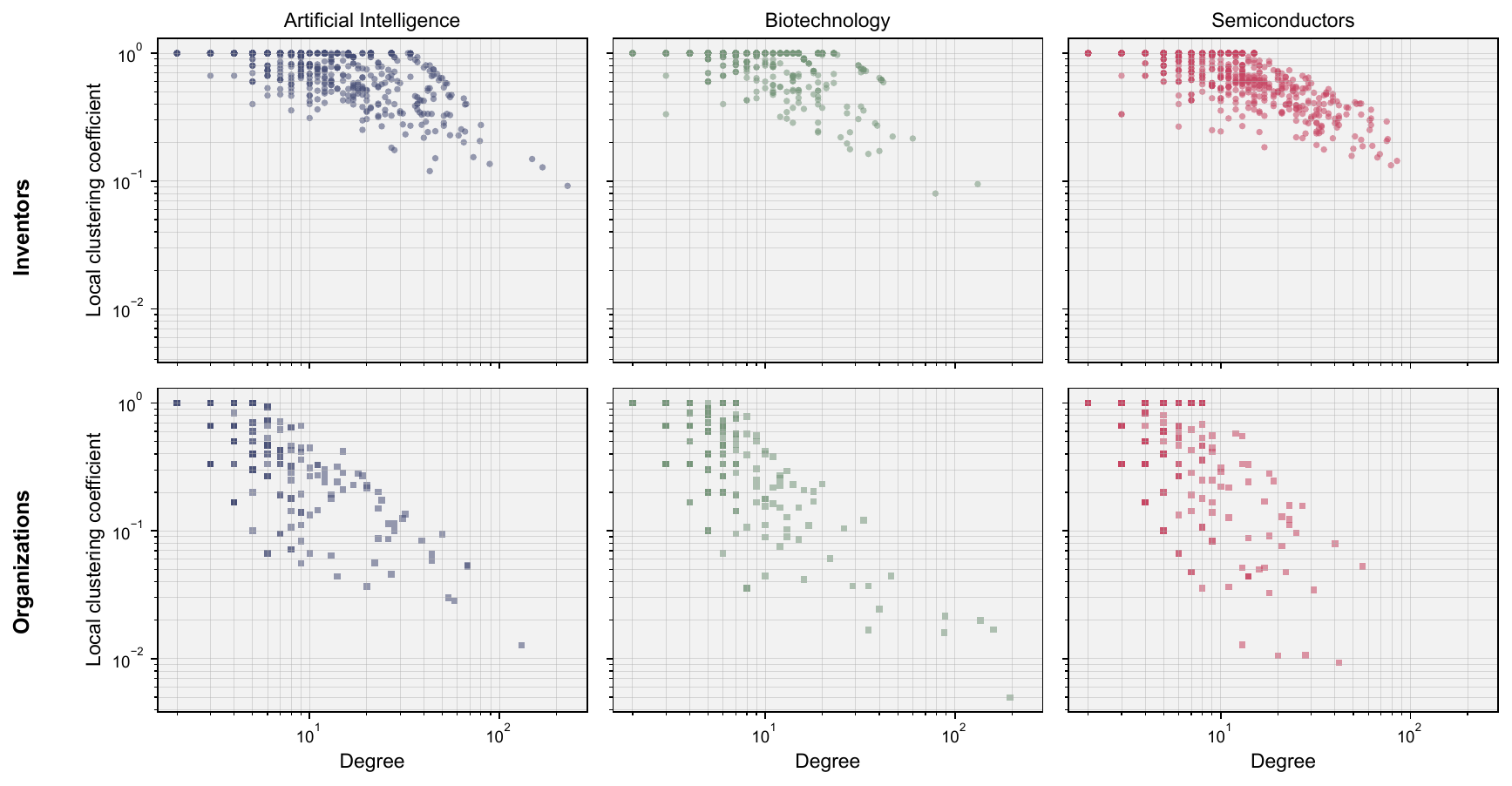}
\caption{\textbf{Local cohesion and node connectivity across sizes.} Clustering patterns of inventor and organization networks in the three, strategic sectors of artificial intelligence, biotechnology and semiconductors (top: only the top-$300$ actors have been considered; bottom: only the top-$700$ actors have been considered). Across all levels and sectors, the patterns are consistent with the ones shown in the main text.}
\label{fig:robustness2}
\end{figure*}

As Figs.~\ref{fig:robustness1} and~\ref{fig:robustness2} show, across all sectors and levels, the CCDDs preserve their heavy-tailed shape, with only minor shifts in slope due to the sample size; inventor networks are more clustered than organization networks, thus confirming that cohesive small-team structures coexist with few, highly-connected nodes; expanding from $300$ to $700$ nodes does not alter the relative patterns among sectors: AI and BT remain the most clustered, while SC exhibits the lowest cohesion. Overall, thus, the main findings are not affected by the selection of the top actors.

For what concerns the analysis of the degree distributions, Table~\ref{tab:bootstrap_pvalues_robustness} reports the bootstrap p-value, $p$, for the goodness-of-fit and the log-likelihood ratio, $R$, to compare the descriptions provided by lognormal and power-law distributions, across levels, sectors and network sizes. Two findings emerge robustly: \emph{i)} the log-likelihood-ratio consistently favors the lognormal over the power-law across all specifications; \emph{ii)} while inventor networks are compatible with lognormal distributions, for organization networks this holds true only for AI and SC.

\begin{table}[t!]
\centering
\small
\begin{tabular}{c|c|c|c|c}
\hline
\hline
\textbf{Level} & \textbf{Sector} & \textbf{Top-300} & \textbf{Top-500} & \textbf{Top-700} \\
 & & $p$\quad\quad\quad\quad$R$ & $p$\quad\quad\quad\quad$R$ & $p$\quad\quad\quad\quad$R$ \\
\hline
\hline
\multirow{3}{*}{Inventors}
& AI   & $0.198$ \quad $-20.28$ & $0.406$ \quad $-22.79$ & $0.072$ \quad $-31.31$ \\
& BT   & $0.038$ \quad $-20.24$ & $0.484$ \quad $-26.29$ & $0.280$ \quad $-29.71$ \\
& SC   & $0.208$ \quad $-17.93$ & $0.222$ \quad $-23.11$ & $0.476$ \quad $-28.45$ \\
\hline
\multirow{3}{*}{Organizations}
& AI   & $0.002$ \quad $-5.36$ & $0.176$ \quad $-3.37$ & $0.000$ \quad $-5.98$ \\
& BT   & $0.000$ \quad $-4.73$ & $0.000$ \quad $-2.58$ & $0.000$ \quad $-5.49$ \\
& SC   & $0.290$ \quad $-3.93$ & $0.148$ \quad $-5.63$ & $0.216$ \quad $-7.01$ \\
\hline
\hline
\end{tabular}
\caption{\textbf{Node connectivity of innovation networks across sizes.} For each level and sector, the table reports the bootstrap p-value, $p$, for the goodness-of-fit and the log-likelihood ratio, $R$, to compare the descriptions provided by lognormal and power-law distributions across the three network sizes; p-values below $0.05$ indicate that the lognormal hypothesis can be rejected; $R<0$ indicates that the lognormal provides a better fit than the power-law.}
\label{tab:bootstrap_pvalues_robustness}
\end{table}

To assess the similarity between partitions, we have computed the Normalized Mutual Information (NMI), the Rand Index (RI) and the Jaccard Index (JI)~\cite{di2025assessing}: the NMI~\cite{Danon2005NMI} quantifies the amount of shared information between the two partitions; the RI~\cite{Rand1971} measures the percentage of node pairs that are grouped together and separated in both partitions; the Jaccard index~\cite{Jaccard1901} solely focuses on the pairs of nodes that belong to the same cluster in both partitions. Each index ranges between $0$ and $1$, with $0$ indicating maximum difference and $1$ indicating perfect similarity.

In standard settings, these indices are used to compare a `test partition' against a `ground-truth' one. Here, however, no ground-truth is available and the indices are employed to compare any two partitions (say, $C$ and $D$) obtained from different algorithms. This naturally raises the question of whether the indices depend on the order in which the two partitions are compared. To clarify this point, it is useful to rewrite the definitions of the indices above in terms of the classical contingency counts between $C$ and $D$. Let \textit{i)} TP denote the number of node pairs placed together in both partitions, \textit{ii)} TN the number of node pairs separated in both, \textit{iii)} FP the number of node pairs that are placed together in $D$ but separated $C$, \textit{iv)} FN the number of node pairs that are put together in $C$ but separated $D$. The Rand Index, then, reads

\begin{equation}
\text{RI}(C,D)=\frac{\text{TP}+\text{TN}}{\text{TP}+\text{TN}+\text{FP}+\text{FN}},
\end{equation}
the Jaccard Index, then, reads

\begin{equation}
\text{JI}(C,D)=\frac{\text{TP}}{\text{TP}+\text{FP}+\text{FN}}
\end{equation}
and the Normalized Mutual Information, then, reads

\begin{equation}
\text{NMI}(C,D)=\frac{2\text{MI}(C,D)}{H(C)+H(D)},
\end{equation}
where $\text{MI}(C,D)=\sum_{i=1}^N\sum_{j=1}^Nh_{ij}\ln(h_{ij}/f_ig_j)$, with $H(C)=-\sum_{i=1}^Nf_i\ln f_i$ and $H(D)=-\sum_{j=1}^Ng_j\ln g_j$. Here, $h_{ij}=n_{ij}/N$ is the (joint) probability that a randomly chosen node belongs to community $i$ in $C$ and to community $j$ in $D$, with $n_{ij}$ being the size of the intersection between these two communities and $N$ being the total number of nodes; correspondingly, $f_i=n_i/N$ and $g_j=n_j/N$ are the marginal probabilities that the same node belongs to community $i$ in $C$ and to community $j$ in $D$, with $n_i$ and $n_j$ denoting their sizes~\cite{di2025assessing}. All indices are symmetric: for what concerns the RI and the JI, exchanging the roles of $C$ and $D$ simply swaps the roles of $FP$ and $FN$; for what concerns the NMI, symmetry follows from noticing that $\text{MI}(C,D)=\text{MI}(D,C)$ and $H(C)+H(D)=H(D)+H(C)$.

\begin{table}[t!]
\centering
\small
\begin{tabular}{c|c|c|ccc|ccc|ccc}
\hline
\hline
\multirow{2}{*}{\textbf{Sector}} & \multirow{2}{*}{\textbf{Level}} & \multirow{2}{*}{\textbf{Top-$N$}} & \multicolumn{3}{c|}{\textbf{dcSBM vs SBM}} & \multicolumn{3}{c|}{\textbf{dcSBM vs $Q$}} & \multicolumn{3}{c}{\textbf{SBM vs $Q$}} \\
\cline{4-12}
& & &
\textbf{NMI} & \textbf{RI} & \textbf{JI} &
\textbf{NMI} & \textbf{RI} & \textbf{JI} &
\textbf{NMI} & \textbf{RI} & \textbf{JI} \\
\hline
\hline
\multirow{6}{*}{AI}
& \multirow{3}{*}{Inventors}
&   300 & 0.874 & 0.949 & 0.543 & 0.806 & 0.907 & 0.414 & 0.803 & 0.904 & 0.403 \\
& & 500 & 0.871 & 0.966 & 0.613 & 0.888 & 0.958 & 0.587 & 0.843 & 0.948 & 0.519 \\
& & 700 & 0.824 & 0.967 & 0.497 & 0.769 & 0.938 & 0.388 & 0.761 & 0.933 & 0.368 \\
\cline{2-12}
& \multirow{3}{*}{Organizations}
&   300 & 0.629 & 0.869 & 0.375 & 0.813 & 0.932 & 0.525 & 0.662 & 0.875 & 0.378 \\
& & 500 & 0.769 & 0.923 & 0.531 & 0.794 & 0.923 & 0.536 & 0.786 & 0.918 & 0.540 \\
& & 700 & 0.622 & 0.889 & 0.302 & 0.812 & 0.954 & 0.615 & 0.633 & 0.886 & 0.291 \\
\hline
\multirow{6}{*}{BT}
& \multirow{3}{*}{Inventors}
&   300 & 0.880 & 0.960 & 0.586 & 0.909 & 0.967 & 0.683 & 0.826 & 0.936 & 0.461 \\
& & 500 & 0.863 & 0.964 & 0.564 & 0.872 & 0.960 & 0.555 & 0.862 & 0.956 & 0.549 \\
& & 700 & 0.897 & 0.975 & 0.652 & 0.857 & 0.960 & 0.528 & 0.822 & 0.951 & 0.424 \\
\cline{2-12}
& \multirow{3}{*}{Organizations}
&   300 & 0.734 & 0.895 & 0.592 & 0.709 & 0.907 & 0.618 & 0.635 & 0.871 & 0.511 \\
& & 500 & 0.650 & 0.910 & 0.500 & 0.744 & 0.932 & 0.593 & 0.707 & 0.914 & 0.544 \\
& & 700 & 0.719 & 0.937 & 0.566 & 0.710 & 0.919 & 0.500 & 0.647 & 0.906 & 0.458 \\
\hline
\multirow{6}{*}{SC}
& \multirow{3}{*}{Inventors}
&   300 & 0.712 & 0.914 & 0.384 & 0.795 & 0.903 & 0.444 & 0.739 & 0.887 & 0.407 \\
& & 500 & 0.797 & 0.950 & 0.505 & 0.794 & 0.935 & 0.499 & 0.746 & 0.916 & 0.397 \\
& & 700 & 0.771 & 0.946 & 0.382 & 0.734 & 0.901 & 0.348 & 0.709 & 0.892 & 0.291 \\
\cline{2-12}
& \multirow{3}{*}{Organizations}
&   300 & 0.735 & 0.906 & 0.534 & 0.810 & 0.928 & 0.546 & 0.761 & 0.906 & 0.469 \\
& & 500 & 0.778 & 0.914 & 0.433 & 0.771 & 0.933 & 0.441 & 0.802 & 0.924 & 0.453 \\
& & 700 & 0.817 & 0.943 & 0.505 & 0.787 & 0.938 & 0.423 & 0.798 & 0.943 & 0.483 \\
\hline
\hline
\end{tabular}
\caption{\textbf{Agreement between partitions across sectors, levels, algorithms and sizes.} For each sector, level, algorithm and size the table reports the Normalized Mutual Information (NMI), the Rand index (RI) and the Jaccard index (JI) between pairs of partitions. Across all settings, NMI and RI are typically larger than $0.70$ while JI lies in the range $[0.30,0.70]$.}
\label{tab:agreement_methods_nmi_ri_jaccard}
\end{table}

\begin{figure*}[t!]
\centering
\includegraphics[width=0.74\linewidth]{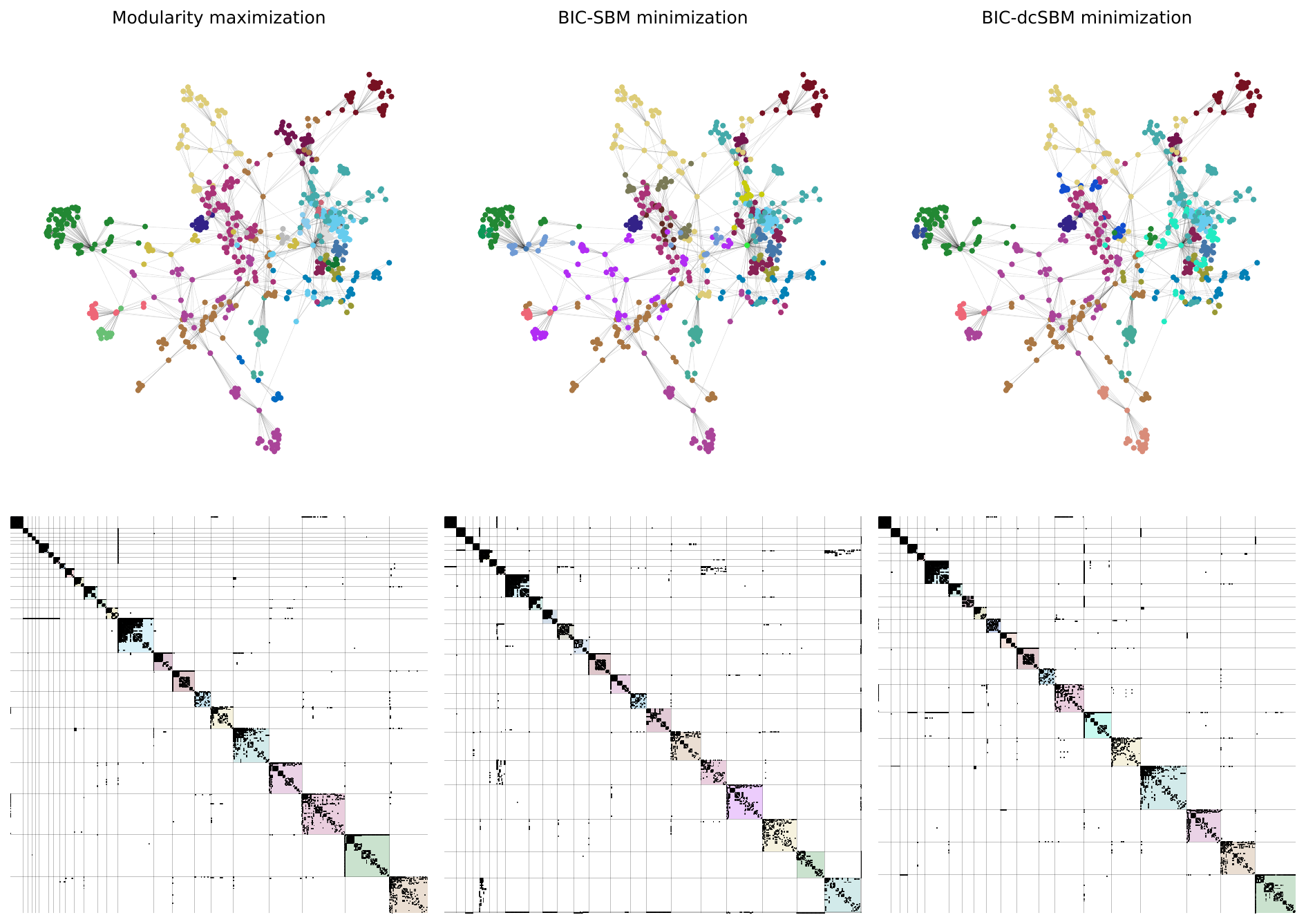}
\caption{\textbf{Mesoscale structure of the co-inventorship network in the biotech sector (top-700 actors).} The higher density and average clustering coefficient characterizing inventor networks reflect into a nested architecture, with dense modules acting as cores of other sparser ones. The patterns are consistent with the ones shown in the main text.}
\label{fig:xxx}
\end{figure*}

\begin{figure*}[t!]
\centering
\includegraphics[width=0.74\linewidth]{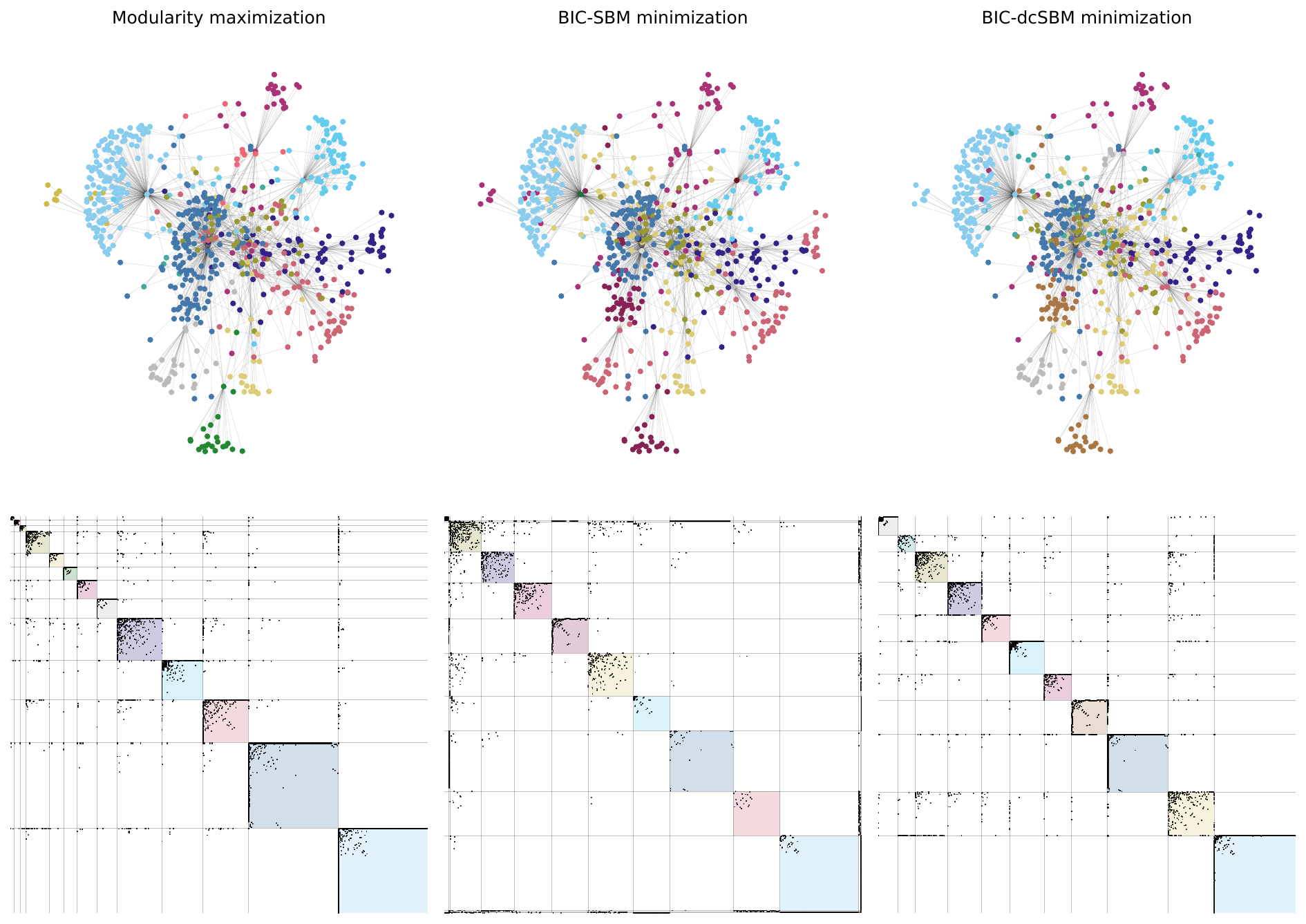}
\caption{\textbf{Mesoscale structure of the co-ownership network in the biotech sector (top-700 actors).} Differently from inventor networks, organization networks are much less modular, appearing as combinations of tree-like structures, mirroring a level of core-periphery -ness that is even more pronounced than the one characterizing inventor networks. The patterns are consistent with the ones shown in the main text.}
\label{fig:yyy}
\end{figure*}

Table~\ref{tab:agreement_methods_nmi_ri_jaccard} reports the similarity scores between the partitions retrieved by our algorithms: \textit{i)} the similarity between partitions is consistently higher at the individual level than at the organization level. Inventor networks typically exhibit RI values above $0.93$ and JI values in the range $[0.45,0.65]$, whereas organization networks frequently display lower agreement, with JI values often between $0.29$ and $0.54$. This confirms that inventor networks possess a more cohesive mesoscale structure, recognized by each algorithm, while organization networks are comparatively more diffuse, hence described differently by different algorithms; \textit{ii)} according to the JI values, the partitions obtained by minimizing the BIC-dcSBM are, overall, more similar to those returned by modularity maximization than to those obtained by minimizing the BIC-SBM. This pattern indicates that accounting for degree heterogeneity mitigates the tendency of the BIC-SBM to `favor' core-periphery structures - more specifically, to split the core-nodes and the peripheral ones, a tendency punished by the JI, that is sensitive to the number of false negatives (i.e. pairs of nodes grouped together in the first partition but separated in the second); \textit{iii)} according to the NMI values, the partitions induced by minimizing the BIC-SBM and those obtained by minimizing the BIC-dcSBM are the most similar ones. For additional evidence of the mesoscale structure of co-inventorship and co-ownership networks across sectors and levels, see Fig.~\ref{fig:xxx}, Fig.~\ref{fig:yyy} and Table~\ref{tab:clusters_bic_q_300_700}.

\begin{figure}[t!]
\centering
\includegraphics[width=\linewidth]{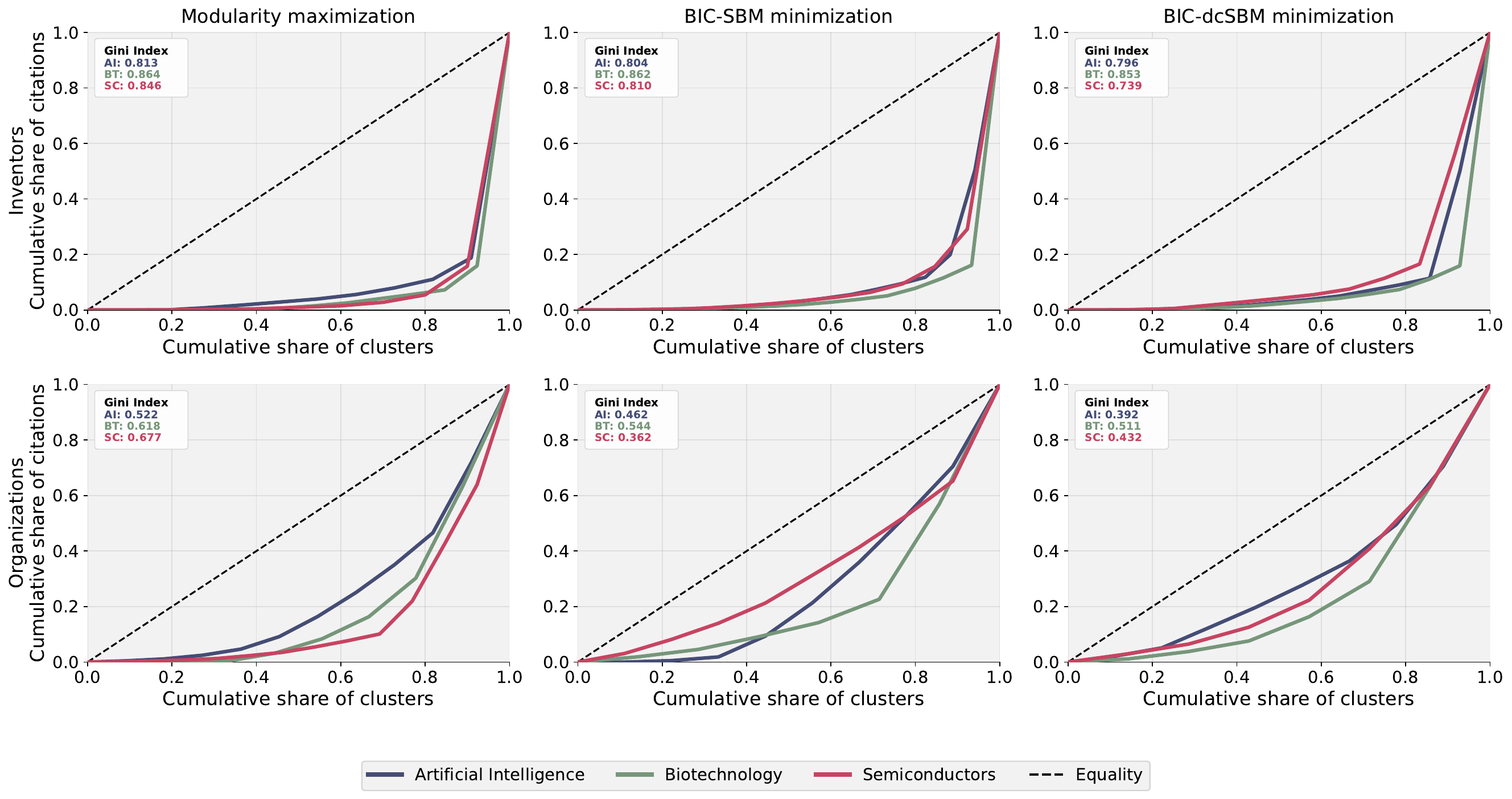}\\
\includegraphics[width=\linewidth]{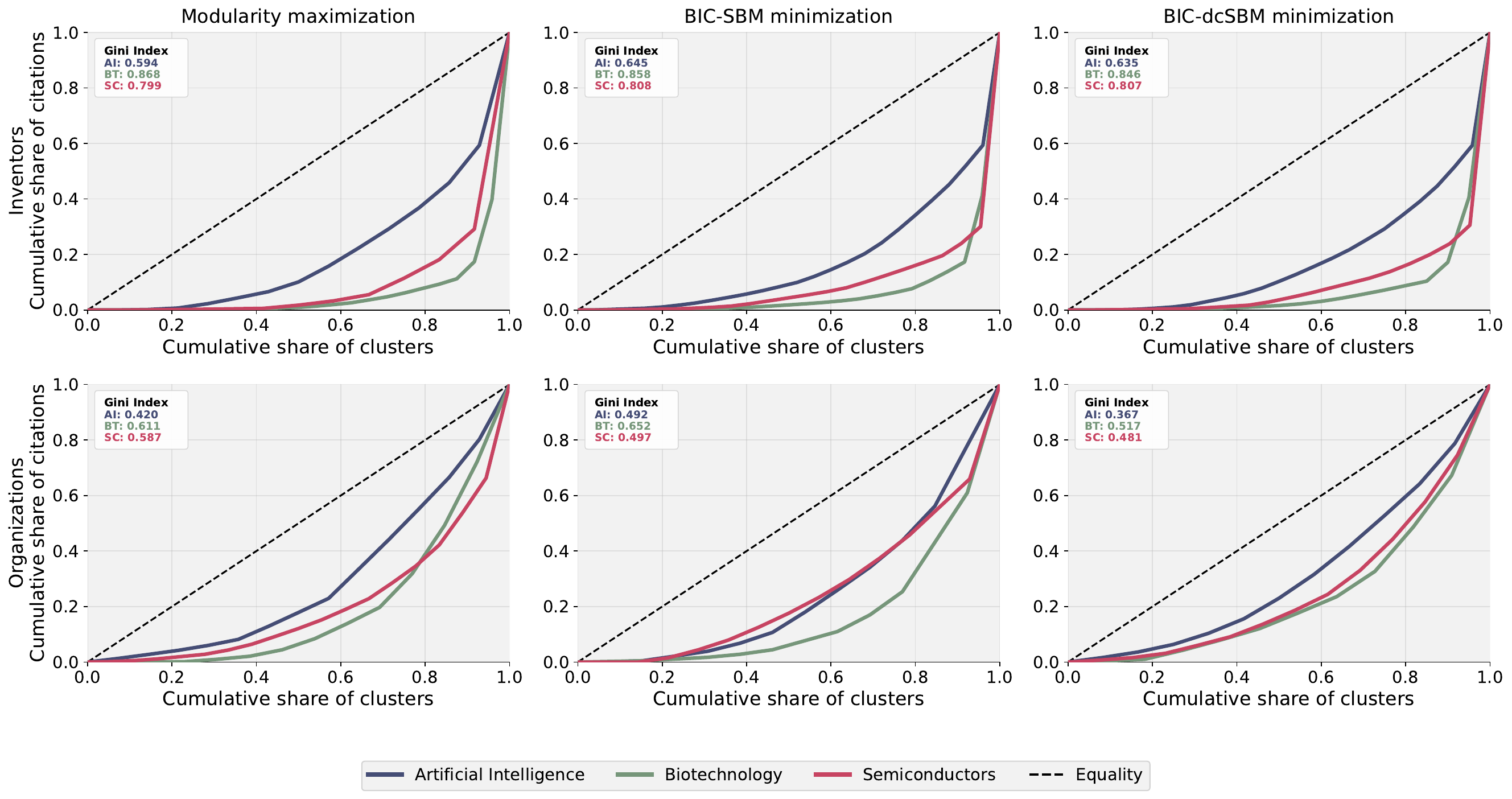}
\caption{\textbf{Inequality of the innovation impact across sizes.} Lorenz curves of forward citations for inventor and organization networks (top panels: only the top-$300$ actors have been considered; bottom panels: only the top-$700$ actors have been considered). Inequality patterns are consistent with the ones shown in the main text: inventor networks show a steadily high inequality across sectors, whereas organization networks are more diverse, with the AI sector exhibiting the highest level of inequality.\\
}
\label{fig:lorenz_robustness_300_700_1}
\end{figure}

Finally, we have repeated the calculation of the Lorenz curves. As Fig.~\ref{fig:lorenz_robustness_300_700_1} shows, the overall patterns closely match the ones shown in the main text: \textit{i)} a limited subset of communities concentrates a large share of forward citations; \textit{ii)} inventor networks are more unequal than organization networks. These patterns are fully consistent with the Gini coefficients reported in Fig.~\ref{fig:lorenz_robustness_300_700_1}.

\clearpage

\clearpage

\begin{table*}[t!]
\centering
\begin{tabular}{c|c|c|l|c|c|c|c|c|c|c}
\hline
\hline
\multicolumn{1}{c|}{\textbf{Sector}} &
\multicolumn{1}{c|}{\textbf{Level}} &
\multicolumn{1}{c|}{\textbf{Top-$N$}} &
\multicolumn{1}{c|}{\textbf{Algorithm}} &
\multicolumn{1}{c|}{\textbf{\# clusters}} &
\multicolumn{1}{c|}{$\bar{k}_{\text{within}}$} &
\multicolumn{1}{c|}{$\sigma_k^{\text{within}}$} &
\multicolumn{1}{c|}{IC/EC} &
\multicolumn{1}{c|}{\textbf{BIC-SBM}} &
\multicolumn{1}{c|}{\textbf{BIC-dcSBM}} &
\multicolumn{1}{c}{\textbf{$Q$}} \\
\hline
\hline
\multirow{12}{*}{AI}
& \multirow{6}{*}{Inv.}
& \multirow{3}{*}{300}
& Modularity maximization & 11 & 7.74 & 3.74 & 15.84 & 13055.38 & 11163.24 & 0.7724 \\
& &
& BIC-SBM minimization & 17 & 6.90 & 1.50 & 3.01 & 5506.26 & - & 0.6810 \\
& &
& BIC-dcSBM minimization  & 14 & 8.26 & 2.74 & 5.01 & - & 8057.47 & 0.7469 \\
\cline{3-11}
& \multirow{6}{*}{}
& \multirow{3}{*}{700}
& Modularity maximization & 17 & 12.78 & 16.35 & 3.61 & 35103.82 & 34159.62 & 0.6571 \\
& &
& BIC-SBM minimization    & 23 & 27.31 &  7.83 & 1.33 & 27060.18 & -        & 0.4988 \\
& &
& BIC-dcSBM minimization  & 21 & 16.61 & 15.28 & 1.56 & -        & 31075.17 & 0.5592 \\
\cline{2-11}
& \multirow{6}{*}{Org.}
& \multirow{3}{*}{300}
& Modularity maximization & 11 &  4.23 &  6.98 & 3.18 &  5653.71 &  6630.15 & 0.6248 \\
& &
& BIC-SBM minimization    &  9 & 1.48 &  1.79 & 0.50 &  4376.08 &  -       & 0.2029 \\
& &
& BIC-dcSBM minimization  &  9 &  3.29 &  4.31 & 2.89 &  -       &  6525.58 & 0.6232 \\
\cline{3-11}
& \multirow{6}{*}{}
& \multirow{3}{*}{700}
& Modularity maximization & 14 &  3.81 &  8.30 & 4.83 & 13871.12 & 17146.06 & 0.7160 \\
& &
& BIC-SBM minimization    & 13 & 17.82 &  4.57 & 0.90 & 11572.17 & -        & 0.3807 \\
& &
& BIC-dcSBM minimization  & 12 &  4.13 &  8.29 & 3.94 & -        & 17060.38 & 0.6973 \\
\hline
\multirow{12}{*}{BT}
& \multirow{6}{*}{Inv.}
& \multirow{3}{*}{300}
& Modularity maximization & 13 &  8.20 &  6.16 & 14.15 &  6734.43 &  8059.83 & 0.8220 \\
& &
& BIC-SBM minimization    & 15 & 10.60 &  4.51 &  3.58 &  5961.45 &  -       & 0.7104 \\
& &
& BIC-dcSBM minimization  & 14 &  9.07 &  5.80 &  7.45 &  -       &  7705.59 & 0.7977 \\
\cline{3-11}
& \multirow{6}{*}{}
& \multirow{3}{*}{700}
& Modularity maximization & 24 &  9.37 &  8.02 & 14.20 & 20116.88 & 23834.52 & 0.8538 \\
& &
& BIC-SBM minimization    & 24 & 19.09 &  4.76 &  5.31 & 17206.82 &        - & 0.7836 \\
& &
& BIC-dcSBM minimization  & 20 & 10.85 &  7.67 & 11.44 &        - & 22769.13 & 0.8506 \\
\cline{2-11}
& \multirow{6}{*}{Org.}
& \multirow{3}{*}{300}
& Modularity maximization &  9 &  4.18 & 12.59 &  4.02 &  5954.62 &  5741.52 & 0.5671 \\
& &
& BIC-SBM minimization    & 8 & 1.67 &  2.65 &  0.60 &  3220.73 &  -       & 0.2214 \\
& &
& BIC-dcSBM minimization  &  7 &  4.22 & 12.61 &  3.74 &  -       &  5672.67 & 0.5519 \\
\cline{3-11}
& \multirow{6}{*}{}
& \multirow{3}{*}{700}
& Modularity maximization & 13 &  4.33 & 12.25 &  3.78 & 15126.48 & 16645.30 & 0.6561 \\
& &
& BIC-SBM minimization    & 14 & 1.67 &  2.83 &  0.64 & 9795.19 & -        & 0.3008 \\
& &
& BIC-dcSBM minimization  & 11 &  4.35 & 12.25 &  2.80 & -        & 16248.27 & 0.6280 \\
\hline
\multirow{12}{*}{SC}
& \multirow{6}{*}{Inv.}
& \multirow{3}{*}{300}
& Modularity maximization & 10 &  8.99 &  9.23 &  6.13 &  9612.46 & 20906.33 & 0.6503 \\
& &
& BIC-SBM minimization    & 13 & 18.22 &  5.20 &  1.30 &  8408.14 & -        & 0.4639 \\
& &
& BIC-dcSBM minimization  & 12 & 10.79 &  9.30 &  2.43 &  -       &  9939.78 & 0.6056 \\
\cline{3-11}
& \multirow{6}{*}{}
& \multirow{3}{*}{700}
& Modularity maximization & 12 & 11.17 & 11.53 &  8.88 & 31407.89 & 50795.70 & 0.6987 \\
& &
& BIC-SBM minimization    & 22 & 16.05 &  7.67 &  1.31 & 27199.28 & -        & 0.5013 \\
& &
& BIC-dcSBM minimization  & 21 & 14.05 & 11.23 &  2.13 & -        & 31659.14 & 0.6078 \\
\cline{2-11}
& \multirow{6}{*}{Org.}
& \multirow{3}{*}{300}
& Modularity maximization & 13 &  2.88 &  4.47 &  8.04 &  4483.67 &  6309.28 & 0.7646 \\
& &
& BIC-SBM minimization    &  9 & 11.03 &  2.71 &  1.37 &  3805.29 &  -       & 0.4484 \\
& &
& BIC-dcSBM minimization  &  7 &  3.24 &  4.48 &  9.38 &  -       &  6091.20 & 0.7473 \\
\cline{3-11}
& \multirow{6}{*}{}
& \multirow{3}{*}{700}
& Modularity maximization & 18 &  3.25 &  4.71 & 11.94 & 11678.41 & 17353.10 & 0.8409 \\
& &
& BIC-SBM minimization    & 14 &  9.71 &  3.66 &  4.54 & 10770.69 &        - & 0.7349 \\
& &
& BIC-dcSBM minimization  & 13 &  3.66 &  4.73 &  9.88 &  -       & 16812.80 & 0.8279 \\
\hline
\hline
\end{tabular}
\caption{\textbf{Clusters diversity across sectors, levels, algorithms and sizes.} For each sector, level, algorithm and size the table reports the number of detected clusters, the BIC instantiated with the SBM and the dcSBM (only the relevant one for the corresponding partition, both for the partition maximizing modularity), the modularity $Q$, the within-cluster average degree $\bar{k}_{\text{within}}=R^{-1}\sum_{r=1}^R\bar{k}_r$, the within-cluster degree standard deviation $\sigma_{k}^{\text{within}}=R^{-1}\sum_{r=1}^R\sigma_k^r$ and the ratio between the partition-specific numbers of intra-cluster and inter-cluster edges IC/EC.}
\label{tab:clusters_bic_q_300_700}
\end{table*}

\clearpage

\end{document}